\documentclass[a4paper,fleqn]{cas-sc}
\usepackage{amsmath,amssymb,graphicx,amsthm,mathtools}

\usepackage{multicol}

\usepackage[numbers]{natbib}

\usepackage{dsfont}
\usepackage{diagbox}

\usepackage{colortbl}

\usepackage{xcolor}

\makeatletter
\newcommand{\Scale}[2][1]{\scalebox{#1}{$\m@th#2$}}
\makeatother

\newcommand{\sblacktriangleright}{%
  \vcenter{\hbox{\Scale[0.55]{\blacktriangleright}}}%
}

\newcommand{\graycell}{\cellcolor{gray!50}}
\definecolor{dargray}{rgb}{0.18, 0.18, 0.18}
\definecolor{darkgreen}{rgb}{0.01,0.6,0.1}
\definecolor{lightrose}{rgb}{0.996,0.75,0.793}
\definecolor{rose}{cmyk}{0.75, 0.75, 0,0}
\definecolor{winered}{rgb}{0.6,0.1,0.1}
\definecolor{lightyellow}{rgb}{1, 1, 0.6}
\definecolor{transparent}{rgb}{1,1,1}
\definecolor{lightlightgray}{rgb}{0.88, 0.88, 0.88}
\definecolor{lightgray}{rgb}{0.8, 0.8, 0.8}
\definecolor{lightblue}{rgb}{0.527,0.805,0.977}
\definecolor{lightgreen}{rgb}{.74,1,0}
\usepackage{pgfplots}
\pgfplotsset{compat=1.3}
\usepackage[framemethod=TikZ]{mdframed}
\usetikzlibrary{arrows.meta,positioning,backgrounds,patterns,calc,fit,shapes,matrix}
\definecolor{darkyellow}{rgb}{.99, .87, 0.04}
\definecolor{lightrose}{rgb}{0.996,0.75,0.793}
\definecolor{lightblue}{rgb}{0.527,0.805,0.977}
\definecolor{lightgreen}{rgb}{.74,1,0}
\tikzset{linemark/.style =   {line width= 4pt, color=#1}}
\mdfdefinestyle{ownframe}{%
    outerlinewidth=.3pt,
    roundcorner=2pt,
    innertopmargin=2pt,
    innerbottommargin=2pt,
    innerrightmargin=5pt,
    innerleftmargin=10pt,
    backgroundcolor=yellow!1!white
 }
 \mdfdefinestyle{challengeframe}{%
   outerlinewidth=.3pt,
   roundcorner=2pt,
   innertopmargin=0pt,
  innerbottommargin=6pt,
   innerrightmargin=8pt,
   innerleftmargin=10pt
 }
\usepackage{paralist}

\newtheorem{theorem}{Theorem}

\newtheorem{definition}{Definition}
\newtheorem{example}{Example}
\newtheorem{lemma}{Lemma}
\newtheorem{observation}{Observation}
\newtheorem{proposition}{Proposition}
\newtheorem{claim}{Claim}
\newtheorem{remark}{Remark}
\newtheorem{open}{Open Question}
\newtheorem{challenge}{Challenge}

\newcommand{\challengeF}[1]{
\mdfsetup{nobreak=true}%
  \begin{mdframed}[style=challengeframe]
    \begin{challenge}
      #1
    \end{challenge}
  \end{mdframed}
}

\newcommand{\decprob}[3]{%
  \medskip
  \begin{minipage}{0.94\linewidth}%
    \textsc{#1}\\
    \textbf{Input:} #2\\
    \textbf{Question:} #3
  \end{minipage}%
  \medskip  
}

\newcommand{\taskprob}[3]{%
  \medskip
  \begin{minipage}{0.94\linewidth}%
    \textsc{#1}\\
    \textbf{Input:} #2\\
    \textbf{Task:} #3
  \end{minipage}%
  \medskip  
}

\ExplSyntaxOn
\keys_set:nn { stm / mktitle } { nologo }
\ExplSyntaxOff

\usepackage{etoolbox}

\AtBeginEnvironment{figure}{}
\AtBeginEnvironment{table}{}

\usepackage{needspace}

\usepackage[linecolor=green!70!black, backgroundcolor=green!10, bordercolor=black, textsize=tiny,textwidth=1cm, disable]{todonotes}

\newcommand{\mypar}[1]{
  \smallskip

  \noindent
  \textbf{#1}
}

\newcommand{\myemph}[1]{{\color{green!40!black}\emph{#1}}}

\usepackage{scrextend}
\usepackage{subeqnarray}

\newcommand{\probname}[1]{\textsc{#1}}

\newcommand{\thetatwop}{\ensuremath{\text{P}^{\text{NP}[\log]}}}
\newcommand{\sigmatwop}{\ensuremath{\Sigma_2^{\text{P}}}}

\usepackage[noend,ruled,linesnumbered]{algorithm2e} 
\usepackage[sort&compress,nameinlink,capitalize]{cleveref}
\crefname{algorithm}{Algorithm}{Algorithms}
\crefname{lemma}{Lemma}{Lemmas}
\crefname{proposition}{Proposition}{Propositions}
\crefname{observation}{Observation}{Observations}
\crefname{figure}{Figure}{Figures}
\crefname{theorem}{Theorem}{Theorem}
\crefname{claim}{Claim}{Claims}
\crefname{table}{Table}{Tables}
\crefname{open}{Open question}{Open questions}
\crefalias{AlgoLine}{line}

\newcommand{\probSWD}{\textsc{Single-Winner Determination}}
\newcommand{\probWD}{\textsc{Winner Determination}}
\newcommand{\probManipulate}{\textsc{Voting Manipulation}}
\newcommand{\probMW}{\textsc{Multi-Winner Determination}}
\newcommand{\probResource}{\textsc{Resource Allocation}}
\newcommand{\probCoalition}{\textsc{Coalition Formation}}
\newcommand{\probLobbying}{\textsc{Lobbying}}
\newcommand{\probHedonic}{\textsc{Hedonic Games}}
\newcommand{\probMatchup}{\textsc{Matching under Preferences}}

\newcommand{\probCC}{\probname{CC-MW}}
\newcommand{\probMonroe}{\probname{Monroe-MW}}
\newcommand{\probPAV}{\probname{PAV-MW}}
\newcommand{\probMAV}{\probname{MAV-MW}}

\newcommand{\PAV}{{{PAV}}}
\newcommand{\MAV}{{{MAV}}}
\newcommand{\CC}{{CC}}
\newcommand{\Monroe}{{Monroe}}

\newcommand{\csize}{\ensuremath{k}}

\newcommand{\aaa}{\ensuremath{\mathcal{A}}}
\newcommand{\vvv}{\ensuremath{\mathcal{V}}}
\newcommand{\ppp}{\ensuremath{\mathcal{P}}}
\newcommand{\RR}{\ensuremath{\mathcal{R}}}

\newcommand{\dumA}{\ensuremath{B}}
\newcommand{\dumB}{\ensuremath{C}}

\newcommand{\enn}{\ensuremath{\hat{n}}}
\newcommand{\emm}{\ensuremath{\hat{m}}}

\newcommand{\appset}{\ensuremath{\mathsf{V}}}

\newcommand{\seq}[1]{\langle #1\rangle}
\newcommand{\rank}{\ensuremath{\mathsf{rk}}}
\newcommand{\sco}{\ensuremath{\mathsf{S}}}
\newcommand{\ballot}{\ensuremath{\mathsf{b}}}
\newcommand{\misr}{\ensuremath{\mathsf{R}}}

\newcommand{\satA}{\ensuremath{\mathsf{score}^{\mathsf{A}}}}
\newcommand{\satPAV}{\ensuremath{\mathsf{score}^{\mathsf{P}}}}
\newcommand{\satMAV}{\ensuremath{\mathsf{score}^{\mathsf{M}}}}
\newcommand{\distB}{\ensuremath{\rho^{\mathsf{B}}}}
\newcommand{\distA}{\ensuremath{\rho^{\mathsf{A}}}}

\newcommand{\distMAV}{\ensuremath{\rho^{\mathsf{M}}}}

\newcommand{\repre}{\ensuremath{\sigma}}

\newcommand{\SP}{\textsc{SP}}
\newcommand{\SC}{\textsc{SC}}

\newcommand{\assign}{\ensuremath{\sigma}}

\newcommand{\lowerB}{\ensuremath{\lfloor n/ \csize \rfloor}}

\newcommand{\betzlercite}{\cite{BetzlerMW2013}}
\newcommand{\azizcite}{\cite{Aziz15}}
\newcommand{\legrandcite}{\cite{LeGrand04}}
\newcommand{\skowroncite}{\cite{SYFE2015}}
\newcommand{\procacciacite}{\cite{procaccia2008complexity}}
\newcommand{\crcite}{\cite{CheRoy2022}}
\newcommand{\yangcite}{\cite{yang2023parameterized}}
\newcommand{\misracite}{\cite{misra2015parameterized}}

\newcommand{\liucite}{\cite{LiuGuo2016}}
\newcommand{\elkindcite}{\cite{EL2015StructureDicho}}
\newcommand{\ourcite}{\cite{chen2023efficient}}
\newcommand{\misraFPTcite}{\cite{MisraSonarVaid2017MW}}
\newcommand{\petersciteSP}{\cite{PetersLackner20}}

\newcommand{\monroekballot}{\textbf{[P\ref{prop:Monroe-W1-k+b}]}}
\newcommand{\PAVXPscoreref}{\textbf{[P\ref{prop:PAV-XP-score}]}}
\newcommand{\CCMWmn}{\textbf{[T\ref{thm:CC-fpt-m/n}]}}
\newcommand{\CCMWR}{\textbf{[T\ref{them:CC-R}]}}
\newcommand{\PAVFPTscoreballotref}{\textbf{[T\ref{prop:PAV-FPT-s+b}]}}
\newcommand{\MAVSPSC}{\textbf{[P\ref{prop:MAV-SPSC}]}}

\newcommand{\NP}{NP}
\newcommand{\NPh}{\NP-hard}
\newcommand{\PLS}{PLS}
\newcommand{\FPT}{FPT}
\newcommand{\FPTlong}{fixed-parameter tractable}
\newcommand{\PP}{P}
\newcommand{\XP}{XP}
\newcommand{\Wone}{W[1]}
\newcommand{\Woneh}{\Wone-hard}
\newcommand{\Wtwo}{W[2]}
\newcommand{\Wtwoh}{\Wtwo-hard}
\newcommand{\conp}{co\NP}

\newcommand{\weaksym}{w}

\newcommand{\tNPh}{{\color{red!60!black}\NP-h}}
\newcommand{\tNPc}{{\color{red!60!black}\NP-c}}
\newcommand{\twNPc}{{\color{red!60!black}\weaksym\NP-c}}

\newcommand{\tFPT}{{\color{green!40!black} \FPT}}
\newcommand{\tPP}{{\color{green!60!black}\PP}}
\newcommand{\tXP}{{\color{blue!60!black}\XP}}
\newcommand{\tWoneh}{{\color{red!50!blue}\Wone-h}}

\newcommand{\tWtwoh}{{\color{red!70!blue}\Wtwo-h}}

\newcommand{\tSigmac}{{\color{red!80!black}$\Sigma^P_2$-c}}

\newcommand{\numdelvot}{\ensuremath{\mathsf{t}_{\mathsf{v}}}}
\newcommand{\numdelalt}{\ensuremath{\mathsf{t}_{\mathsf{a}}}}

\newcommand{\goodG}{\ensuremath{\mathcal{F}}}

\newcommand{\FE}{\textsc{HG-FE}}
\newcommand{\ADD}{\textsc{HG-ADD}}

\newcommand{\singles}{\ensuremath{V_{\mathsf{S}}}}
\newcommand{\nonsingles}{\ensuremath{V_{\mathsf{NS}}}}

\newcommand{\Bns}{\ensuremath{B_{\mathsf{NS}}}}
\newcommand{\hgoodG}{\ensuremath{\hat{\goodG}}}

\newcommand{\hG}{\hat{G}}

\newcommand{\newresult}{\ensuremath{\heartsuit}}
\newcommand{\litresult}{$^*$}

\newcommand{\friends}{friend-appreciation}
\newcommand{\enemies}{enemy-aversion}

\newcommand{\friend}{friend}
\newcommand{\enemy}{enemy}

\newcommand{\friendapp}{\friends\ model}
\newcommand{\enemyav}{\enemies\ model}

\newcommand{\core}{\textsc{Core}\xspace}

\newcommand{\score}{\textsc{SCore}\xspace}
\newcommand{\Nash}{\textsc{Nash}\xspace}
\newcommand{\IndS}{\textsc{IndS}\xspace}

\newcommand{\IS}{\textsc{Independent Set}\xspace}

\newcommand{\Clique}{\textsc{Clique}\xspace}
\newcommand{\verif}{\textsc{Verif}}

\newcommand{\exist}{\textsc{Exist}}
\newcommand{\find}{\textsc{Search}}

\newcommand{\fas}{\ensuremath{\mathsf{f}}}
\newcommand{\maxdeg}{\ensuremath{\Delta}}

\newcommand{\maxcoal}{\ensuremath{\kappa}}
\newcommand{\fixcoal}{\ensuremath{d}}
\newcommand{\numcoal}{\ensuremath{|\Pi|}}

\newcommand{\tw}{\ensuremath{\mathbf{tw}}}

\newcommand{\AddPrefs}{Additive Preferences}
\newcommand{\addprefs}{additive preferences}

\newcommand{\agents}{\ensuremath{\mathcal{V}}}
\newcommand{\ag}[1]{\ensuremath{v_{#1}}}
\newcommand{\util}[1]{\ensuremath{\mu}_{#1}}

\newcommand{\coal}{\ensuremath{A}}
\newcommand{\bcoal}{\ensuremath{B}}

\newcommand{\aG}{\ensuremath{G^{\mathcal{A}}}}

\newcommand{\nrutil}{\ensuremath{|\util{}|}}

\usepackage{enumitem}

\tikzstyle{pn} = [draw=gray, circle, inner sep=1.5pt,fill=gray]
\tikzstyle{edc} = [arrows = {-Stealth[length=7pt, inset=2pt]}, thick]
\tikzstyle{fc} = [green!60!black, edc]
\tikzstyle{labelst} = [fill=white, inner sep = 1pt]
\tikzstyle{gadget} = [draw=black, fill=blue!20!white,rounded corners = 2pt]

\begin{document}

\title[mode = title]{Computational Social Choice:\\ Parameterized Complexity and Challenges}  
\shorttitle{Computational Social Choice: Parameterized Complexity and Challenges}

\shortauthors{Chen et al.}

\author[1]{Jiehua Chen}%

\cormark[1]

\ead{jiehua.chen@ac.tuwien.ac.at}

\ead[url]{www.ac.tuwien.ac.at/jchen}

\credit{}

\author[1]{Christian Hatschka}%

\ead{chatschka@ac.tuwien.ac.at}

\author[1]{Sofia Simola}

\ead{ssimola@ac.tuwien.ac.at}

\credit{}

\affiliation[1]{organization={Institute for Logic and Computation, TU Wien, Austria},
            city={Vienna},
            postcode={1040}, 
            country={Austria}}

\cortext[1]{Corresponding author}

\begin{abstract}
   We survey two key problems--\probMW{} and \probHedonic{}--in Computational Social Choice,
   with a special focus on their parameterized complexity,
   and propose some research challenges in the field. 
\end{abstract}


\maketitle

\section{Introduction}\todo{TODONOTES ARE ON}
Computational Social Choice (COMSOC) delves into the computational and algorithmic aspects of social choice issues arising from collective decision making.
Positioned at the intersection of computer science, economics, political and social sciences, and artificial intelligence, COMSOC exemplifies interdisciplinary research.
This breadth is reflected in its diverse publication venues.
COMSOC research regularly appears in theoretical computer science conferences (e.g., SODA, ICALP, ESA, MFCS, STACS),
artificial intelligence and multi-agent systems venues (AAAI, IJCAI, ECAI, NeurIPS, AAMAS, JAAMAS),
economics and social choice outlets (EC, SAGT, ADT, Journal of ACM-EC,  Journal of Economic Theory, Economic Theory, Social Choice and Welfare,  Review of Economic Design), 
as well as prestigious multidisciplinary journals spanning computer science and economics (Mathematics of Operations Research, Games and Economic Behavior, Artificial Intelligence, Journal of Artificial Intelligence Research, Theoretical Computer Science), to name just a few.
This remarkable diversity of research venues underscores both the theoretical richness and practical significance of computational social choice problems across multiple disciplines.

Typical problems in COMSOC include:
\begin{compactitem}
  \item \probWD: Aggregating voters' preferences to reach a consensus outcome, whether it is a single winner, multiple winners, or a consensus ranking.
  \item \probManipulate: Manipulating preferences to achieve a more favorable voting outcome.
  \item \probCoalition: Optimally dividing agents into disjoint groups (called coalitions) based on their preferences, with \probMatchup{} and \probHedonic{} being among the most popular and widely studied special cases. 
  \item \probResource: Fairly allocating a set of items (divisible or indivisible) among agents.
\end{compactitem}

These COMSOC problems naturally arise in many real-world scenarios—from political elections to resource allocation in computer networks—but solving them optimally often poses significant computational challenges. Many of these problems are NP-hard or even harder. 
For instance, \probSWD\ under Dodgson's\footnote{Named after Charles L.\ Dodgson, better known as Lewis Carroll, who wrote the famous children's book \emph{Alice's Adventures in Wonderland}.} rule and Kemeny's\footnote{Named after John Kemeny, an American mathematician and co-developer of the BASIC programming language.} rule is NP-hard~\cite{BarTovTri1989} and has been shown to be complete for the complexity class~$\thetatwop$~\cite{HHR97}.\footnote{The complexity class~$\thetatwop$ includes decision problems solvable in polynomial time with a logarithmic number of NP-oracle queries. It encompasses NP and coNP, and is contained in~$\sigmatwop$.}
The winner determination of \probLobbying, which is to determine an optimal way to manipulate multi-issue referenda, is NP-complete~\cite{CFRS07}.
Determining the existence of a diverse and stable matching is complete for the complexity class~$\sigmatwop$~\cite{ChenGanianHamm2020ijcai-diversestable}.
Determining the existence of a core stable partition in \probHedonic{} with additive preferences is $\sigmatwop$-complete as well by~\citet{WOEGINGER2013101}. %
Determining the existence of an envy-free allocation of indivisible goods is NP-complete since this problem generalizes the NP-complete \textsc{3-Partition} problem~\cite{GJ79}.

When facing such computational intractability, a natural question arises:
What aspects of these problems contribute to their hardness, and can we identify meaningful parameters that, when bounded, make these problems tractable? This question
invites the exploration of the parameterized complexity of COMSOC problems, 
which offers a more nuanced analysis than classical complexity theory by examining how specific problem parameters affect computational difficulty.

Mike Fellows and colleagues~\cite{CFRS07lobbying} pioneered this exploration with their work on \probLobbying, demonstrating it to be \Wtwo{}-complete with respect to the number of voters to lobby.
This landmark study marked the first foray into the parameterized complexity of COMSOC problems.
They also noted in their paper that the same work~\cite{BarTovTri1989} demonstrating the NP-hardness of Dodgson's rule also provides an ILP formulation, indirectly confirming its fixed-parameter tractability with respect to the number of alternatives.
This was in the late 90s, the same period when Rod Downey and Mike Fellows founded the theory of Parameterized Complexity~\cite{DowneyF99, DF13}. By now many books on parameterized complexity are available, such as those by \citet{Nie06,FlumG06}, and \citet{CyFoKoLoMaPiPiSa2015}. %

Since then, parameterized complexity analysis has flourished as a standard tool for the algorithmic complexity study of COMSOC problems. 
The first survey on fixed-parameter algorithms for COMSOC problems, specifically voting problems, was presented by Lindner and Rothe in 2008~\cite{LR08}.
Subsequently, Betzler et al.~\cite{BetBreCheNie2012} offered a more comprehensive survey on parameterized techniques for winner determination and voting manipulation, dedicated to Mike Fellows on his 60th birthday.
In honor of the 60th birthday of Jianer Chen --- another prolific researcher in parameterized complexity --- Bredereck et al.~\cite{BreCheFalGuoNieWoe2014} summarized nine research challenges in the parameterized complexity of COMSOC problems, many of which have since been addressed.
In 2016, Dorn and Schlotter co-authored a chapter~\cite{DS17} inviting more COMSOC researchers to delve into parameterized complexity.
In 2018, Gupta et al.~\cite{GRSZ18} provided a first survey on the parameterized study of matching under preferences, particularly NP-hard \textsc{Stable Marriage} variants.
In 2023, Chen and Manlove introduced kernelization to the economic community in a chapter on algorithmics for matching under preferences~\cite{ChenManlove2022}.

This article is dedicated to our friend Mike Fellows in celebration of his 70th birthday.
We take a parameterized complexity perspective on two active research topics in computational social choice (COMSOC):
\probMW\ and \probHedonic{} -- the latter being a distinctive and compelling variant of the broader \probCoalition\ problem.
We focus on these topics due to the substantial research attention they have received in recent years.
For \probMW, see the work of \citet{%
  BetzlerMW2013,yang2023parameterized,misra2015parameterized,LiuGuo2016,chen2023efficient}.
For \probHedonic, see the work of \citet{peters2016graphical,CheCsaRoySim2023Verif,durand2024enemies,hanaka2024core,hanaka2022revisited}.

Our focus on \probMW\ and \probHedonic{} is motivated by several key considerations:
\begin{compactenum}[(1)]
  \item \textbf{Research momentum:} These two areas have seen particularly intense research activity in recent years, with numerous breakthrough results that showcase diverse parameterized complexity techniques.
  The rapid development makes them ideal case studies for demonstrating %
  how parameterized complexity methods effectively address computational challenges in social choice.
  
  \item \textbf{Complementary complexity landscapes:}
  These problems are complete for different complexity classes within the polynomial hierarchy:
  While \probMW\ typically yields NP-complete optimization problems, \probHedonic{} frequently involves existence questions that reach $\Sigma_2^P$-completeness.
  By studying both problems, we demonstrate that parameterized approaches can provide tractability regardless of a problem's position in the polynomial hierarchy.

  \item \textbf{Practical significance:} Both problems have immediate applications to real-world systems. Multi-winner determination directly models committee selection and representation systems in governance, while hedonic games capture coalition formation in economic, political, and social networks where agents have preferences over group membership.
  
  \item \textbf{Structural insights:} The parameters we study for these problems—such as committee size, preference restrictions, and graph-theoretic measures—provide structural insights that extend beyond these specific domains to other COMSOC problems.
\end{compactenum}

\paragraph{Survey structure.} This survey is organized into two main parts: \probMW{} and \probHedonic.  %
\begin{compactitem} %
  \item For each of our two focus areas, we first present the formal framework and foundational concepts (Sections~\ref{sec:MWPrelim}--\ref{sec:MWRestricted} and Sections~\ref{sec:HGstab}--\ref{sec:HGClassic}).
  \item We then identify the key parameters of interest (Sections~\ref{sec:MWparam} and~\ref{sec:HGparam}).
  \item Next, we present selected parameterized complexity results with proof techniques (Sections~\ref{sec:MWParamC} and~\ref{sec:HGparamC}).
  \item Finally, we discuss additional topics, research challenges, and future directions (Sections~\ref{sec:FurtherMW} and~\ref{sec:FurtherHG}).
\end{compactitem}
Tables~\ref{table:mw},~\ref{table:additive}, and~\ref{table:FE} summarize the current state of knowledge in these domains.

\section{Multi-Winner Determination}\label{sec:mw}

Multi-winner determination (also known as committee or multi-winner election) concerns the selection of a fixed-size subset of alternatives that best reflects voters' preferences according to a specified notion of optimality; for a broader introduction to multi-winner elections, we refer the reader to \citet{faliszewski2017multiwinner} and \citet{LacknerSkownron23ABC}.
This problem arises naturally in many practical scenarios, from selecting parliamentary representatives to forming committees or shortlisting projects.
As in single-winner elections, a variety of voting rules define what ``optimal'' means, capturing principles such as proportionality, diversity, or welfare maximization. These rules often lead to significant computational challenges.

In this section, we survey the parameterized complexity of multi-winner determination problems, with a particular focus on voting rules based on misrepresentation functions and scoring methods; see \cref{table:mw} for an overview.
This section is structured as follows. 

\cref{sec:MWPrelim} introduces the formal terminology and foundational concepts, accompanied by illustrative examples.
\cref{sec:MWRestricted} discusses two prominent preference structures (single-peaked and single-crossing) that are known to reduce computational complexity. 
\cref{sec:MWparam} defines the parameters under investigation.
\cref{sec:MWParamC} presents selected proofs and proof sketches, emphasizing the techniques used to establish parameterized complexity results.
Finally, \cref{sec:FurtherMW} explores additional topics related to multi-winner determination, including other voting rules and preference structures.

\subsection{Preliminaries}\label{sec:MWPrelim}

In this subsection, we introduce the formal framework for multi-winner determination.
We define the basic concepts of alternatives, voters, preference profiles, committees, and misrepresentation functions.
We also provide illustrative examples to clarify the terminology and motivate the later complexity results.

Given a non-negative integer~$t\in \mathds{N}$, let $[t]$ denote the set~$\{1,2, \cdots, t\}$.
\paragraph{Preference Profiles.}
Let $\aaa=\{a_1,a_2,\cdots,a_m\}$ denote a finite set of $m$ alternatives, and $\vvv=\{v_1,v_2,\cdots,v_n\}$ a finite set of $n$ voters.
A \myemph{preference profile} specifies the preference orders of each voter~$v_i$ from~$\vvv$ over~$\aaa$, denote as~$\succ_i$.
We distinguish between linear and approval preferences.
In the linear preference model, every voter's preference order~$\succ$ is a \myemph{linear preference list} (aka.\ \myemph{ranking}) over the alternatives~$\aaa$.
For instance, a voting having linear preference list~$a_3\succ a_4 \succ a_2 \succ a_1$ means that he strictly prefers $a_3$ to $a_4$, $a_4$ to $a_2$, and $a_2$ to $a_1$.
In the approval preference model, every voter either approves or disapproves of each alternative, 
so his preference order~$\succ$ is a subset~$\appset_i\subseteq \aaa$ %
of alternatives that he approves of, called \myemph{approval set}. 
Note that an approval set can also be considered as a dichotomous weak order, so having approval set~$\{2,3\}$ is equivalent to $\{2,3\}\succ \{1,4\}$,
and we say that $2$ and $3$ are preferred to~$1$ and $4$, which $2$ is tied with $3$ and $1$ is tied with $4$.

Formally, a \myemph{preference profile} (\myemph{profile} in short) is a tuple~\myemph{$\ppp=(\aaa,\vvv,\RR)$} consisting of a finite set~$\aaa$ of $m$ alternatives,
a finite set~$\vvv$ of $n$ voters,
and a collection~\myemph{$\RR$} of $n$ preference orders~$\RR=(\succ_1, \succ_2, \cdots, \succ_n)$ such that $\succ_i$ denotes the preference order of voter~$v_i$, $i\in [n]$;
note that all preference orders are either linear preference lists or approval sets.

When discussing the multi-winner determination topic, if not stated otherwise, we will use $m$ to refer to the number of alternatives and $n$ to the number of voters, unless stated otherwise. 
For $x\succ_i y$, we say that voter~$v_i$ \myemph{prefers} $x$ to~$y$. For the sake of clarity, we will call a preference profile an \myemph{approval preference profile} if the preference orders are approval sets; otherwise it is a \myemph{linear preference profile}.

Given a voter~$v_i \in \vvv$ and an alternative~$a \in \aaa$, we define the rank of~$a$ in the preferences of~$v_i$ as the number of alternatives that are preferred to~$a$, i.e., $\rank_i(a) = |\{b\in \aaa \colon b \succ_i a\}|$.

\begin{example}\label{ex:profile}
Let $\aaa=\{a_1,a_2,a_3,a_4\}$ and $\vvv=\{v_1,v_2,v_3\}$.

\noindent The following is an example of a linear preference profile for~$\aaa$ and $\vvv$.
\begin{align*}
  v_1 \colon  a_3\succ a_4 \succ a_2 \succ a_1,\quad v_2 \colon  a_2\succ a_3 \succ a_4 \succ a_1,\quad  v_3 \colon  a_1\succ a_2 \succ a_3 \succ a_4.
\end{align*}
The rank of alternative~$a_4$ in the preference order of voter~$v_2$ is $\rank_2(a_4) = 2$ since there are two alternatives that are preferred to $a_4$, namely $a_2$ and $a_3$.

In the approval preference model, a voter's preferences can be the subset $\{a_2,a_3\}$, which means that he approves of~$a_2$ and $a_3$, and disapproves of the rest. 
The following is an example of an approval preference profile:
\begin{align*}
  v_1 \colon  \{a_2,a_3\}, \quad v_2 \colon  \{a_1\}, \quad  v_3 \colon \{a_3, a_4\}.
\end{align*}
\end{example}

\paragraph{Voting Rules.}
A (multi-winner) voting rule is a function that given a preference profile $\ppp$ returns a $\csize$-sized subset of~$\aaa$, called a \myemph{committee}, which is in some sense optimal.
Before we formally define the multi-winner determination problem, we introduce four approval and two linear voting rules. We choose these rules as they have been studied extensively from both classical and parameterized complexity perspectives.

We begin with two rules designed for linear preference profiles—the \Monroe\ rule and the Chamberlin-Courant (\CC) rule—both of which aim to minimize a notion of misrepresentation between voters and their assigned representatives.
\begin{definition}[Two voting rules for linear preference profiles]\label{def:linear-rules}
  Let $\ppp=(\aaa,\vvv, \RR)$ be a linear preference profile with $n$ voters,
  and let $\csize$ be a number, called the committee size. 
  Given a committee~$W\subseteq \aaa$ and an assignment~$\assign\colon \vvv \to W$, the \myemph{misrepresentation}~${\distB(W, \assign)}$ of $(W,\assign)$ is defined as the sum of the ranks of the alternatives assigned to the voters\footnote{The symbol~``$\mathsf{B}$'' refers to the first letter in the single winner voting rule \myemph{Borda count} since the misrepresentation function is equivalent to the reverse Borda score. Borda count itself is  named after the French mathematician Jean-Charles de Borda (1733--1799).}:
  \begin{align*}
      \myemph{$\distB(W, \assign)$} \coloneqq\sum_{v_i\in \vvv} \rank_i(\assign(v_i)).
  \end{align*}
  \begin{description}
    \item[\Monroe.]
    The \myemph{Monroe} rule selects a size-$\csize$ committee~$W\subseteq \aaa$ and a \myemph{proportional} assignment~$\assign\colon \vvv \to W$ such that $\distB(W, \assign)$ is minimized.
    
  Here, $\assign$ is called \myemph{proportional} if each alternative is assigned the same number of voters (up to rounding), i.e.,
    $\lowerB \le |\{v\in V\colon \repre^{-1}(a)\}|\le \lceil{n/k\rceil}$ holds for all~$a\in W$.

    \item[{Chamberlin-Courant} (CC).] The \myemph{\CC} rule simply selects a size-$\csize$ committee~$W\subseteq \aaa$ and an assignment~$\assign\colon \vvv \to W$ such that $\distB(W, \assign)$ is minimized.
  \end{description}
\end{definition}

While the \Monroe\ and \CC\ rules introduced just now can be adapted for approval preferences, 
we also consider two additional rules specifically designed for the approval setting. 
The Minimax Approval Voting (\MAV) rule aims to minimize the maximum disagreement between 
any voter's preferences and the committee, while the Proportional Approval Voting (\PAV) rule 
implements a form of proportional representation through a diminishing returns scoring function. 
Together, these four rules represent different approaches to aggregating approval preferences, 
each with distinct computational and representational properties.

\begin{definition}[Four voting rules for approval preference profiles]\label{def:approval-rules} %
  Let $\ppp=(\aaa,\vvv, \RR)$ be an approval preference profile with $m$ alternatives and $n$ voters,
  and let $\csize$ be a number, called the committee size.
  Given a committee~$W\subseteq \aaa$ and an assignment~$\assign\colon \vvv \to W$, the \myemph{misrepresentation} ${\distA(W, \assign)}$ of $(W,\assign)$ is defined as the number of voters that do not approve of their assigned alternatives:
  \begin{align*}
   \myemph{$\distA(W, \assign)$} \coloneqq |\{v_i\in \vvv \colon \assign(v_i)\notin \appset_i\}|.
  \end{align*}
  Consequently, the \myemph{representation} \myemph{score}~$\satA(W, \assign)$ is defined as the number of voters that approve of their assigned alternatives:
   \begin{align*}
   \myemph{$\satA(W, \assign)$} \coloneqq |\{v_i\in \vvv \colon \assign(v_i)\in \appset_i\}|.
  \end{align*}
  Observe that $\distA(W,\assign) + \satA(W, \assign) = n$.
  
  \begin{description}
    \item[\Monroe\ and \CC.] The \myemph{\Monroe} and \myemph{\CC} rules for approval preference profiles are analogous to their linear counterparts.
  \Monroe\ selects a size-$\csize$ committee~$W\subseteq \aaa$ and a \myemph{proportional} assignment~$\assign\colon \vvv \to W$ such that $\distA(W, \assign)$ is minimized.
  The \CC\ rule selects a size-$\csize$ committee~$W\subseteq \aaa$ and an assignment~$\assign\colon \vvv \to W$ such that $\distA(W, \assign)$ is minimized.

  Equivalently, \Monroe\ selects a pair~$(W, \assign)$ such that $\assign$ is proportional and ${\satA(W, \assign)}$ is maximized,
  while \CC\ rule selects a pair~$(W,\assign)$ such that ${\satA(W, \assign)}$ is maximized,
  
  \item[Minimax Approval Voting (\MAV).] The \myemph{\MAV} rule selects a size-$\csize$ committee~$W\subseteq \aaa$ such that the maximum of the \emph{Hamming distance}s between $W$ and any input approval set is minimized, where the \emph{Hamming distance} between two sets~$X$ and $Y$ is defined as the size of the symmetric difference between~$X$ and $Y$. Formally, define
  \begin{align*}
    \myemph{$\distMAV(W)$}\coloneqq \max_{v_i\in \vvv}\big(|\appset_i\setminus W| + |W\setminus \appset_i|\big).
  \end{align*}
  Then, \MAV\ selects a committee~$W$ such that $\distMAV(W)$ is minimized. 
   Equivalently, \MAV\ selects a size-$\csize$ committee~$W\subseteq \aaa$ such that the smallest sum of the number of approved alternatives in the committee and disapproved alternatives not in the committee,  denoted as $\satMAV(W)$, is maximized, where
   \begin{align*}
     \myemph{$\satMAV(W)$}\coloneqq \min_{v_i\in \vvv}|\appset_i\cap W|+|(\aaa\setminus \appset_i) \cap (\aaa\setminus W)|.
   \end{align*}
   Observe that $\displaystyle\distMAV(W)+\satMAV(W)=m$.
  
  \item[{Proportional Approval Voting} ({\PAV}).] Unlike the previously discussed rules, the \myemph{\PAV} rule does not aim to minimize the misrepresentation, but rather aims to maximize some score.
  First, let \myemph{$h(x)$}{}$\coloneqq\sum_{j=1}^{x}\frac{1}{j}$ denote the sum of the first~$x$ elements of the harmonic sequence.
  The \emph{\PAV} rule selects a size-$\csize$ committee~$W\subseteq \aaa$ such that
  the sum of partial harmonic sequences is maximized:
  \begin{align*}
    \myemph{$\satPAV(W)$} \coloneqq \sum_{v_i\in\vvv}h(|\appset_i\cap W|). %
  \end{align*}
\end{description}
\end{definition}

\begin{remark}\label{remark:CC-MaxCover}
  For approval preferences, finding an optimal size-$\csize$ committee under the \CC\ rule is equivalent to the \textsc{Maximum Covering} problem~\cite{AS1999,BPS2016,SF2015}, where the goal is to select a given number of subsets to cover a maximum number of elements.
  The latter problem has been studied extensively from the approximation and parameterized perspectives.

  Further, finding an optimal size-$\csize$ committee under the \MAV\ rule is equivalent to a special variant of the \textsc{Closest String} problem~\cite{FraLit1997,LiMaWa2002,CheHerSor2019pnorm}, where the goal is to find a binary vector with fixed number of ones so that the largest Hamming distance is minimized.
\end{remark}

\begin{example}
  Consider the following linear preference profile with alternatives~$\aaa=\{a_1, a_2, a_3, a_4, a_5\}$, voters~$\vvv=\{v_1, v_2, v_3, v_4\}$, and linear preference lists:
  \begin{align*}
    v_1 \colon  a_1 \succ a_2 \succ a_3 \succ a_4 \succ a_5, \quad
    v_2 \colon  a_4 \succ a_3 \succ a_2 \succ a_5 \succ a_1,\\
    v_3 \colon  a_4 \succ a_3 \succ a_5 \succ a_2 \succ a_1, \quad
    v_4 \colon  a_4 \succ a_5 \succ a_3 \succ a_2 \succ a_1.
  \end{align*}
  Suppose we are looking for a committee of size~$\csize=2$.
  \begin{compactitem}[--]
    \item Under \Monroe, there are two optimal committees~$W_1$ and $W_2$, both with misrepresentation~$\distB(W_1,\assign_1)=\distB(W_2,\assign_2)=3$:
  \begin{align*}
    W_1=\{a_2,a_4\} \text{ with }   \assign_1(v_1)=\assign_1(v_2)= a_2\text{,  } \assign_1(v_3)=\assign_1(v_4) = a_4;\\
    W_2 =\{a_3, a_4\} \text{ with }  \assign_2(v_1)=\assign_2(v_2)= a_3\text{,  } \assign_2(v_3)=\assign_2(v_4) = a_4.
  \end{align*}
  \item Under \CC, there is a unique optimal committee~$W_3=\{1,4\}$ with misrepresentation zero. 
\end{compactitem}

  \noindent Let us turn to approval preferences and consider the following profile with~$\aaa=\{a_1, a_2, a_3, a_4, a_5\}$, $\vvv=\{v_1, v_2, v_3, v_4, v_5\}$, and approval preferences:
  \begin{align*}
    \appset_1=\{a_1, a_2, a_3\}, \quad \appset_2=\{a_1,a_2, a_4\}, \quad
    \appset_3=\{a_1,a_2,a_5\}, \quad \appset_4= \{a_1,a_2\}, \quad \appset_5=\{a_5\}. 
  \end{align*}
  Suppose we are looking for a committee of size~$\csize=2$ again.
  \begin{compactitem}[--]
    \item[--] Under both \Monroe\ and \CC, there are two optimal committees:
  \begin{align*}
    W_4=\{a_1,a_5\} \text{ with~ } \assign_4(v_1) = \assign_4(v_2) =  \assign_4(v_4) = a_1, \text{ and ~} \assign_4(v_3)=\assign_4(v_5)=a_5,\\
    W_5=\{a_2,a_5\} \text{ with~ } \assign_5(v_1) = \assign_5(v_2) =  \assign_5(v_4) = a_2, \text{ and ~} \assign_5(v_3)=\assign_5(v_5)=a_5.
  \end{align*}
 The misrepresentation is $\distA(W_4,\assign_4)=\distA(W_5,\assign_5)=0$. Note that for \CC, there are multiple valid assignments for the committees.

 \item Under \MAV, any committee containing alternative $a_1$ or alternative $a_2$ is an optimal committee.
 \noindent The largest Hamming distance for these committees is~$3$.

 \item Under \PAV, committee $W_{6}=\{a_1,a_2\}$ is the unique optimal committee, with~$\satPAV(W_{5})=6$.
 \end{compactitem}
\end{example}

\paragraph{Problem Definitions.}
The decision variant of winner determination under the \Monroe\ rule is defined as follows.

\begin{mdframed}[style=ownframe]
  \decprob{\probMonroe}
{A preference profile~$\ppp=(\aaa,\vvv,\RR)$, a number~$\csize$, and a misrepresentation bound~$\misr\in \mathds{R}_0^+$.}
{Is there a size-$\csize$ committee~$W\subseteq \aaa$ and a \emph{proportional} assignment~$\assign\colon \vvv \to W$ such that $\distB(W,\assign)\le \misr$ for linear preference profiles and
   $\distA(W,\assign)\le \misr$ for approval preference profiles?}
\end{mdframed}

Similarly, we define \probCC\ for linear and approval preferences and \probMAV\ by using the respective misrepresentation function~$\distB$, $\distA$, and $\distMAV$.
For \probPAV, the misrepresentation bound in the input is replaced with a score bound~$\sco$ and we ask whether there exists a size-$\csize$ committee~$W$ with $\satPAV(W)\ge \sco$.

\subsection{Profiles with (Nearly) Single-Peaked and Single-Crossing Preferences}\label{sec:MWRestricted} %

We now introduce two important structured preference domains that often yield tractability results: single-peaked and single-crossing preferences. These structures capture scenarios where preferences follow specific patterns that reflect real-world decision-making contexts. As we will see in \cref{remark:spsc}, these mathematical structures have natural interpretations in political and economic settings.

\begin{definition}[Single-peaked and single-crossing preferences]
  Let $\ppp=(\aaa, \vvv, \RR)$ be a profile with either linear preference lists or approval preferences.

  \noindent We say that $\ppp$ is \myemph{single-peaked} (\myemph{\SP}) with respect to a linear order~$\rhd$ of the alternatives~$\aaa$ if for \emph{each voter $v_i \in \vvv$} the following holds:
\begin{compactitem}[--]
  \item If $\ppp$ is a linear profile, then
   for \emph{each} three alternatives~$a,b,c \in \aaa$ with
   $a\rhd b \rhd c$ or $c\rhd b \rhd a$ it holds that  
   ``$a \succ_i b$'' implies ``$b\succ_i c$''.
   \item If $\ppp$ is an approval profile,
   then the alternatives approved by~$v_i$, i.e., in~$\appset_i$, form a continuous interval along the order~$\rhd$. 
\end{compactitem}
\noindent We say that~$\ppp$ is \myemph{single-crossing} (\myemph{\SC}) with respect to a linear order~$\sblacktriangleright$ of the voters~$\vvv$ if the following holds:
\begin{compactitem}[--]
  \item If $\ppp$ is a linear profile, then
   for \emph{each} pair~$\{x,y\}\subseteq \aaa$ of alternatives and each three voters~$v_i, v_j, v_k\in \vvv$ with $v_i\sblacktriangleright v_j \sblacktriangleright v_k$ it holds that
``$x\succ_i y$ and $x \succ_k y$'' implies ``$x\succ_j y$''.
   \item If $\ppp$ is an approval profile,
   then for each alternative, the voters that approve of it form a continuous interval along the order~$\sblacktriangleright$. 
 \end{compactitem}
 We say that $\ppp$ is \myemph{single-peaked} (resp.\ \myemph{single-crossing})
 if there exists a linear order~$\rhd$ of the alternatives (resp.\ a linear order~$\sblacktriangleright$ of the voters)
 such that $\ppp$ is single-peaked (resp.\ single-crossing) with respect to~$\rhd$ (resp.\ $\sblacktriangleright$).
\end{definition}

\begin{figure}
\centering
\begin{tabular}{ c c c }
  \begin{tikzpicture}[black]
    \begin{axis}[
      xtick = {1,2,3,4},
      xticklabels = {$a_1$, $a_2$, $a_3$, $a_4$},
      xlabel shift=-1ex,
      y dir=reverse,
      ylabel shift=-.5ex,
      ylabel style={align=center}, 
      ylabel={Positions in\\preferences},
      ytick=data,
      label style={font=\normalsize},
      tick label style={font=\normalsize},
      width=5cm, height=8cm,
      x = 1.2cm,
      y = .7cm, scale only axis=true, 
      xmin=0.7,                  %
      xmax=4.6,                  %
      ymin=0.7,                  %
      ymax=4.4,                  %
      legend style={draw=none,nodes={scale=1.2, transform shape},at={(0,1.5)},anchor=north west}
      ]
      \addplot[color=green!60!black,mark=*,line width=1.2pt] coordinates {
        (1,  4)
        (2,  3)
        (3,  1)
        (4,  2)
      };
      \node[green!60!black, font=\footnotesize] at (axis cs:4.3,2) {$v_1$};
      
      \addplot[color=red,mark=triangle*,line width=1.2pt] coordinates {
        (1,  4)
        (2,  1)
        (3,  2)
        (4,  3)
      };
      \node[red, font=\footnotesize] at (axis cs:4.3,3) {$v_2$};
      
      \addplot[color=blue,mark=square*,line width=1.2pt] coordinates {
        (1,  1)
        (2,  2)
        (3,  3)
        (4,  4)
      };
      \node[blue, font=\footnotesize] at (axis cs:4.3,4) {$v_3$};
      \node at (axis cs:4.3,5) {d};
    \end{axis}

  \begin{scope}[xshift=8cm,yshift=2cm]
    \node at (0,0) (s) {\phantom{1}};
    \matrix (scprofile) [below = -1pt of s, matrix of math nodes, row sep=4pt, column sep=2pt] {
      {\color{green!60!black}v_1} \colon & a_3 & \succ & a_4  & \succ & a_2 & \succ & a_1\\
      {\color{red}v_2} \colon & a_2 & \succ & a_3 & \succ & a_4 & \succ & a_1\\
      {\color{blue}v_3} \colon & a_1 & \succ & a_2 & \succ & a_3 & \succ & a_4\\
    };
    \begin{pgfonlayer}{background}
  \draw[linemark=green!80!black,opacity=.5] 
  (scprofile-1-2.north) -- (scprofile-1-2.south) -- 
  (scprofile-2-4.north) -- (scprofile-2-4.south) -- 
  (scprofile-3-6.north) -- (scprofile-3-6.south); %

  \draw[linemark=yellow!70,opacity=.6] 
  (scprofile-1-4.north) -- (scprofile-1-4.south) -- 
  (scprofile-2-6.north) -- (scprofile-2-6.south) -- 
  (scprofile-3-8.north) -- (scprofile-3-8.south);

  \draw[linemark=red!50,opacity=.9] 
  (scprofile-1-6.north) -- (scprofile-1-6.south) -- 
  (scprofile-2-2.north) -- (scprofile-2-2.south) -- 
  (scprofile-3-4.north) -- (scprofile-3-4.south);

  \draw[linemark=lightblue,opacity=.5]%
  (scprofile-1-8.north) -- (scprofile-1-8.south) -- 
  (scprofile-2-8.north) -- (scprofile-2-8.south) -- 
  (scprofile-3-2.north) -- (scprofile-3-2.south);

  \node[draw,inner sep=-1pt, fit=(scprofile)] {};
\end{pgfonlayer}

\
\end{scope}
\end{tikzpicture}
\end{tabular}
\caption{Left: Illustration of a single-peaked order of the alternatives for the linear preference profile given in \cref{ex:profile}.
  Right: Illustration of a single-crossing order of the voters for the same profile (ordered from top to down).}\label{fig:SCSP}
\end{figure}

\begin{example}\label{ex:spsc}
  One can check that the linear preference profile (resp.\ the approval preference profile) given in \cref{ex:profile} is \SP\ with respect to the linear order~$a_1\rhd a_2 \rhd a_3 \rhd a_4$.
  Similarly, the linear preference profile (resp.\ the approval preference profile) given in \cref{ex:profile} is \SC\ with respect to the linear order~$v_1\rhd v_2 \rhd v_3$ (resp.\ $v_1 \rhd v_3 \rhd v_2$). The \SP\ and \SC\ orders for the linear preference profiles are illustrated in \cref{fig:SCSP}.

  The following approval preference profile with four alternatives~$\aaa=\{a_1,a_2,a_3,a_4\}$ and $5$ voters~$\vvv=\{v_1,v_2,v_3,v_4,v_5\}$ is neither \SP\ nor \SC:
  \begin{align*}
    \appset_1 = \{a_1, a_2\},\quad
    \appset_2 = \{a_2, a_3\},\quad
    \appset_3 = \{a_3, a_4\},\quad
    \appset_4 = \{a_1, a_3\},\quad
    \appset_5 = \{a_1, a_4\}.
  \end{align*}
  If we delete two voters, e.g., $v_4$ and $v_5$, then the resulting profile becomes \SP\ with respect to the order~$1\rhd 2 \rhd 3 \rhd 4$; it also becomes \SC\ with respect to the voter order~$v_1\rhd v_2 \rhd v_3$.
  If we delete alternative~$1$, then the resulting profile becomes \SP\ with respect to the order~$2 \rhd 3 \rhd 4$; it also becomes \SC\ with respect to the voter order~$v_1\rhd v_2 \rhd v_4\rhd v_3 \rhd v_5$.
\end{example}

\begin{remark}\label{remark:spsc}
  For the linear model, single-peakedness describes the situation when each voter has an ideal single peak on the order~$\rhd$ such that his preferences towards all alternatives to the left (resp.\ right) of the peak are decreasing when traversing along the order,
  while single-crossingness requires that each pair of alternatives can divide the voter order~$\sblacktriangleright$ into at most \myemph{two continuous} suborders such that in each suborder all voters have the same relative order over this pair.

  For the approval model, single-peakedness and single-crossingness are equivalent to the consecutive ones property in a binary matrix by representing the approval preferences via a binary matrix.
\end{remark} %

Determining single-peakedness and single-crossingness is easy.

\begin{theorem}[\cite{BarTri1986,DoiFal1994,EscLanOez2008,BaHa2011,BCW12,BLConsecutiveOnes76}]
 It is polynomial-time solvable to determine whether a given profile (either linear or approval) is single-peaked (resp.\ single-crossing).
\end{theorem}

We now consider profiles that are ``nearly structured''--those that become single-peaked or single-crossing after removing a small number of voters or alternatives.
This notion of ``distance to tractability'' provides a natural parameterization that can lead to fixed-parameter tractable algorithms.

\begin{definition}
  We say that a preference profile is \myemph{$\numdelvot$-\SP} (resp.\ \myemph{$\numdelalt$-\SP}) if there exists a subset of at most $\numdelvot$ voters (resp.\ $\numdelalt$ alternatives) deleting which makes the profile \SP.
  Similarly, we define \myemph{$\numdelvot$-\SC} and \myemph{$\numdelalt$-\SC} profiles.
\end{definition}

We have the following results on determining the distance to being \SP\ (resp.\ \SC).

\begin{theorem}[\cite{BreCheWoe2016,ELP17NSP}]
  Let $\ppp$ be a \emph{linear} preference profile.
  It is NP-complete to determine whether $\ppp$ is $\numdelvot$-\SP, and polynomial-time solvable to determine whether it is $\numdelalt$-\SP.

  It is polynomial-time solvable to determine whether $\ppp$ is $\numdelvot$-\SC,
  and NP-complete to determine whether it is $\numdelalt$-\SC.

  Determining $\numdelvot$-\SP{ness} (resp.\ $\numdelalt$-\SC{ness}) is fixed-parameter tractable with respect to $\numdelvot$ (resp.\ $\numdelalt$).
\end{theorem}

By \cref{remark:spsc}, we can infer the following from hardness results of the \textsc{Consecutive Ones Submatrix} problem~\cite{HGCOSM2002,NS15}.

\begin{theorem}[\cite{HGCOSM2002,NS15}]
  Let $\ppp$ be an \emph{approval} profile.
  It is NP-complete to determine whether $\ppp$ is $\numdelvot$-\SP, $\numdelalt$-\SP, $\numdelvot$-\SC, or $\numdelalt$-\SC.
Determining $\numdelvot$-\SP{ness} (resp.\ $\numdelalt$-\SC{ness}) is fixed-parameter tractable with respect to $\numdelvot$ (resp.\ $\numdelalt$).
\end{theorem}

The following complexity remains open.
\begin{open}
  Is it  fixed-parameter tractable with respect to $\numdelalt$ (resp.\ $\numdelvot$) to determine whether a given approval preference profile is $\numdelalt$-\SP\ (resp.\ $\numdelvot$-\SC)?
\end{open}

\subsection{Parameters for Multi-Winner Determination}\label{sec:MWparam}
\noindent Multi-winner determinations offer many parameters to study.
In the following, we list the most commonly studied ones.
\begin{compactitem}[--]
  \item The number of alternatives~$m$.
  \item The number of voters~$n$.
  \item The committee size~$\csize$.
  \item For \Monroe, \CC, and \MAV, we have the misrepresentation bound~$\misr$.
  \item For approval preference profiles, we have the score bound~$\sco$ and the maximum size of an approval set~$\displaystyle\ballot\coloneqq \max_{v_i\in V}|\appset_i|$.
\end{compactitem}
We also consider two parameters that measure the distance of a preference profile to being single-peaked (resp.\ single-crossing).
Note that they can be considered as a parameter of ``distance to tractability''.
\begin{compactitem}[--]
  \item The number~$\numdelvot$ denotes the minimum number of voters to delete to make the input profile \SP\ (resp.\ \SC).
  \item The number~$\numdelalt$ denotes the minimum number of alternatives to delete to make the input profile \SP\ (resp.\ \SC). 
\end{compactitem}

\cref{table:mw} summarizes the state of the art of parameterized complexity results with respect to the listed parameters.

\begin{table}[T!]
  \caption{Result overview on \probMW.
    A gray cell means that the parameter is not relevant for the corresponding problem. For results presented in \cref{sec:MWParamC}, we add the corresponding statement in the table. Results which have only a reference to the survey paper are new results by us.}\label{table:mw}
  \setlength{\tabcolsep}{2pt}
\renewcommand{\arraystretch}{1.2}

\resizebox{\textwidth}{!}{
\begin{tabular}{@{}r @{\quad}cc c cc  c cc c cc c cc c cc@{}}
  \toprule
 Parameters &  \multicolumn{11}{c}{Approval preference profiles} && \multicolumn{5}{c}{Linear preference profiles} \\\cline{2-12}\cline{14-18}
            & \multicolumn{2}{c}{\probMonroe} &&
                                             \multicolumn{2}{c}{\probCC}
            & & \multicolumn{2}{c}{\probMAV}  & & \multicolumn{2}{c}{\probPAV}
            & &
                \multicolumn{2}{c}{\probMonroe} & & \multicolumn{2}{c}{\probCC}
  \\
  \cline{2-3}  \cline{5-6}  \cline{8-9} \cline{11-12} \cline{14-15}  \cline{17-18}
  \#alter.\ $m$ &
                  \tFPT & \betzlercite && \tFPT & \betzlercite/\CCMWmn & & \tFPT & \misracite && \tFPT & \yangcite
        & &
            \tFPT & \betzlercite   & & \tFPT & \betzlercite
  \\
  \#voters $n$ &
                 \tFPT & \betzlercite && \tFPT & \betzlercite/\CCMWmn & & \tFPT & \misracite && \tFPT & \yangcite
              & &
                  \tFPT & \betzlercite &  & \tFPT & \betzlercite
  \\
  Cmt.\ size $\csize$ &
                          \tWtwoh/\tXP & \betzlercite && \tWtwoh/\tXP & \betzlercite & & \tWtwoh/\tXP & \misracite && \tWoneh/\tXP & \azizcite
              &&
                 \tWtwoh/\tXP & \betzlercite   && \tWtwoh/\tXP & \betzlercite
  \\
  Misrepr.~$\misr$ &
                     \tNPh & \betzlercite & & \tNPh & \betzlercite/\CCMWR && \tFPT & \misracite & & \graycell & \graycell
             & &
                 \tWoneh/\tXP & \crcite/\betzlercite & & \tWoneh/\tXP &\crcite/\betzlercite/\CCMWR
  \\
  Max app.~$\ballot$ &
                       \tNPh & \procacciacite && \tNPh & \procacciacite  && \tNPh & \legrandcite && \tNPh & \azizcite
            &&
               \graycell & \graycell && \graycell & \graycell
  \\
  Score~$\sco$ &
                   ? & ? && \tFPT & \yangcite &&  ? & ?  && ?/\tXP & ?/\PAVXPscoreref
              & &
                   \graycell & \graycell && \graycell & \graycell
    \\
  \midrule 
  $\csize+\misr$ &
                   \tWtwoh/\tXP & \betzlercite & & \tWtwoh/\tXP & \betzlercite && \tFPT & \misracite & & \graycell & \graycell
            & &
                \tFPT &\betzlercite& & \tFPT &\betzlercite
  \\
  $\csize+\ballot$ &
                       \tWoneh/\tXP & \monroekballot & & \tWoneh/\tXP & \yangcite && \tFPT & \yangcite & & \tWoneh/\tXP & \azizcite
                                               & &
                       \graycell & \graycell & & \graycell & \graycell
    \\
  $\csize+\sco$ &
                  ? & ? & & \tFPT & \yangcite && ? & ?  & & ?/\tXP & ?
                                            & &
                  \graycell &\graycell& & \graycell &\graycell
  \\
  $\ballot+\misr$ &
                    \tNPh & \procacciacite & & \tNPh & \procacciacite && \tFPT & \yangcite & & \graycell & \graycell
                                             & &
                    \graycell &\graycell& & \graycell &\graycell
  \\
  $\ballot+\sco$ &
                   ? & ? & & \tFPT & \yangcite &&  ? & ? & & \tFPT & \yangcite/\PAVFPTscoreballotref
                                             & &
                   \graycell &\graycell& & \graycell &\graycell
  \\
  \midrule 
  \SP\ &
         \tPP & \betzlercite  && \tPP & \betzlercite && \tPP & \liucite && \tPP & \petersciteSP
                                             &  &
         ? & ? & & \tPP & \betzlercite
  \\
  $\numdelvot$-\SP\ &
                      \tFPT &  \ourcite &&  \tFPT &  \ourcite &&  ? &  ? &&  ? &  ?
            &&
               ? &  ? &&  \tFPT &  \ourcite
  \\
  $\numdelalt$-\SP\ &
                      ?/\tXP &  \ourcite &&  \tFPT &  \misraFPTcite &&  \tFPT &  \MAVSPSC &&  ?  & ?
                                           &&
                      ? &  ? && \tFPT &  \misraFPTcite
  \\[1ex]

  \SC\ &
         \tPP & \ourcite & &  \tPP & \elkindcite & &   \tPP & \liucite & & ?& ?
                                             &&
         \tNPh & \skowroncite & & \tPP & \skowroncite
  \\
  $\numdelvot$-\SC\ &
                      \tFPT &  \ourcite &&  \tFPT &  \ourcite && ? &  ? &&  ? & ?
                                           &&
                      \tNPh &  \skowroncite &&  \tFPT &  \ourcite
  \\
  $\numdelalt$-\SC\ &
                      ?/\tXP &  \ourcite &&  \tFPT &  \misraFPTcite &&  \tFPT &  \MAVSPSC && ?  & ?
                                           &&
                      \tNPh &  \skowroncite &&  \tFPT &  \misraFPTcite
  \\\bottomrule
\end{tabular}}
\end{table}

\subsection{Parameterized Complexities}\label{sec:MWParamC}

This section explores the parameterized complexity landscape of multi-winner determination problems through representative results and proof techniques. We present selected theorems that illustrate both tractability frontiers and hardness barriers, highlighting the algorithmic methods that characterize each result.

We organize our presentation to showcase a variety of proof techniques central to parameterized complexity analysis, progressing from simpler approaches to more sophisticated methods:

\begin{description}
  \item[Brute force approaches (Theorem~\ref{thm:CC-fpt-m/n}):] Exploiting bounded search spaces when certain parameters are small.

  \item[Parameterized reductions (Theorem~\ref{them:CC-R}):] Establishing hardness and providing \XP\ algorithms by restricting the search space.

  \item[{Win/win strategies} (Proposition~\ref{prop:PAV-XP-score}):]
  Showing that either the parameter bounds other relevant parameters or the instance admits a trivial solution.\footnote{This technique was named by \citet{Fellows2003} and has proven effective in various contexts, including deriving \FPT-algorithms.}

  \item[{Kernelization} (Theorem~\ref{prop:PAV-FPT-s+b}):]
  A powerful \FPT-technique that reduces the problem size to a function of the parameter, thereby establishing membership in \FPT. 

  \item[{Structural parameters} (Proposition~\ref{prop:MAV-SPSC}):]
  Exploiting properties of structured preferences and generalizing a dynamic programming approach originally designed for single-peaked (resp. single-crossing) profiles to obtain an \FPT-algorithm for the  $\numdelalt$-SP (resp. $\numdelalt$-SC) case.
\end{description}

We use the symbol~(\newresult) to indicate original contributions and a star with citation (e.g., [cite]\litresult) to denote results that follow directly from existing work.

\paragraph{}

We now present selected parameterized complexity results that illustrate key techniques in the field.
We begin with a straightforward fixed-parameter tractability result that leverages the bounded search space when either the number of alternatives or voters is small.

\begin{theorem}[\betzlercite]\label{thm:CC-fpt-m/n}
\probCC\ parameterized by the number~$m$ of alternatives (resp.\ the number~$n$ of voters) is fixed-parameter tractable. 
\end{theorem}

\begin{proof}
  The proof for the parameter~$m$ is straightforward by brute-force searching for all possible committees of size~$\csize$; there are $m^{\csize}$ i.e., $O(2^m)$ many.
  For each size-$\csize$ committee~$W$, we find in polynomial time an optimal assignment~$\assign$ by mapping each voter~$v_i$ to the highest ranked alternative from~$W$ (interpreting approval preferences as a weak order) and computing the corresponding misrepresentation. 
  We return a committee and the associated assignment which induces the smallest misrepresentation.
  In this way, we can find an optimal committee and assignment with minimum misrepresentation in $O(2^{m}\cdot n\cdot m)$ time, showing that \probCC\ is fixed-parameter tractable with respect to~$m$.

  The idea for the parameter~$n$ is to check all possible partitions of voters.
  Each subset in a partition is assigned to the same alternative in the committee.
  There are at most $n^n$ ways to partition the voters; we can assume that the committee size is $\csize < n$ as otherwise we can simply return an optimal committee that contains the highest ranked alternative of each voter with $\misr=0$.  
  For each subset~$V_j$ in a given partition~$(V_1,V_2,\cdots,V_{p})$, it suffices to compute an alternative that minimizes the misrepresentation for~$V_j$.
  This can be done in polynomial time by simply checking the misrepresentation induced by each singular alternative for~$V_j$.
  This approach has a running time of $n^n\cdot n\cdot m$.
  Hence, the problem is  fixed-parameter tractable with respect to~$n$.
\end{proof}

The previous result established fixed-parameter tractability through straightforward brute forcing.
We now turn to another parameter that allows both hardness limitations and algorithmic possibilities.
The following result establishes \Wone-hardness of \probCC\ with respect to the misrepresentation bound $\misr$, via a parameterized reduction from the \Wone-complete \Clique{} problem.
It also demonstrates membership in \XP, showing that while the problem is unlikely to be fixed-parameter tractable for this parameter, it remains solvable in polynomial time for each fixed value of $\misr$.
Finally, \cref{remark:CC-MaxCover} immediately implies NP-hardness for  \probCC\ with approval preferences. %

\begin{theorem}[\betzlercite,\crcite]\label{them:CC-R}
  For linear preference profiles, \probCC\ parameterized by the misrepresentation bound~$\misr$ is \Wone-hard\ and in \XP.
  For approval preference profiles, \probCC\ remains NP-hard even if the misrepresentation bound is $\misr=0$.
\end{theorem}
\begin{proof}[Proof sketch.]
  The \Wone-hardness follows via a parameterized reduction from the \Wone-complete problem \Clique\ parameterized by the size~$h$ of the clique.
  Let $I=(G=(U,E),h)$ be an instance of \Clique\ with $U=\{u_1,u_2,\cdots, u_{\enn}\}$ being the vertex set and $E=\{e_1,e_2,\cdots,e_{\emm}\}$ being the edge set.
  The idea is to create a linear preference profile and set the misrepresentation bound in such a way that only alternatives that are ranked in one of the first positions can be selected to a committee.
  By the bound on the committee size, we make sure that alternatives that are selected to the committee must correspond to vertices and edges of a size-$h$ clique.
  
  Formally, we create an instance $I'=(\ppp=(\aaa,\vvv,\RR),\csize)$ of \probCC\ as follows. %
\begin{compactitem}[--]
  \item For each vertex~$u_i\in V$, we create two \myemph{vertex-alternative}s \myemph{$a_i$ and $b_i$}, and a set~$\dumA_i$ of $\enn$ dummy alternatives.
  The dummy alternatives serve as a \emph{blocker} to prevent alternatives that ranked beyond the second position from being selected.
  \item For each edge~$e_j\in E$, we add an \myemph{edge-alternative $c_j$} and a set~$\dumB_j$ of $\emm$ dummy alternatives.
  Again, the dummy alternatives serve as a \emph{blocker} to prevent alternatives that ranked beyond the second position from being selected.
\end{compactitem}
 Now, we describe the voters and their preferences; below ``$\cdots$'' denotes to an arbitrary but fixed linear order of the remaining alternatives that are not mentioned explicitly while ``$\seq{X}$'' denotes an arbitrary but fixed linear order of the alternatives in~$X$.
\begin{compactitem}[--]
  \item For each vertex~$u_i\in U$, we create a voter~$v_i$ with the following linear preference list $b_i\succ_i a_i\succ_i \seq{\dumA_i}\succ_i\cdots$. %
  \item For each edge $e_j\in E$ with $e_j=\{u_i, u_s\}$, we create $2h$ voters.
  The first $h$ voters~$v_j^1,v_j^2, \cdots, v_j^h$ have the same linear preference list as follows:
  $c_j\succ a_i \succ \seq{\dumB_j}\succ\cdots$.
  The remaining $h$ voters~$v_j^{h+1}, v_j^{h+2},\cdots, v_j^{2h} $ have the same linear preference list as follows: $c_j\succ a_s\succ \seq{\dumB_j}\succ\cdots$. 
\end{compactitem}
Overall, the alternatives and the voters are $\aaa=\displaystyle\bigcup_{u_i\in U}\left(\{a_i,b_i\}\cup \dumA_i\right) \cup \bigcup_{e_j\in E}\left(\{c_j\}\cup \dumB_j\right)$
and
$\vvv=\{v_i\mid i\in [n]\}\cup \{v_j^1,v_j^2,\cdots, v_j^{2n}\mid j\in [\emm]\}$.
The committee size is set to $\csize=\emm-\binom{h}{2}+\enn$.
The misrepresentation bound is set to $\misr=h+h\cdot2\binom{h}{2}$.

This completes the description of the construction.
One can verify that for a size-$\csize$ committee~$W$ and an assignment~$\assign\colon \vvv \to W$ to stay within the misrepresentation bound~$\misr$ every voter must be assigned to an alternative that is ranked either in the first or the second position.
Using this observation, it is fairly straightforward to show that there are exactly $\binom{h}{2}$ edge-alternatives that are not selected to~$W$ and they must correspond to the edges of a size-$h$ clique. 

The intuition for the \XP-membership with respect to parameter~$\misr$ is to guess a subset~$V'\subseteq V$ of at most~$\misr$ voters,
each of whom is \emph{not} assigned to his most-preferred alternative and hence contributes misrepresentation of at least one.
For each voter in~$V'$, we further guess an alternative with rank~$\le \misr$ that he is assigned to; there are $\le \misr$ choices since alternatives with ranks larger than $\misr+1$ will induce misrepresentation of at least $\misr+1$.
Overall, there are at most $\misr^\misr$ many ways of assigning voters in $V'$.
For a correct guess, we need to assign each voter in $V\setminus V'$ to his most-preferred alternative.
Finally, we check whether the assignment gives rise to a size-$\csize$ committee and has misrepresentation not exceeding~$\misr$.
This approach has a running time of $n^{O(\misr)}\cdot (n\cdot m)$, as desired. 

For approval preference profiles, since the \CC\ rule corresponds to \textsc{Maximum Covering} (see \cref{remark:CC-MaxCover}), NP-hardness follows immediately.
\end{proof}

Next, we show that when the score~$\sco$ is very small, \probPAV\ is trivial, which helps to design \XP-algorithm for parameter~$\sco$.

\begin{lemma}[\newresult]\label{lem:PAV-score-very-small}
  If $\sco \le \min(\csize,n)$, then every \probPAV-instance admits a solution, which can be found in polynomial time.
\end{lemma}

\begin{proof}
  Without loss of generality, we assume that every voter approves at least one alternative because voters that approve \emph{no} alternatives do not affect the score of any committee.
  We claim to be able to greedily find a size-$\sco$ committee that contributes a score of at least $\sco$:
  As long as the committee~$W$ has size less than $\csize$, we do the following:
  If there exists an alternative from~$\aaa\setminus W$ that is not approved by any voter that approves of at least one alternative in~$W$, we find one and add it to~$W$; otherwise we add an arbitrary alternative to~$W$.

  It is straightforward that the thus found committee~$W$ has score either~$n$ (if $\csize \ge n$) or $\csize$ (if $\csize < n$). In any case, the score is at least $\sco$ by our assumption. 
  This approach clearly can be done in polynomial time.
\end{proof}

The above lemma immediately implies the following \XP\ result, using win/win-strategy. 

\begin{proposition}[\newresult]\label{prop:PAV-XP-score}
\probPAV\ parameterized by the score~$\sco$ is in \XP.
\end{proposition} 
\begin{proof}
  By \cref{lem:PAV-score-very-small}, we can assume that $\sco \ge \min(\csize, n)$. %
  If $\sco \ge n$, then we can use the \FPT\ algorithm for~$n$~\cite{yang2023parameterized} to obtain fixed-parameter tractability for $\sco$.
  If $\sco \ge \csize$, then we can brute-force search all size-$\csize$ committees and solve the problem in \XP\ time with respect to~$\sco$.
\end{proof}
Yang and Wang~\cite{yang2023parameterized} provide a branching algorithm to show that \probPAV\ is fixed-parameter tractable with respect to the combined parameter~$(\ballot,\sco)$.
Below, we provide a kernelization algorithm to obtain an exponential-size kernel for ($\ballot$, $\sco$).

\begin{theorem}[\newresult]\label{prop:PAV-FPT-s+b}
\probPAV\ admits a kernel for the combined parameter max.\ size~$\ballot$ of an approval set  and the score~$\sco$.
\end{theorem}
\begin{proof}[Proof sketch.]
First, let us assume that no alternative is approved by $\sco$ or more voters, as in this case any committee containing that alternative would have a score of at least $\sco$.
Second, by \cref{lem:PAV-score-very-small}, we can assume that $\sco \ge \min(\csize, n)$.
If $\sco \ge n$, then the problem has a trivial polynomial-size kernel: The number of voters is already bounded by~$\sco$, and the number of alternatives is at most $\ballot \cdot \sco$ since each voter approves at most $\ballot$ alternatives.

In the remainder of the proof, we may assume that $\sco \ge \csize$.
It remains to bound the number of alternatives.
To achieve this, we partition the alternatives into three disjoint subsets based on the number of voters they are approved by and upper-bound the size of each subset accordingly; for the sake of readability, we assume that $\sco/\ln(\csize)$ and $\sco/\csize$ are integral. %
Let $\myemph{\aaa(x)}\coloneqq \{v_i\in \vvv \mid x\in \appset_i\}$ denote the set of all voters that approve of~$x$.
\begin{align*}
  L \coloneqq \{a\in\aaa \colon |\aaa(a)|\geq\frac{\sco}{\ln(\csize)}\}, 
  M  \coloneqq \{a\in\aaa \colon \frac{\sco}{\csize}\leq|\aaa(a)|<\frac{\sco}{\ln(\csize)}\}, 
  T \coloneqq \{a\in\aaa \colon |\aaa(a)|<\frac{\sco}{\csize}\}.
\end{align*}
We observe that for $|L|\geq \csize$, there must exist a committee with score $\sco$ since by picking $\csize$ alternatives from $L$, we achieve a score of at least $\frac{\sco}{\ln(\csize)}\cdot\sum_{i=1}^\csize\frac{1}{i}\geq\frac{\sco}{\ln(\csize)}\cdot\ln(\csize)=\sco$; this would be the case if all alternatives in $L$ are approved by the same $\frac{\sco}{\ln(\csize)}$ voters.

To upper-bound $|M|$ and $|T|$, we assume that for each alternative there are at most $\csize-1$ other alternatives that are approved by exactly the same of set of voters. 
In other words, $|\{b \in \aaa : \aaa(b)=\aaa(a)\}| \le \csize$ holds for all $a\in \aaa$.
This is valid since the committee size is $\csize$ and two alternatives of the same ``type'' are exchangeable. 

We now upper-bound~$|M|$. 
We partition $M$ into $M_{\frac{\sco}{\csize}}\cup M_{ \frac{\sco}{\csize} + 1} \cup \cdots\cup M_{ \frac{\sco}{\ln(\csize)}-1}$. For each~$\ell\in \{\frac{\sco}{\csize},\cdots, \frac{\sco}{\ln(\csize)}-1\}$, let $M_{\ell}\coloneqq \{a\in M : |\aaa(a)|=\ell\}$.
In other words, $M_\ell$ consists of all alternatives that are approved by exactly~$\ell$ voters.

\begin{claim}\label{cl:M}
  If $|M_{\ell}| \ge \csize\cdot(\ballot-1)\cdot\ell + \csize$,
  then $M_{\ell}$ contains $\csize$ alternatives that are approved by pairwisely disjoint subsets of voters.
\end{claim}
\begin{proof}\renewcommand{\qedsymbol}{(of \cref{cl:M})~$\diamond$}
Since each voter approves at most $\ballot$ alternatives,
for all alternatives~$a\in M_\ell$,
at most $(\ballot-1)\cdot\ell$ alternatives in~$M_\ell$ can be approved by a voter that also approves of~$a$, i.e.,
$|\{b \in M_{\ell} : \exists v_i \in \vvv \text{ with } \{a,b\} \subseteq \appset_i\}| \le (\ballot-1)\cdot \ell$.
Hence, if $|M_\ell| \ge \csize\cdot(\ballot-1)\cdot\ell + \csize$,
then we can find $\csize$ alternatives as follows.
We greedily pick an alternative~$a$ from~$M_{\ell}$ and delete all alternatives~$c$ in $M_{\ell}$ that may share with~$a$ some same voter, i.e., $\aaa(a)\cap \aaa(c)\neq \emptyset$.
In each step, we delete at most $(\ballot-1)\cdot \ell$ alternatives in addition to $a$.
Hence, after $\csize$ steps, we have picked $\csize$ alternatives that pairwisely do not share any voters.
\end{proof}

By \cref{cl:M}, we can either find $\csize$ alternatives in~$M_{\ell}$ that together yield a score of at least $\csize\cdot \frac{\sco}{\csize}=\sco$ or upper-bound~$M$ in a function of~$\sco+\ballot$ since the range of $\ell$ is bounded in~$\sco$.

For the last subset~$T$, we again partition it into $T_1\cup\cdots\cup T_{\frac{\sco}{\csize}-1}$, where $T_j \coloneqq \{a\in T : |\aaa(a)|=j\}$ for $j\in [\frac{\sco}{\csize}-1]$.
Note that any committee of score~$\sco$ must contain at least one alternative from $M\cup L$ because the largest score that any $\csize$ alternatives in $T$ can induce is less than $\csize\cdot \frac{\sco}{\csize}$. 

For each~$T_j$, $j = \frac{\sco}{\csize}-1, \frac{\sco}{\csize}-2, \cdots, 1$,
let \myemph{$V[T_j]$}~$\coloneqq \{v_i\in\vvv\mid \appset_i\cap(L\cup M\cup T_j\cup T_{j+1}\cup \cdots \cup T_{\frac{\sco}{\csize}-1})\neq\emptyset\}$, i.e., $V[T_j]$ consists of voters whose approval ballots contain an alternative that is approved by at least~$j$ voters.
We aim to upper-bound~$|T_j|$ by partitioning $T_j$ into two disjoint subsets~$X_{j}$ and $Y_j$.
The set $X_j$ consists of all alternatives that are approved by at least one voter in~$V[T_{j+1}]$ and
$Y_j$ the remaining alternatives, that is, the ones that are only approved by voters in~$\vvv\setminus V[T_{j+1}]$.

Since each alternative is approved by at most~$\sco$ voters, we have that $|V[T_{j+1}]|\le \sco \cdot (|L|+|M| + \sum_{\ell=j+1}^{\frac{\sco}{\csize}-1}|T_{\ell}|)$.
Since each voter approves at most~$\ballot$ voters, it holds that $|X_j|\le |V[T_{j+1}]|\cdot (\ballot-1) \le  \sco \cdot (|L|+|M| + \sum_{\ell=j+1}^{\frac{\sco}{\csize}-1}|T_{\ell}|)$. 

To upper-bound~$|Y_j|$, we use a reasoning similar to the one for \cref{cl:M}.
\begin{claim}\label{cl:Tell}
  If $|Y_j| \ge \csize\cdot(\ballot-1)\cdot j + \csize$,
  then $Y_j$ contains $\csize$ alternatives that are approved by pairwisely disjoint subsets of voters, and no size-$\csize$ committee with maximum score contains any alternatives from~$T_{1}\cup \cdots \cup T_{j-1}$.
\end{claim}
\begin{proof}\renewcommand{\qedsymbol}{(of \cref{cl:Tell})~$\diamond$}
  The proof for the first part of the statement is the same as the one for \cref{cl:M}.
  For the second part, let $a_1, \cdots, a_{\csize}$ be the $\csize$ alternatives approved by pairwisely disjoint subsets of voters.
  Suppose, for the sake of contradiction, that $W$ is a committee with maximum score and contains exactly $k'$ alternative from $T_{1}\cup \cdots T_{j-1}$ with $k' > 0$, called $c_1, \cdots, c_{k'}$.
  Then, we can replace all these~$k'$ alternatives with $a_1,\cdots, a_{k'}$ to achieve a higher score because
  each alternative~$c_z$, $z\in [k']$, is approved by at most~$j-1$ voters, while each alternative~$a_z$, $z\in [\csize]$, is approved by exactly~$j$ distinct voters, all of them from~$\vvv \setminus \{V[T_{j+1}]\}$.
  In other words,~$c_1,\cdots, c_{k'}$ contribute a net score of at most $k'\cdot (j-1)$ while
  $a_1,\cdots, a_{k'}$ contribute a net score of exactly~$k'\cdot j$.
  Hence the replacement increases the score, a contradiction.
\end{proof}
Recall that $\sco \ge \csize$.
Since we have upper-bounded both~$X_j$ and $Y_j$ by a function of~$\sco+\ballot$,
we can upper-bound the size of $T_j$, $j=\frac{\sco}{\csize}-1, \cdots, 1$ by a function of $\sco+\ballot$, as desired.
Thus we have upper-bounded $T$ by a function of $\sco + \ballot$.
Since we have also upper bounded the size of $L$ and $M$ by a function of $\sco + \ballot$ we have bounded the number of alternatives.
This concludes the proof of the kernelization.
\end{proof}

We observe that the reduction by Aziz et al.~\azizcite\ which is from \IS, also shows that \probPAV\ parameterized by $(\csize+\ballot)$ is \Woneh.
In fact, the same construction can also straightforwardly adapted to show parameterized intractability for the \Monroe\ rule.
The only difference is that the misrepresentation bound needs to be set to $\deg(G)\cdot (n-\csize)$, where $\deg(G)$ is the maximum degree of an vertex in the \IS\ instance.
\begin{proposition}[\azizcite\litresult]\label{prop:Monroe-W1-k+b}
\probMonroe\ parameterized by the committee size~$\csize$ and the max.\ size of an approval set~$\ballot$ is \Woneh.
\end{proposition}

Besides the open questions indicated by the questions marks in the result table, our first challenge is concerned with matching parameterized complexity upper bounds.
\challengeF{\label{challenge:matching-upper-bound}%
  Obtain a matching complexity parameterized upper bound for every W-hardness result given in \cref{table:mw}. For instance, is \probMW{} parameterized by the committee size~$\csize$ contained in \Wtwo\ for \probMonroe, \probCC, and \probMAV and \Wone\ for \probPAV?
}

We now turn to structured preferences.
Liu and Guo~\liucite\ provide a polynomial-time algorithm to solve \probMAV\ when the preferences are \SP\ (resp.\ \SC).
We adapt their algorithm to show that \probMAV\ is \FPT\ when the profile is close to being \SP\ (resp.\ \SC); this idea has been proven useful for \Monroe\ and \CC\ already~\cite{MisraSonarVaid2017MW,chen2023efficient}. 

\begin{proposition}[\newresult]\label{prop:MAV-SPSC}
For $\numdelalt$-\SP\ profiles, \probMAV\ parameterized by the number~$\numdelalt$ of alternatives to delete is \FPT.
For $\numdelalt$-\SC\ profiles, \probMAV\ parameterized by the number~$\numdelalt$ of alternatives to delete is \FPT.
\end{proposition}

\begin{proof}[Proof Sketch.]
The approach for the $\numdelalt$-\SP\ case and the $\numdelalt$-\SC\ case is the same.
We extend the polynomial-time algorithm by Liu and Guo~\liucite\ by brute-force searching for all possible subsets of the $\numdelalt$ deleted alternatives that can be part of an optimal committee~$W$. %
Let $\aaa'$ be a set of $\numdelalt$ alternatives that need to be deleted for the profile to be \SP\ (resp.\ \SC); there are $2^{\numdelalt}$ subsets of $\aaa'$.
For a fixed subset, we assume that it is contained in~$W$. For the remaining alternatives from $\aaa\setminus\aaa'$, we can use the algorithm by Liu and Guo. This approach has a running time of $2^{\numdelalt}\cdot|I|^{O(1)}$, which is \FPT\ parameterized by $\numdelalt$.
\end{proof}

Our next challenge is concerned with the distance to tractability parameter.

\challengeF{\label{challenge:distance-tractable}
  Resolve the parameterized complexity of \probMonroe\ and \probPAV\ with respect to $\numdelvot$ and~$\numdelalt$, respectively.}

\section{Hedonic Games}\label{sec:hedonic}

Hedonic games are a specialized form of the coalition formation problem, in which a set of agents each has preferences over the \emph{coalitions} (i.e., subsets of agents) they may belong to.
The objective is to partition the agents into disjoint coalitions that satisfy some notion of stability.
Hedonic games naturally generalize preference-based matching problems~\cite{Manlove2013}. For broader introductions to hedonic games, we refer the reader to the survey by Woeginger~\cite{WOEGINGER2013101} and the book chapter by Aziz and Savani~\cite{AS2016HedonicGamesChapter}.

In this section, we survey the parameterized complexity of verification and existence questions related to hedonic games under two types of widely studied preference representations.
See \cref{table:additive,table:FE} for an overview. 

The structure of this section is similar to that for the multi-winner determination section.
\cref{sec:HGstab} formally defines hedonic games, several commonly studied stability concepts and their relations, and the central algorithmic questions.
\cref{sec:HGCompact} reviews two types of widely studied compact preference representations that make the preference elicitation process polynomial in the number of agents and provide examples for illustration. 
\cref{sec:HGClassic} examines the classical computational complexity of the associated problems.
\cref{sec:HGparam} defines the parameters considered in this survey.
\cref{sec:HGparamC} provides an overview of relevant parameterized complexity results, accompanied by selected proofs that illustrate a range of proof techniques.
Finally, \cref{sec:FurtherHG} explores topics in hedonic games that extend beyond the core scope of this survey.

\newcommand{\ourunpub}{\cite{durand2024enemies}}
\newcommand{\cliqueres}{{\footnotesize $\triangle$}}
\renewcommand{\arraystretch}{1.15}
\addtolength{\tabcolsep}{-0.4em}

\newcommand{\checksource}{{\large \color{red!50!black}\textbf{!}}}
\begin{table}[T!]
  \caption{Parameterized complexity results for Hedonic Games with \AddPrefs.
    ``\twNPc'' and \tSigmac$^*$ mean that the corresponding hardness relies on exponential utility values; 
    all other hardness results hold when the utilities are polynomial on $n$. %
    For results presented in \cref{sec:HGparamC}, we add the corresponding statement in the table.
Some of the results for parameter $\nrutil$ are derived from a restriction of Hedonic Games with \AddPrefs\ called Hedonic Games with Friends, Enemies, \emph{and Neutrals}. In this restriction, the possible utility values are $\{-1, 0, n\}$, and hence $\nrutil = 3$. }
  \label{table:additive}

 \begin{tabular}{@{}r @{\;\;\;} cc c cc c cc c cc@{}}

  \toprule
   & \multicolumn{2}{c}{\core- \& \score-\verif} & ~~~~ &  \multicolumn{2}{c}{\core- \& \score-\exist} & ~~& \multicolumn{2}{c}{\Nash-\exist} & ~~& \multicolumn{2}{c}{\IndS-\exist}\\
   \cline{2-3} \cline{5-6}\cline{8-9} \cline{11-12}
Max deg.~\maxdeg &  \tNPc & \cite{CheCsaRoySim2023Verif} && \tSigmac & \cite{peters2017precise}  && \tNPc & \cite{CheCsaRoySim2023Verif}  && \tNPc & \cite{CheCsaRoySim2023Verif}\\
   \# diff.\ util.~\nrutil &\tNPc & \cite{sung2007core,aziz2013computing} &&  \tSigmac & \cite{WOEGINGER2013101,peters2017precise}  & & \tNPc & \cite{sung2010computational} && \tNPc & \cite{sung2010computational}\\
   FAS~\fas &  \tNPc & \cite{CheCsaRoySim2023Verif} && ? & ? & & \tNPc & \cite{CheCsaRoySim2023Verif} & & \tNPc & \cite{CheCsaRoySim2023Verif}\\
   \tw &  \twNPc & \cite{hanaka2024core},\textbf{[P\ref{prop:twaddsverif}]}  && \tSigmac$^*$ \& ? & \cite{hanaka2024core} & & \tNPc & \cite{hanaka2022revisited} & & ? &?\\
\midrule
   $\maxdeg + \nrutil$ &\tNPc & \cite{CheCsaRoySim2023Verif}  & & \tSigmac & \cite{peters2017precise}  & & \tNPc & \cite{CheCsaRoySim2023Verif} & & \tNPc & \cite{CheCsaRoySim2023Verif}\\ 
   $\maxdeg + \tw$ &\tFPT & \cite{peters2016graphical}  && \tFPT & \cite{peters2016graphical}  && \tFPT & \cite{peters2016graphical}  && \tFPT & \cite{peters2016graphical} \\
   $\maxdeg + \nrutil + \fas$ & \tNPc & \cite{CheCsaRoySim2023Verif} && ?  & ? && \tNPc & \cite{CheCsaRoySim2023Verif} && \tNPc & \cite{CheCsaRoySim2023Verif}  \\
   \midrule
$\maxdeg + \fas + \nrutil + \maxcoal$  & \tNPc & \cite{CheCsaRoySim2023Verif} && \graycell & \graycell && \graycell & \graycell  && \graycell & \graycell \\
$\maxdeg + \fas + \nrutil + \numcoal$  &  \tNPc & \cite{CheCsaRoySim2023Verif} && \graycell & \graycell && \graycell & \graycell  && \graycell & \graycell\\
   \bottomrule
  \end{tabular}
\end{table}

\newcommand{\fessymb}{\ensuremath{^\dagger}}
\begin{table}[T!]
\centering

\caption{%
  Parameterized complexity results for Hedonic Games with Friends and Enemies.
  Note that the maximum initial coalition size~$\maxcoal$ together with the number~$\numcoal$ of coalitions bounds the instance size.
  Thus, we omit parameter combinations that contain both~$\maxcoal$ and $\numcoal$. 
  The symbol \fessymb\ means we consider the parameter feedback edge set rather than feedback arc set.
For results presented in \cref{sec:HGparamC}, we add the corresponding statement in the table. Results which have only a reference to the survey paper are new results by us.}
\label{table:FE}
 \resizebox{\textwidth}{!}
{
 \begin{tabular}{@{}r @{\;\;} cc c cc c cc c cc c cc c cc}
  \toprule
   & \multicolumn{5}{c}{\core- \& \score-\verif} & & \multicolumn{2}{c}{\core-\find{}} & & \multicolumn{2}{c}{\score-\exist{}} & & \multicolumn{5}{c}{\Nash-\exist{}}\\ \cline{2-6} \cline{8-9} \cline{11-12} \cline{14-18}
   & \multicolumn{2}{c}{\friend} & &  \multicolumn{2}{c}{\enemy}  & & \multicolumn{2}{c}{\enemy} & & \multicolumn{2}{c}{\enemy} & &\multicolumn{2}{c}{\friend} & &\multicolumn{2}{c}{\enemy}\\\cline{2-3} \cline{5-6} \cline{8-9} \cline{11-12}\cline{14-15}\cline{17-18}
 
   Max deg.~$\maxdeg$
   &
     \tNPc & \cite{CheCsaRoySim2023Verif}
   &&
       \tFPT  & \ourunpub
   &&
      \tFPT  & \ourunpub
   &&
      \tNPc & \ourunpub
   &&
      \tNPc &   \cite{CheCsaRoySim2023Verif}
   &&
\tNPc & \cite{brandt2022single},\textbf{[P\ref{prop:maxdegen}]}\\

   FAS/FES~$\fas$  &
      \tNPc & \cite{CheCsaRoySim2023Verif}
   &&
      \tFPT\fessymb & \ourunpub  
   &&
      \tFPT\fessymb  & \ourunpub
   & &
       \tFPT\fessymb &  \ourunpub & &
      \tNPc &  \cite{CheCsaRoySim2023Verif} & &
                 \tFPT &  \textbf{[P\ref{prop:nashfas}]} \\
  $\maxdeg + \fas$ & 
       \tNPc & \cite{CheCsaRoySim2023Verif} &&
      \tFPT\fessymb  & \ourunpub & &
         \tFPT\fessymb  & \ourunpub
   & &  \tFPT\fessymb & \ourunpub &  &    \tNPc & \cite{CheCsaRoySim2023Verif}  & & \tFPT &  \textbf{[P\ref{prop:nashfas}]}\\
  \midrule
   Max coal.\ size~$\maxcoal$ & \tWoneh/\tXP & \cite{CheCsaRoySim2023Verif},\textbf{[P\ref{thm:maxcoalxp}]} &&
                                                                               \tWoneh/\tXP & \cite{sung2007core}/\cite{hanaka2022revisited}  & &
\graycell &\graycell  & & \graycell &\graycell & & \graycell &  \graycell & & \graycell &  \graycell\\
 \#Coal.~$\numcoal$ & \tNPc & \cite{CheCsaRoySim2023Verif} &&  \tNPc & \ourunpub   & & \graycell &\graycell & & \graycell &\graycell & & \graycell &  \graycell & & \graycell &  \graycell \\
  $\maxdeg + \maxcoal$ & \tFPT & \cite{CheCsaRoySim2023Verif}&&   \tFPT  & \ourunpub  & & \graycell &\graycell & &  \graycell &\graycell  &&\graycell &  \graycell & & \graycell &  \graycell\\
  $\maxdeg + \numcoal$ & \tNPc & \cite{CheCsaRoySim2023Verif}&&    \tFPT  & \ourunpub  & & \graycell &\graycell & & \graycell &\graycell  && \graycell &  \graycell & & \graycell &  \graycell \\
   $\fas + \maxcoal$ & \tFPT & \cite{CheCsaRoySim2023Verif},\textbf{[T\ref{hed:thm:maxcoalfas}]}&&  \tFPT\fessymb & \ourunpub & & \graycell & \graycell & & \graycell & \graycell  &&  \graycell &  \graycell & & \graycell &  \graycell\\
    $\maxdeg + \fas + \numcoal$ & \tNPc & \cite{CheCsaRoySim2023Verif} && \tFPT\fessymb  & \ourunpub  & & \graycell &\graycell & &  \graycell &\graycell & & \graycell &  \graycell & & \graycell &  \graycell\\
   \bottomrule
 \end{tabular}}
\end{table} 

\subsection{Preliminaries}\label{sec:HGstab}

Formally, the input of a hedonic game consists of a tuple~\myemph{$(\agents, (\succeq_i)_{v_i\in \agents})$}, where $\agents$ denotes a finite set of $n$ agents with $\agents=\{v_1,v_2,\cdots,v_n\}$ and each~$\succeq_i$, $v_i\in \agents$, represents the preferences of agent~$v_i$ over coalitions~$C \subseteq \agents$ that contain~$v_i$.
Such preferences are complete, reflexive, and transitive.
For instance~$C_1\succeq_i C_2$ means that agent~$\ag{i}$ \myemph{weakly prefers}~$C_1$ over $C_2$.
 From a preference list~$\succeq_i$, we derive the \myemph{symmetric~$\sim_i$} and \myemph{asymmetric} part~\myemph{$\succ_i$} as follows. %
We say that agent~$\ag{i}$ \myemph{strictly} prefers~$C_1$ to $C_2$, written as \myemph{$C_1 \succ_i C_2$},  if $C_1 \succeq_i C_2$, but $C_2 \not{\succeq_i} C_1$. 
Accordingly, we say that agent~$\ag{i}$ is \myemph{indifferent between} $C_1$ and $C_2$, written as \myemph{$C_1 \sim_i C_2$},  if $C_1 \succeq_i C_2$ and $C_2 \succeq_i C_1$.

We call the entire agent set~$\agents$ the \myemph{grand coalition}.
A \myemph{partition}\footnote{Aka.\ \myemph{coalition structure} in cooperative games.}~$\Pi$ of $\agents$ is a division of $\agents$ into disjoint coalitions, i.e., the coalitions in~$\Pi$ are pairwisely disjoint and $\bigcup_{\coal\in \Pi} \coal = \agents$.
Given a partition~$\Pi$ of $\agents$ and an agent~$\ag i\in \agents$, let~\myemph{$\Pi(\ag i)$} denote the coalition which contains~$\ag i$.

There are many stability concepts studied in the literature.
In the following, we review a few prominent ones.

\begin{definition}
  A coalition~$\bcoal$ is \myemph{blocking} a partition~$\Pi$ if every agent~$\ag i \in \bcoal$ strictly prefers~$\bcoal$ to $\Pi(\ag i)$, and it is \myemph{weakly blocking}~$\Pi$ if every agent~$\ag i \in \bcoal$ weakly prefers~$\bcoal$ to $\Pi(\ag i)$ and at least one agent~$\ag i\in \bcoal$ strictly prefers~$\bcoal$ to $\Pi(\ag i)$.

  A coalition~$\bcoal$ is \myemph{blocking} a partition~$\Pi$ if every agent~$\ag i \in \bcoal$ strictly prefers~$\bcoal$ to their current coalition $\Pi(\ag i)$.
  This represents a group of agents who would all benefit by deviating together.

  Similarly, $\bcoal$ is \myemph{weakly blocking}~$\Pi$ if every agent~$\ag i \in \bcoal$ weakly prefers~$\bcoal$ to $\Pi(\ag i)$ and at least one agent~$\ag i\in \bcoal$ strictly prefers~$\bcoal$ to $\Pi(\ag i)$. In this case, some agents strictly improve while others remain indifferent to the change.

  \begin{compactitem}[--]
    \item $\Pi$ is \myemph{core stable} (resp.\ \myemph{strictly core stable}) if \emph{no} coalition is strictly (resp.\ weakly) blocking~$\Pi$.
    \item $\Pi$ is \myemph{Nash stable} if no agent \myemph{envies} any other coalition in $\Pi$; an agent~$\ag{i}$ \myemph{envies} a coalition~$\bcoal$ if he strictly prefers~$\bcoal \cup \{\ag i\}$ to $\Pi(\ag{i})$. %
    \item $\Pi$  is \myemph{individually stable} if no agent~$\ag i$ and coalition $\bcoal\in  \Pi \cup \{ \emptyset\}$ form a \myemph{blocking tuple}; $(\ag{i}, \bcoal)$ is \emph{blocking} if
    $\ag i$ strictly prefers $\bcoal \cup \{\ag i\}$ to $\Pi(\ag{i})$ and each agent~$\ag j\in \bcoal$ weakly prefers $\bcoal\cup \{\ag i\}$ to $\Pi(\ag{j})$.
  \end{compactitem}
\end{definition}

By definition, we obtain the following relations among the just defined solution concepts.

\begin{flushleft}

\begin{minipage}{0.5\textwidth}

\begin{observation}\label{obs:stabilityrelation}
The stability concepts have the relations presented on the right:
Implication (1) holds since the agent in a blocking tuple must envy the coalition in the tuple.
Implication (2) holds since a blocking tuple forms a weakly blocking coalition.
Implication (3) holds since a blocking coalition is also a weakly blocking coalition.

\end{observation}
\end{minipage}
\begin{minipage}{0.45\textwidth}
\def \xx {0.7}
   \def \xy {0.3}

{\centering

 \begin{tikzpicture}[scale=2, black]
 \node[gadget] at (0,0) (a) {Individual stability};
 \node[gadget] at (0, \xy*2.5) (b) {Nash stability};
 \node[gadget] at (\xx*2.5, \xy*0) (c) {Core stability};
 \node[gadget] at (\xx*2.5, \xy*2.5) (d) {Strict core stability};
  \draw[edc] (b) edge node[midway, labelst] {\small \emph{implies} (1)} (a);
  \draw[edc] (d) edge node[midway, labelst] {\small \emph{implies} (2)} (a);
  \draw[edc] (d) edge node[midway, labelst] {\small \emph{implies} (3)} (c);
  \end{tikzpicture}
  
  }

\end{minipage}

\end{flushleft}

\subsubsection{Problem Definitions}\label{sec:HGProb}
The central algorithmic questions %
revolving around the above solution concepts are to decide whether a given partition satisfies one of the stability concepts,
to decide whether a desirable partition exists,
and to find a desirable partition.
The formal definitions of these three types of problems regarding core stability are given below.

\begin{mdframed}[style=ownframe]
  \decprob{\core-\verif}{
    A set of agents \agents, and preferences $(\succeq_i)_{\ag i \in \agents}$, and a partition~$\Pi$ of~$\agents$.
}{
  Is there a \emph{strictly blocking} coalition for~$\Pi$?
}
\end{mdframed}

\begin{mdframed}[style=ownframe]
  \decprob{\core-\exist}{
    A set of agents \agents, and preferences $(\succeq_i)_{\ag i \in \agents}$.}{
    Is there a core stable partition?
  }
\end{mdframed}

\begin{mdframed}[style=ownframe]
  \taskprob{\core-\find}{
    A set of agents \agents, and preferences $(\succeq_i)_{\ag i \in \agents}$.}{
    Find a core stable partition if it exists.
  }
\end{mdframed}

Analogously, we can define problems for strict core stability, Nash stability, and individual stability,
and call them
 \score-\verif, \score-\exist, \score-\find, \Nash-\verif, \Nash-\exist,  \Nash-\find,  \IndS-\verif, \IndS-\exist, \IndS-\find, respectively.

\begin{remark}\label{remark:hedonic-complexity-upperbound}
  By definition, \core-\verif\ and \score-\verif\ are contained in \NP,
  while \Nash-\verif\ and \IndS-\verif\ are in \PP.
  Hence, \core-\exist\ and \score-\exist\ are in $\sigmatwop$ while  \Nash-\exist\ and \IndS-\exist\ are in \NP\ as we can check every agent-coalition pair in polynomial time.

  Clearly, if an existence problem is computationally hard, then its search variant is as well.
  Hence, when discussing complexity results, we only talk about the search variant if its existence counterpart is in P.
\end{remark}

\subsection{Compact Representations of Agents' Preferences}\label{sec:HGCompact}
Since the number of possible coalitions that contain an agent is exponential in the number of agents, it may not be feasible to explicitly express a preference list.
In this section, we survey two models that compactly represent agents' preferences:
Hedonic Games with Additive Preferences~(\ADD) and 
Hedonic Games with Friends and Enemies~(\FE).

\paragraph{Hedonic Games with \AddPrefs~(\ADD).}

Additive preferences allow agents to give arbitrary utilities (as integral values) to other agents.
The preferences over a coalition are derived directly from the sum of the utilities of the agents in the coalition.

\begin{definition}[Additive preferences (\ADD)]
Let $\agents = \{\ag 1, \ag 2, \cdots, \ag n\}$ be a set of agents.
Every agent $\ag i \in \agents$ has a \myemph{utility function} $\util i \colon \agents \setminus \{\ag i\} \to \mathds{Z}$.
For each agent~$\ag i \in \agents$, the preference order~$\succeq_i$ is derived as follows:
For two coalitions~$S$ and~$T$ containing \ag i, agent \ag i \myemph{(strictly) prefers}~$S$ to~$T$, written as $S \succ_i T$, if $\sum_{v_j \in S} \util i (\ag j) > \sum_{v_j \in T} \util i (\ag j)$.
Agent \ag i is \myemph{indifferent} between $S$ and $T$, written as $S \sim_i T$, if $\sum_{v_j \in S} \util i (\ag j) = \sum_{v_j \in T} \util i (\ag j)$.
Hence, agent \ag i \myemph{weakly prefers}~$S$ to~$T$ if $S \succ_i T$ or $S \sim_i T$.
\end{definition}

\begin{flushleft}
\begin{minipage}{0.85\textwidth}
\begin{example}
Let $\agents = \{\ag 1, \ag 2, \ag 3, \ag 4\}$.
The utilities of the agents are presented as the arcs in the figure on the right: An arc $(\ag i, \ag j)$ with a value $x$ signifies that $\util i (\ag j) = x$. If there is no arc $(\ag i, \ag j)$, then $\util i (\ag j) = 0$.

The instance admits a core stable partition $\Pi_1=\{\{\ag 1\}, \{\ag 2\}, \{\ag 3\}, \{\ag 4\}\}$, where every agent obtains utility $0$.
This is however not individually stable, as $(\ag 2, \{\ag 4\})$ forms a blocking tuple, and by \cref{obs:stabilityrelation} neither strictly core stable nor Nash stable. 
The partition $\Pi_2=\{\{\ag 1, \ag 2, \ag 4\}, \{\ag 3\}\}$ is strictly core stable, as no coalition weakly blocks it.
By \cref{obs:stabilityrelation}, it is thus also individually stable and core stable. 
It is also Nash stable. %

\end{example}
\end{minipage}
\begin{minipage}{0.13\textwidth}
  \def \xx {.75}
  \def \xy {1.2}
  { \centering
\begin{tikzpicture}[black]
    \foreach \i / \j / \n in {3/3/1,2/2/2,4/2/3,3/1/4} {
      \node[pn] at (\i*\xx, \j*\xy) (\n) {};
    }
    \foreach \n / \p / \l in {1/above/-1, 2/left/-2, 3/right/-2, 4/below/-1} {
      \node[\p = \l pt of \n] {$\ag \n$}; 
    }
    \foreach \s / \t / \x in {1/3/-2,
    3/1/1,3/4/-1} {
      \draw[edc] (\s) edge[bend left=30] node[midway, labelst] {\x} (\t);
    }
        \foreach \s / \t / \x in {1/2/2,2/4/1,2/3/-1} {
      \draw[edc] (\s) edge[bend right=20] node[midway, labelst] { \x} (\t);
    }
\end{tikzpicture}

}
\end{minipage}
\end{flushleft}

\paragraph{Hedonic Games with Friends and Enemies~(\FE).}

\citet{dimitrov2006simple} introduce a simple yet natural restriction of the \addprefs\ model,
where each agent divides other agents into two categories, \emph{friends} and \emph{enemies}, and his preferences are based on the numbers of friends and enemies in the prospective coalition.
There are two variants, the \friends\ and \enemies\ models. 
In the \myemph{\friendapp}, the agents prefer a coalition with more friends to one with fewer friends. If two coalitions have the same number of friends, the one with fewer enemies is better. 
In the \myemph{\enemyav}, the agents instead prioritize a low number of enemies: The agents prefer a coalition with fewer enemies to one with more enemies. Subject to that, they prefer a coalition with more friends.
Such hedonic preferences can be compactly represented via a directed graph.
To this end, recall that for a directed graph~$G$ and a vertex~$v\in V(G)$, the sets~\myemph{$N^+_G(v)$} and \myemph{$N^-_G(v)$} denote the out- and in-neighborhood of~$v$.

\begin{definition}[Preferences from friends and enemies (\FE)]
  Let $\goodG$ be a directed graph on the agent set~$\agents$, called \myemph{friendship} graph, such that an agent~$\ag i$ regards another agent~$\ag j$ as a friend whenever $\goodG$ contains the arc~$(\ag i,\ag j)$; otherwise $\ag i$ regards $\ag j$ as an enemy.
  That is, $N^{+}_{\goodG}(\ag{i})$ consists of all friends while $\agents\setminus (N^{+}_{\goodG}(\ag{i})\cup \{\ag{i}\})$ all enemies of agent~$\ag{i}$.
  
  Under the \myemph{\friendapp}, the preferences are derived as follows:
  For each agent~$\ag i\in \agents $, %
  each  two coalitions~$S$ and $T$ containing~$\ag i$, agent $i$ \myemph{(strictly) prefers}~$S$ to~$T$, written as \myemph{$S \succ_i T$}, if 
\begin{compactenum}[(i)]
  \item either $|N^+_{\goodG}(\ag i)\cap S| > |N^+_{\goodG}(\ag i)\cap T|$, 
  \item or $|N^+_{\goodG}(\ag i)\cap S| = |N^+_{\goodG}(\ag i)\cap T|$ and $|S \setminus N^+_{\goodG}(\ag i)| < | T \setminus N^+_{\goodG}(\ag i)|$. 
\end{compactenum}
Agent~$\ag i$ is \myemph{indifferent} between $S$ and $T$, written as \myemph{$S \sim_i T$},
if  $|N^+_{\goodG}(i)\cap S| = |N^+_{\goodG}(i)\cap T|$ and $|S \setminus N^+_{\goodG}(i)| = | T \setminus N^+_{\goodG}(i)|$. 
Agent~$i$ \myemph{weakly prefers}~$S$ to~$T$ if $S\succ_i T$ or~$S\sim_i T$. 

Under the \myemph{\enemyav}, the preferences are instead derived as follows:
For two coalitions~$S$ and $T$ containing~$\ag{i}$, agent $\ag{i}$ \myemph{(strictly) prefers}~$S$ to~$T$, written as \myemph{$S \succ_i T$}, if 
\begin{compactenum}[(i)]
  \item either $|S \setminus N^+_{\goodG}(\ag i)| < | T \setminus N^+_{\goodG}(\ag i)|$,  
  \item or $|S \setminus N^+_{\goodG}(\ag i)| = | T \setminus N^+_{\goodG}(\ag i)|$ and $|N^+_{\goodG}(\ag i)\cap S| > |N^+_{\goodG}(\ag i)\cap T|$. 
\end{compactenum}
Agent~$\ag i$ is \myemph{indifferent} between $S$ and $T$, written as \myemph{$S \sim_i T$},
if $|S \setminus N^+_{\goodG}(\ag i)| = | T \setminus N^+_{\goodG}(\ag i)|$ and $|N^+_{\goodG}(\ag i)\cap S| = |N^+_{\goodG}(\ag i)\cap T|$.
Agent~$\ag i$ \myemph{weakly prefers}~$S$ to~$T$ if $S\succ_i T$ or $S\sim_i T$.
\end{definition}

\begin{remark}\label{remark:FEN->Add}
  Note that both models above are special cases of the \addprefs\ model:
  In the \friends\ case, set $\util{x}(y) = n$ if $(x,y)$ is an arc in the friendship graph, and $\util{x}(y)=-1$ otherwise,
  while in the \enemies\ case,   set $\util{x}(y) = 1$ if $(x,y)$ is an arc in the friendship graph, and $\util{x}(y)=-n$ otherwise.
\end{remark}

\begin{flushleft}

\begin{minipage}{0.75\textwidth}
\begin{example}\label{hg:ex1}
Let $\agents = \{\ag 1, \ag 2, \ag 3, \ag 4\}$ and the friendship graph be depicted on the right.
Under the \friendapp, the partition consisting of only the grand coalition~$\agents$ is stable according to all four discussed stability concepts.
The partition $\{\{\ag 1,\ag 2,\ag 3\}, \{\ag 4\}\}$ is strictly core stable and
hence core stable and individually stable, but it is not Nash stable since agent~\ag{4} envies coalition $\{\ag 1, \ag 2, \ag 3\}$.
\end{example}

\end{minipage}
\begin{minipage}{0.23\textwidth}

\def \xx {1}
   \def \xy {0.5}
  { \centering
\begin{tikzpicture}[black]
    \foreach \i / \j / \n in {3/3/1,2/2/2,4/2/3,3/1/4} {
      \node[pn] at (\i*\xx, \j*\xy) (\n) {};
    }
    \foreach \n / \p / \l in {1/above/-1, 2/left/1, 3/right/1, 4/below/-1} {
      \node[\p = \l pt of \n] {$\ag \n$}; 
    }
    \foreach \s / \t in {1/2, 2/1,1/3,3/1,4/2,3/4} {
      \draw[fc] (\s) edge[bend left=20] (\t);
    }
\end{tikzpicture}

}
\end{minipage}
\end{flushleft}
\vspace{-0.2cm}
\noindent \textit{
Under the \enemyav, the grand coalition is not stable, as some agents have enemies. The partition $\{\{\ag 1,\ag 2\}, \{\ag 3\}, \{\ag 4\}\}$ is core stable, but not strictly core stable, because $\{\ag 1,\ag 3\}$ blocks it weakly. As a matter of fact, there is no strictly core stable partition. The partition $\{\{\ag 1,\ag 2\}, \{\ag 3\}, \{\ag 4\}\}$ is individually stable, because no agent can join an other coalition without the agents in that coalition disimproving. It is not Nash stable, because \ag 3 prefers $\{\ag 3, \ag 4\}$ to $\{\ag 3\}$. As a matter of fact, there is no Nash stable partition.}

\subsection{Classical Complexities}\label{sec:HGClassic}

\cref{remark:hedonic-complexity-upperbound} already observes some complexity upper bounds for the verification and existence problems.
In the following, we provide some complexity lower bounds as well.

\paragraph{\ADD.}
Under \addprefs, \core-\exist\ is $\sigmatwop$-complete by~\citet{WOEGINGER2013101}
and  \score-\exist\ is $\sigmatwop$-complete by \citet{peters2017precise}.
\citet{Olsen09NSsymmetricadd} shows that \Nash-\exist\ is weakly \NP-complete.
Later, \citet{sung2010computational} show that \Nash-\exist\ and \IndS-\exist\ are strongly \NP-complete.

A special case of the \addprefs\ model is when the utility functions of the agents are \myemph{symmetric}.
Formally, it means that for every $\ag i, \ag j \in \agents$ we have that $\util{i}(\ag j) = \util{j}(\ag i)$.
\citet{peters2017precise} shows that \core-\exist\ and \score-\exist\ remain $\sigmatwop$-complete even when the preferences are symmetric.
\citet{bogomolnaia2002stability} show that for symmetric preferences, Nash stable and hence individually stable partitions always exist.
Finding one however may be difficult since the utility values may be exponentially large. 
\citet{gairing2019computing} show that \Nash-\find\ and \IndS-\find\ are indeed \PLS-complete.\footnote{
The complexity class Polynomial Local Search~(\PLS)~\cite{johnson1988easy} is located between \PP\ and \NP. 
\PLS\ contains local search problems for which local improvements can be found in polynomial time, but reaching a locally optimal solution might require an exponential number of such local improvements.
  A canonical \PLS-complete problem is finding a locally optimal truth assignment for \textsc{Max2SAT}, i.e., an assignment such that no single change of variable from true to false or vice versa can increase the number of satisfied clauses.} 
In contrast, additionally requiring the utilities to be polynomially bounded leads to polynomial-time solvability by \citet{bogomolnaia2002stability}: %

\begin{proposition}[\cite{bogomolnaia2002stability}\litresult]\label{prop:symmaddp}
  If the utilities are additive and symmetric, and have values at most $n^{O(1)}$,
  then every instance admits a Nash stable (and hence individually stable) partition, which can be found in polynomial time. 
\end{proposition}

\begin{proof}
  \citet{bogomolnaia2002stability} observed that for additive and symmetric utilities,
  by assigning an agent to another existing coalition~$B\in \Pi$ that he envies,
  we obtain a new partition which has a 
  strictly higher utilitarian welfare than the original one~$\Pi$,
  where %
  the \myemph{utilitarian welfare} is defined as the sum of the utilities of all agents towards their coalitions. 
  Since such an agent-coalition pair can be found in polynomial time, we can start from a partition with welfare $0$ (everyone alone), and the utilitarian welfare is at most $n^{O(1)}$, we must reach a Nash stable partition in polynomially many steps in $n$. This partition must also be individually stable by \cref{obs:stabilityrelation}.
\end{proof}

As for the verification problem, \citet{sung2007core} show that \core-\verif\ is \NP-complete under the \enemyav, which by \cref{remark:FEN->Add} implies that \core-\verif\ is also strongly \NP-complete under \addprefs.
\citet{aziz2013computing} observe that their reduction also works for \score-\verif.
\IndS-\verif\ and \Nash-\verif\ are clearly in \PP\ under all the discussed preference models; see \cref{remark:hedonic-complexity-upperbound}.

\paragraph{\FE.}
For \friends\ preferences, it is known that a strictly core stable partition
always exists and such a solution can be found in polynomial time: The strongly connected components of the friendship graph form a strictly core stable partition~\cite{dimitrov2006simple}.
By \cref{obs:stabilityrelation}, we obtain that \core-\exist, \IndS-\exist, and the corresponding search variants are in \PP\ as well.
\Nash-\exist\ is shown to be \NPh\ by \citet{brandt2022single}.
\citet{CheCsaRoySim2023Verif} prove that \core- and \score-\verif\ are \NP-hard, demonstrating a rare case where the existence and search variant is computationally easier than the verification problem.

For \enemies\ preferences, \citet{dimitrov2006simple} show that a core stable partition always exists (i.e.\ \core-\exist\ is in \PP), although finding one is \NP-hard.
\score-\exist, on the other hand, is beyond NP~\cite{rey2016toward}. 
\Nash-\exist\ is difficult as well: \citet{brandt2022single} show it is \NP-complete.
\IndS-\exist, on the other hand, is in \PP~\cite{brandt2022single}.

As mentioned earlier, both \core-\verif\ and \score-\verif\ are \NP-complete~\cite{sung2007core,aziz2013computing}.

\subsection{Parameters}\label{sec:HGparam}

\paragraph{\ADD.} We introduce a directed graph, called \myemph{preference graph} $\aG = (\agents, E)$, where there is an arc from $\ag{i}$ to $\ag{j}$ if $\util{i}(\ag j) \neq 0$. 
Consequently, one can study the following parameters for \addprefs. 
\begin{compactitem}[--]
\item Max degree~$\displaystyle\myemph{\maxdeg}\coloneqq \max_{\ag i \in \agents} |N^+_{\aG}(\ag i) \cup N^-_{\aG}(\ag i)|$, i.e., the maximum number of agents an agent may have a non-zero utility towards and who may have a non-zero utility towards him.
\item The number of different values~$\myemph{\nrutil}$ the utility function may take.
\item Feedback arc set number~$\myemph{\fas}$ of the preference graph~$\aG$, we use this typesetting to distinguish from the typical function name $f$. %
\item Treewidth $\myemph{\tw}$ of the underlying \emph{undirected graph} of~$\aG$. 
\end{compactitem}

\paragraph{\FE.} The following parameters have been studied in Hedonic Games with Friends and Enemies:
\begin{compactitem}[--]
  \item Max degree~$\myemph{\maxdeg}\coloneqq \max_{i\in V} |N^+_{\goodG}(i)\cup N^-_{\goodG}(i)|$, i.e., the maximum number of friends an agent may have plus the maximum number of agents who consider him a friend. 
  \item Under the \friendapp\ and \enemyav\ for Nash stability, we look at the feedback arc set number~$\myemph{\fas}$ in the friendship graph.

  Under the \enemyav\ for stability concepts other than Nash stability, we look at the feedback edge set number~$\myemph{\fas}$ in the underlying undirected graph of the friendship graph 
  since we can assume that the friendship arcs are bidirectional. 
  The reason is that for \enemies, no agent would weakly prefer to be in a coalition with at least one enemy.
\end{compactitem}

\paragraph{Verification.} For the verification problem, we additionally have the following parameters:
\begin{compactitem}[--]
  \item Max.\ coalition size~$\myemph{\maxcoal} \coloneqq \max_{\coal\in \Pi}|A|$, i.e., the size of the largest coalition in the partition from the input.
  \item The number of initial coalitions~$\myemph{\numcoal}\coloneqq |\{\coal\in \Pi\}|$.
\end{compactitem}

\subsection{Parameterized Complexities}\label{sec:HGparamC}

This section presents parameterized complexity results for hedonic games, summarized in \cref{table:additive,table:FE}. 
Again to ensure completeness, we also provide several new results previously absent from the literature, marked with (\newresult). 
Established results are referenced as ([cite]\litresult).

We organize the results according to three primary techniques:

\begin{description}
  \item[Brute force approaches (\cref{thm:maxcoalxp,prop:nashfas}):]
  Exploiting small parameter values to exhaustively search bounded solution spaces, often after simple preprocessing.

  \item[Color-coding (\cref{hed:thm:maxcoalfas}):]
     Applying (de)randomized color-coding to detect sparse substructures of bounded size within coalition networks.

    \item[Hardness reductions (\cref{prop:maxdegen,prop:twaddsverif}):]  
    Strengthening existing NP-hardness proofs by adapting reductions from the literature. 

\end{description}

\newcommand{\bsize}{\ensuremath{\kappa'}}

The following straightforward preprocessing result bounds the size of ``interesting'' blocking coalitions. Such preprocessing steps are often an integral part of parameterized algorithms.
\begin{lemma}[\cite{CheCsaRoySim2023Verif}]\label{cl:largecoalfriend}
Under the \friendapp, checking a given partition~$\Pi$ admits a (weakly) blocking coalition $\bcoal$ with $|\bcoal| > \maxcoal$ can be done in polynomial time; recall that $\maxcoal$ denotes the size of the largest coalition in~$\Pi$.
\end{lemma}

\begin{proof}%
  Let $B$ be such a blocking coalition with $|B| > \maxcoal$. 
  Then, every agent~$v_i$ in~$B$ must have more friends in~$B$ than in~$\Pi(v_i)$ as otherwise $v_i$ will have more enemies and will not weakly prefer $B$ to $\Pi(v_i)$.
  This implies we can start with the grand coalition~$S\coloneqq \agents$ and recursively delete an agent~$v_i$ from~$S$ for which the number of friends in his initial coalition~$\Pi(v_i)$ is no less than the number of friends that remain in~$S$. 
  We stop when no more agent is deleted from~$S$.
  Clearly, if $S$ is not empty, then it is the desired blocking coalition.
  Otherwise, we can return that no such blocking coalition exists.
  The procedure can clearly be done in polynomial time. 
\end{proof}

\cref{cl:largecoalfriend} immediately implies the following \XP\ result for~$\csize$.
\begin{proposition}[\cite{CheCsaRoySim2023Verif}]\label{thm:maxcoalxp}

Under the \friendapp, \core-\verif\ and \score-\verif\ parameterized by~$\maxcoal$ is in \XP.
\end{proposition}

\begin{proof}
By \cref{cl:largecoalfriend} we only need to consider potential blocking coalitions of size at most $\maxcoal$ (resp.\ $\maxcoal+1$).
We can iterate through all subsets of size at most $\maxcoal$ in $O(n^{\maxcoal}\cdot n\cdot m)$ time, which is \XP\ when parameterized by $\maxcoal$.
\end{proof}

The following proof is based on color coding.
This technique has been applied to design \FPT\ algorithms for many COMSOC problems~\cite{DS2012SwapBribery,misra2015parameterized,Gupta20,CCR2023OptimalSeat}. %

\begin{theorem}[\cite{CheCsaRoySim2023Verif}]\label{hed:thm:maxcoalfas}
Under the \friendapp, \core-\verif\ parameterized by $(\maxcoal, \fas)$ is fixed-parameter tractable.
\end{theorem}

\begin{proof}[Proof Sketch]
  Let $\goodG = (\agents,E)$ be a friendship graph and $\Pi$ an initial partition.
  The algorithm has two phases.
  First,
  we preprocess the instance so that each non-trivial blocking coalition has at most $\maxcoal$ agents and the reduced instance excludes some undesired cycles.
  Second, we further reduce the friendship graph to one which is acyclic and observe that any non-trivial blocking coalition must ``contain'' an in-tree of size $O(\maxcoal^2)$.
  Hence, for each possible in-tree we can use color-coding~\cite{Alon1995Colorcoding} to check whether it exists in \FPT-time.
  
  We call an agent $\ag i \in \agents$ a \myemph{singleton} if $\Pi(\ag i) = \{\ag i\}$; otherwise he is a \myemph{non-singleton}.
  Let $\singles$ and $\nonsingles$ denote the sets of singleton and non-singleton agents, respectively.
  
 \mypar{Phase one.} We perform the following polynomial-time steps, which verify that there are no ``trivial'' blocking coalitions:
  \begin{compactenum}[\textbf{(P}1\textbf{)}]
    \item Check whether $\Pi$ contains a coalition~$\coal$ such that $\goodG[\coal]$ is not strongly connected.
    If yes, then return NO since the strongly connected subgraph corresponding to the sink component in~$\goodG[U]$ is strictly blocking~$\Pi$. 
    \item If $\goodG[\singles]$ contains a cycle, then the singleton agents on the cycle are strictly blocking $\Pi$, so return NO.  \label{fe-core-fpt-k-delta_prep_2}%
    \item\label{P3} Use the algorithm behind \cref{cl:largecoalfriend} to check in polynomial time whether there is a blocking (resp.\ weakly blocking) coalition of size greater than $\maxcoal$.
\end{compactenum}
Since every non-singleton coalition in $\Pi$ is a connected component, each of them must contain at least one feedback arc.
Recall that each initial coalition has at most~$\maxcoal$ agents.
Thus, the number of non-singletons is at most $\fas \cdot \maxcoal$.
Moreover, by P\ref{P3}, any blocking coalition is of size at most~$\maxcoal$.

Now that we have ensured there are no trivial blocking coalitions, we move to the second phase.
\mypar{Phase two.}
  For each subset~$\Bns\subseteq \nonsingles$ of $k'\le \maxcoal$ non-singletons and each size~$b$ with $k' \le b\le \maxcoal$,
  we check whether there exists a blocking coalition of size~$b$ which contains all non-singletons from $\Bns$, no other non-singletons, and exactly $b-|\Bns|$ singletons; note that after phase one, we only need to focus on coalitions of size at most $\maxcoal$ and can assume that $|\nonsingles|\le \maxcoal\cdot \fas$.
  We return YES if and only if no pair~$(\Bns, b)$ can be extended to a blocking coalition.

  Given $(\Bns, b)$, the task reduces to searching for the $b-|\Bns|$ missing singleton agents, assuming that such an extension is possible.
  To achieve this, we reduce to searching for an in-tree of size $O(\maxcoal^2)$ in a directed acyclic graph~(DAG)~$H$, which using color coding, can be done in $g(\maxcoal)\cdot |H|^{O(1)}$ time.
  First of all, if $|\Bns| = b$, then we check whether $\Bns$ is blocking (resp.\ weakly blocking) in polynomial time and return $\Bns$ if this is the case; otherwise we continue with a next pair~$(\Bns, b)$.
  In the following, let $B$ be a hypothetical blocking coalition of size~$b> |\Bns|$ which consists of~$\Bns$ and $b-|\Bns|$ singleton agents. 
  The search has two steps.

  \begin{compactenum}[\textbf{(C}1\textbf{)}]
    \item \textbf{Construct a search graph~$\hgoodG$ from $\goodG$.} Based on $\goodG$ we construct a DAG~$\hgoodG$, where
  we later search for the crucial part of the blocking coalition.
  We compute the minimum number~$r(a_i)$ of singleton-friends each non-singleton agent~$\ag i$ in $\Bns$ should obtain from $B$ by checking how many friends he has in $\Bns$, and how many friends and enemies he had initially.
  
  Afterwards, we check whether some non-singleton~$\ag i\in \Bns$ satisfies $r(\ag i) > b - |\Bns|$.
  If $\ag i$ is such an agent, then he will not weakly prefer~$B$ to~$\Pi(\ag i)$ since there are not enough friends for him, so we continue with a next pair~$(\Bns,b)$.

  Now, we construct~$\hgoodG$ from $\goodG$ as follows.
  First, we remove all non-singletons (and their incident arcs) that are \emph{not} in $\Bns$ and add an artificial sink vertex~$t$.
  Second, we replace each remaining non-singleton~$a_i\in \Bns$ with $r(\ag i)$-many copies~$a_i^{z}$, $z\in [r(\ag{i})]$
  and replace every arc~$(a_i, s)$ from~$a_i$ to a singleton with $r(\ag i)$ arcs~$(a^z_i, s)$. 
  Third, we replace every arc~$(s, a_i)$ from a singleton~$s$ to a non-singleton with
  an arc~$(s,t)$ from $s$ to the sink~$t$.
  Formally, the arcs in $\hgoodG$ are~\myemph{$\{(a^z_i, s) \mid (a_i, s) \in E(\goodG)
  \cap (\Bns\times \singles), z\in [r(a_i)]\}
  \cup  E(\goodG[\singles]) \cup \{(s, t) \mid (s, a_i) \in E(\goodG) \cap (\singles \times \Bns)\}$}.
  Note that $\hgoodG$ is acyclic since by (P\ref{fe-core-fpt-k-delta_prep_2}) no singleton agents induce a cycle.

  \item \textbf{Search for a tree structure in $\hgoodG$.}
  Observe that in a potential blocking coalition, each non-singleton~$\ag i\in \Bns$ has at least $r(\ag i)$ singleton friends and each singleton agent has at least one friend.
As we translate this blocking coalition to $\hG$, it must correspond to an in-tree rooted at $t$ with the following properties:
\begin{compactenum}[(T1)]
  \item \label{tree1} Every copy of an agent in $\Bns$ has exactly one out-neighbor that is a singleton, and no other copy of the agent has the same out-neighbor.
  \item \label{tree2} Every singleton vertex has exactly one out-neighbor and this out-neighbor is either the root~$t$ or some singleton vertex.
\end{compactenum}
\end{compactenum}

The copies of non-singletons are fixed, but we do not know which singleton-agents are a part of this potential blocking coalition. 
However, as the number of vertices is bounded, the number of different structures this tree can have is also bounded.
We can guess the search tree structure and search for it in \FPT-time by applying the color-coding algorithm inspired by Alon et al.~\cite{Alon1995Colorcoding}.
\end{proof}

Under the \enemyav, \core-\find\ and \score-\verif\ are \FPT\ with respect to~$\fas$~\cite{durand2024enemies}, so there is no need to combine it with $\maxcoal$.

Next we move to discuss \Nash-\exist. The problem is \conp-complete under the \friendapp~\cite{CheCsaRoySim2023Verif} even when the feedback arc set number~$\fas$ is a constant.
However, under the \enemyav, the parameter~\fas\ bounds the number~$n$ of agents who can be in coalitions with other agents, and thus trivially yields an \FPT-algorithm.
Note that we do not need to consider the special case when the preferences are symmetric since then a Nash stable partition can be found in polynomial time by~\cref{prop:symmaddp} and the fact that the utilitarian welfare is at most $n^2$.

\begin{proposition}[\newresult]\label{prop:nashfas}
Under the \enemyav, \Nash-\exist\ parameterized by the feedback arc set number \fas\ is \FPTlong.
\end{proposition}

\begin{proof}[Proof sketch.]
In a Nash stable partition, every agent must be in a coalition that contains only friends. Since every mutual friendship is 2-cycle, the number of agents $\ag i \in \agents$ who can be in a coalition other than $\{\ag i\}$ is bounded above by $2 \fas$. We can exhaustively try every possible partition of these agents -- the other agents are in their own coalitions -- and verify whether any of these partitions is  Nash stable.
\end{proof}

We finish the section by extending two known results from the literature with straightforward modifications.
First, we observe that the \NP-hardness reduction for \Nash-\exist\ under the \enemyav\ from \citet{brandt2022single} can be modified to have a bounded degree.

\begin{proposition}[\cite{brandt2022single}\litresult]\label{prop:maxdegen}
Under the \enemyav, \Nash-\exist\ is \NP-complete even when the maximum degree~$\maxdeg$ is a constant.
\end{proposition}

\begin{proof}[Proof sketch]
\citet{brandt2022single} shows that \Nash-\exist\ is \NP-complete under the \enemyav. 
They reduce from \textsc{Exact Cover by 3-Sets}. %
By \citet{Gonzalez1985}, \textsc{Exact Cover by 3-Sets} is NP-hard even when each element appears in exactly three sets, which lets us bound the degree of all the agents in the reduction except an artificial sink~$c$. We bound the degree of $c$ by copying it so that every agent who has an edge to $c$ has his own copy of $c$.
\end{proof}

Next, we show that minor modifications extend the hardness result for \core-\verif\ under \ADD\ from \citet{hanaka2024core} to \score-\verif.
Note that since the minimum vertex cover size of a graph upper-bounds the treewidth, the result below implies that \score-\verif\ is weakly \NP-complete on graphs with constant treewidth. %

\begin{proposition}[\cite{hanaka2024core}\litresult]\label{prop:twaddsverif}
Under the \ADD-model, \score-\verif\ is weakly \NP-complete even on graphs that admit a vertex cover of size four.
\end{proposition}

\begin{proof}[Proof sketch]
\citet[Theorem 2]{hanaka2024core} show that \core-\verif\ is weakly \NP-complete even on graphs that admit a vertex cover of size two under \ADD.
They reduce from the \textsc{Partition} problem. 
We modify their reduction by adding edges with high negative weights, for example $-5B$ from the vertices $x'$ and $y'$ to every other vertex, except respectively to vertices $x$ and $y$, where $B$ is the total sum of the integers in the  \textsc{Partition} instance.
Now, $\{x, y, x', y'\}$ is a vertex cover of the modified instance.
Moreover, the agents $x'$ and $y'$ cannot join a weakly blocking coalition, because any coalition other than their initial coalition has a large negative utility.
From here on, the proof proceeds analogously to \citet{hanaka2024core}.
\end{proof} %

Our result tables leave behind several open questions for parameters regarding feedback arc set number and treewidth.
The complexity of \core-\exist\ and \score-\exist\ with respect to~$\maxdeg + \nrutil + \fas$ is also open. We conjecture this to remain $\sigmatwop$-hard even for constant parameter values, because the corresponding verification problems remain \NP-hard.
Moreover, parameterized complexity results that are obtained so far suggest that hedonic games remains computationally intractable for many canonical parameters with constant values.
Hence, a major challenge lies in searching for tractable yet well motivated parameterizations.

\challengeF{
  Identify meaningful and practically motivated parameters which give rise to parameterized algorithms for hedonic games. 
}

\section{Further Research in \probMW{} and \probHedonic}
We give a brief discussion of additional voting rules and preference structures in the study of multi-winner determination, and additional preference representation and solution concepts in the study of hedonic games.

\subsection{Multi-Winner Determination}\label{sec:FurtherMW}
\paragraph{Other voting rules.} We list here a few more interesting voting rules. %
\begin{compactitem}[--]
\item For approval preferences, the voting rules \CC\ and \PAV\ are special cases of the \myemph{Thiele voting rules}\footnote{Also referred to as weighted \PAV\ rules or the generalized approval procedures.}. Sornat et al.~\cite{Sornat2022} proved that under the Thiele rules \probMW{} with $\numdelalt$-\SP\ profiles can be solved in~$O^*(2^{\numdelalt})$~time\footnote{As common, the $O^*()$ notation suppresses the polynomial factor.} and that this running time is tight under ETH. 
\item Gupta et al.~\cite{GuptaJain23} study the parameterized complexity of \probMW\ for voting rules, for which a voter's score for a committee depends on his $\lambda$ highest-ranked alternatives in the committee, for some constant $\lambda$. They also developed an $O^*(2^n)$-time algorithm for \probCC\ and showed that it is optimal assuming the \textsc{Set Cover} Conjecture holds. 
\item Ayadi et al.~\cite{Ayadi19} present \FPT-algorithms (parameterized by $n$ and $m$, respectively) for \probMW\ for the \myemph{single transferable vote (STV)} rule for the case where each voter only ranks his $\ell$ favorite alternatives, for some given constant $\ell$.
\item A \myemph{Gehrlein stable committee} is a committee where each alternative in the committee is preferred to each alternative not in the committee by at least half of the voters. Gupta et al.~\cite{GuptaJain20} study the parameterized complexity of finding a fixed-size Gehrlein stable committee using the parameters number~$n$ of voters, $k$, the number of missing arcs in the majority graph, and the size of the tournament vertex deletion set for the majority graph. 
\end{compactitem}

\paragraph{Other structured preferences and frameworks.}
We present other preference structures and frameworks that do not constrict viable committees only by size.
\begin{compactitem}[--]
  \item \myemph{Single-peaked width} is a parameter that can be seen as a distance to single-peakedness, similarly to treewidth measuring the distance to a tree. Cornaz et al.~\cite{Cornaz12} developed an \FPT\ algorithm for \probCC\ parameterized by the single-peaked width,
  by modifying the dynamic programming approach of Betzler et al.~\cite{BetzlerMW2013}.
\item Peters et al.~\cite{PetersLackner20} study \probMW\ with generalized single-peaked preferences: \myemph{single-peaked on a circle}, or \myemph{a tree}.
\item Faliszewski et al.~\cite{Faliszewski20} study voting rules for which the committee size is not fixed. They provided classical and parameterized complexity results.
\item \MAV\ can be considered as a voting rule that minimizes the maximum distance under $p$-norm with $p=1$.
One can also instead look at other $p$-norms~\cite{Sivarajan2018MAVpnorm}.
Chen et al.~\cite{CheHerSor2019pnorm} study \MAV\ for all $p$-norms and for committees that do not have fixed size.
They provide fixed-parameter algorithms for the number of voters and alternatives, respectively.
\item Yang and Wang~\cite{Yang18} investigate voting rules for which the committees are not constrained by size but by other constraints. In their model, alternatives are represented by vertices in a graph, and only committees that have a certain graph structure are valid.
They study the classical and parameterized complexity of \probCC, \probPAV\ and \probMW\ for several voting rules, where an agent's score for a committee depends on how many of their approved alternatives are in the committee and not in the committee. 
\end{compactitem}

\paragraph{Strategic Behavior in Multi-Winner Elections.}

Multi-winner elections have been extensively studied through the lens of strategic behavior, including manipulation~\cite{RPRZ2008MWManipulation}, control~\cite{YangMWManipulationControl2019}, and bribery~\cite{FST2017MWBribery}.

For approval-based multi-winner voting rules, \citet{YangMWManipulationControl2019} provided a comprehensive parameterized complexity analysis of both manipulation and control of voting rules including \PAV\ and \MAV.
He identified several fixed-parameter tractable cases for the canonical parameters ``number of alternatives'', ``number of voters'', and ``number of modifications to the original profile'' in the control case.
Similarly, \citet{BKN2021MWManipulationn} examined parameterized complexity of manipulation of a simple voting rule called $\ell$-approval, for the linear preferences case.

The bribery problem—where an external agent attempts to influence the election outcome by changing voters' preferences—has also received significant attention.
\citet{FST2017MWBribery} investigated the parameterized complexity of constructive bribery of approval-based voting rules, including \CC\ and \PAV,
while \citet{Yang2020MWDestructiveBribery} focused on destructive bribery, where the goal is to prevent specific candidates from being elected.
\citet{BFNT2021SB} systematically investigated the parameterized complexity of a simple variant of bribery problem, called \emph{shift bribery}.

 In a recent paper, \citet{tech-Yang2023DControl} initiated the parameterized study of destructive control of the voting rules he investigated in his constructive control paper~\cite{YangMWManipulationControl2019}.

%
%
%
%

%
%

\subsection{Hedonic Games}\label{sec:FurtherHG}

\paragraph{Other preference representations.}

There are many ways to compactly present agents' preferences over the coalitions in the literature. We present some common ones below:

\begin{compactitem}[--]
\item \myemph{Hedonic Games with Friends, Enemies, and Neutrals}: This extension of Hedonic Games with Friends and Enemies introduced by \citet{ohta2017core} allows agents to be neutrals towards other agents.
Neutral agents do not affect agents preferences towards a coalition.
This model has also been extend to allow agents prefer having more or fewer neutral agents by \citet{barrot2019unknown}.
  \item \myemph{$\mathcal{B}$-} and \myemph{$\mathcal{W}$}-preferences~\cite{cechlarova2001stability}: The agents have preferences over the other agents, and the coalitions are ranked according to the best (i.e., most-preferred) or worst (i.e., least-preferred) agent, respectively.
\item \myemph{Hedonic coalition nets}~\cite{elkind2009hedonic}: Preferences are based on logical formulas.
\item \myemph{Fractional hedonic games}~\cite{aziz2019fractional}: The utility of the coalition is the average value given to the coalition.

\item \myemph{Graphical hedonic games}~\cite{peters2016graphical}: The preferences of an agent only depend on his neighbors on a graph.
\end{compactitem}

\paragraph{Coalitions with fixed or bounded sizes.}

Sometimes, we need to partition the agents into coalitions of the same fixed size~$\fixcoal$,
called \myemph{$\fixcoal$-dimensional matching}.
For instance, for $\fixcoal=2$, we are looking for a matching.
It is straightforward that in this case, core stability corresponds to the well known \myemph{stable matching}.
There are two famous problem variants for the study of stable matching: \textsc{Stable Marriage} (the bipartite case) and \textsc{Stable Roommates} (the non-bipartite case)~\cite{GaleShapley1962}.
Since the latter is NP-complete~\cite{Ronn1990}, \core-\exist\ remains NP-complete even if we require each coalition to be of size~$\fixcoal=2$.

There has been research into restricted preferences such as \myemph{metric} preferences~\cite{woeginger2013core}, \myemph{Euclidean} preferences~\cite{arkin2009geometric,Chen2022Euclid2dSR}, narcissistic, single-peaked, and single-crossing preferences~\cite{BarTri1986,BreCheFinNie2020-spscSM-jaamas}, as well as preferences with master lists~\cite{BredHeeKnoNie2020multidimensional}.

When the coalitions are of size \emph{at most} $\fixcoal$, \core-\exist\ is NP-complete on \ADD\ for every $d \geq 3$, even when the preferences are symmetric, and positive, and \score-\exist\ is NP-complete even when the preferences are symmetric, binary, and positive~\cite{levinger2024bounded}.
\citet{bilo2022hedonic} study the case where there is a list $(n_1, \cdots, n_{\ell})$ of coalition sizes, and we must find a partition $\Pi =\{P_1, \cdots P_{\ell}\}$ such that $|P_i| = n_i$ for every $i \in [\ell]$. They focus on a stability concept called swap-stability.
\citet{li2023partitioning} focus on the case when $n$ agents must be partitioned into coalitions of size $\lceil \frac{n}{k} \rceil$ or $\lfloor \frac{n}{k} \rfloor$, where $k \in [n]$ is a constant, and show some approximation guarantees.
As this is a generalization of \textsc{Stable Roommates} (consider $k = n/2$, $n$ is even), determining the existence of a core stable partition in this case remains NP-hard.

\paragraph{Other solution concepts.}
The stability concepts discussed in the previous sections are not the only ones studied in the literature.
Here, we discuss a few more stability concepts that have been studied in the context of \ADD.
Unless stated otherwise, all results hold for \ADD.

Many of these solution concepts have not been studied from the parameterized complexity viewpoint, opening many possible new research directions.

\begin{description}[leftmargin=1ex]

\item[Partitions that maximize social welfare.]
Not all solution concepts are related to the existence of (subsets of) agents that might wish to deviate. There are also solution concepts that wish to maximize either the social welfare, or the welfare of the worst-off agent.
\begin{compactitem}[--] 
\item Partitions with \myemph{maximum utilitarian welfare}: Find a partition $\Pi$ that maximizes $\sum_{P \in \Pi} \sum_{\ag i \in P} \util{i}(P)$.
Clearly, such a partition always exists.
Finding one however is NP-hard and the verification problem is NP-complete~\cite{aziz2013computing}. \citet{hanaka2019computational} study the parameterized complexity regarding treewidth and the number of apexes. 

\item Partitions with \myemph{maximum egalitarian welfare}: Find a partition $\Pi$ that maximizes $\min_{P \in \Pi} \min_{\ag i \in P} \util{i}(P)$.
Again, such a partition always exists.
Finding one however is NP-hard and the verification problem is NP-complete~\cite{aziz2013computing}. \citet{hanaka2019computational} study the parameterized complexity regarding treewidth, the number of apexes and the vertex cover number. 
\end{compactitem}

\item[{Global stability concepts.}]
The following two stability concepts compare different partitions with each other, and require that either everyone or at least half of the agents prefer the other partition:
\begin{compactitem}[--]
  \item \myemph{Pareto optimal}:
  A partition $\Pi$ is Pareto optimal if there is no other partition $\Pi'$, where
  every agent weakly prefers~$\Pi'$ to $\Pi$ and at least one agent strictly prefers~$\Pi$ to~$\Pi'$.
  It is straightforward that a partition with maximum utilitarian welfare is also Pareto optimal.
  \citet{aziz2013computing} show that verifying whether a given partition is Pareto optimal is however NP-complete and that finding a partition that is both Pareto optimal and individually rational (no agent is better of leaving his partition alone) is NP-hard.
  \citet{bullinger2020pareto} shows that under some restricted cases, such as symmetric preferences, a Pareto optimal partition can be computed in polynomial time. \citet{elkind2020price} study the welfare lost by enforcing that a partition is Pareto optimal.
\citet{peters2016graphical} show that both finding and verifying a Pareto-optimal partition are \FPT\ parameterized by the combined parameter $\tw + \maxdeg$.
\item \myemph{Popular}: No other partition wins in a pairwise election. By \citet{aziz2013computing}, these are hard to both find and verify. \citet{brandt2022finding} extend these results to symmetric preferences and for a related condition called strong popularity.
\end{compactitem}

  \item[Other stability concepts.]
Contractual stability concepts extend already known stability concepts so that if an agent or agents wish to leave a coalition, the coalition that is left must not disimprove.
Wonderful stability enforces that every agent is in a coalition containing only friends, and is in the largest friendship clique available to him.
\begin{compactitem}[--]

\item \myemph{Contractual strict core}: Let $\bcoal$ a coalition that weakly blocks a partition $\Pi$. Then~$\bcoal$ blocks $\Pi$ under contractual strict core only if there are no agents $\ag i \in \agents$,  $ \ag j \in \Pi(\ag i)$ such that $\Pi(\ag j) \succ_j \Pi(\ag j ) \setminus \bcoal$, i.e., \ag j disimproves if the agents in~\bcoal\ leave their coalitions in $\Pi$. \citet{sung2007myopic} show that a contractually strictly core stable partition must exist, and provide an algorithm that runs in the worst case in exponential time. \citet{aziz2013computing} show that verifying contractual strict core is weakly NP-hard, even if the initial partition is the grand coalition.

\item \myemph{Contractually Nash stable}:
Let $\ag i$ be an agent who envies another coalition in a partition~$\Pi$.
He blocks $\Pi$ under contractual Nash stability only if no agent in $\Pi(\ag i)$ disimproves if $\ag i$ leaves. \citet{sung2007myopic} show that a contractually Nash stable partition does not necessarily exist, unless the preferences satisfy a condition called weak mutuality. \citet{bullinger2022boundaries} shows that deciding whether one exists is NP-hard. Under the friends and enemies model, there is always a contractually Nash stable partition, which can be found in polynomial time under both the \friendapp\ and  \enemyav~\cite{brandt2022single}.

\item \myemph{Contractually individually stable}: Let $(\ag i, \coal)$ be a blocking tuple for a partition $\Pi$. The tuple  blocks $\Pi$ under contractual individual stability only if no agent in $\Pi(\ag i)$ disimproves if $\ag i$ leaves.
\citet{aziz2013computing} prove that there is always a contractually individually stable partition, which can be found in polynomial time.  \citet{bogomolnaia2002stability} discuss relations between contractual individual stability and other solution concepts.

\item \myemph{Wonderfully stable}~\cite{woeginger2013core}:
This solution concept is proposed by Woeginger~\cite{woeginger2013core} and is defined for the friends and enemies model.
A partition is wonderfully stable if every agent is in a coalition that corresponds to the largest clique that contains him. 
\citet{woeginger2013core} shows that under \enemyav, determining whether a wonderfully stable partition exists is contained in $\Theta^P_2$, and \citet{rey2016toward} show that it is DP-hard.
\end{compactitem}
\end{description}

Combinations of the above properties have also been studied, see for example~\cite{aziz2013computing,bullinger2020pareto,barrot2019stable}.

\section{Conclusion}\label{sec:conclusion}

In this survey, we have examined the parameterized complexity landscape of two central problems in computational social choice: \probMW{} and \probHedonic.
Our analysis uncovers several salient patterns:

\begin{compactitem}[--]
    \item \textbf{Parameter Effectiveness:}
    Natural parameters such as the number of alternatives or voters/agents often admit straightforward \FPT\ algorithms via brute-force enumeration. In contrast, parameters like committee size~$\csize$ or maximum degree~$\maxdeg$ in the friendship graph frequently induce parameterized intractability (\Woneh, \Wtwoh, or even \NP-hardness).

    \item \textbf{Combined Parameters:}
    While many problems remain hard when parameterized by a single parameter, combined parameterizations often yield tractability. This emphasizes the inherently multivariate nature of computational difficulty in COMSOC problems.

    \item \textbf{Structured Preferences:}
    Restricting preferences to structured domains (e.g., single-peaked or single-crossing profiles) typically leads to polynomial-time algorithms.
    Moreover, parameterizations that measure ``distance to tractability''--quantifying deviation from ideal structures--have proven both theoretically robust and practically meaningful.
    
    \item \textbf{Proof Techniques:}
    The study of parameterized complexity in COMSOC leverages a wide spectrum of techniques,
    ranging from brute-force enumeration to kernelization, color-coding, and dynamic programming.
    Beyond theoretical significance, these methods could offer a foundation for practical algorithmic implementations.
  \end{compactitem}

Beyond voting and hedonic games, two additional subfields—\probMatchup{} and \probResource{}—have seen significant progress in parameterized complexity.

For \probMatchup{}, foundational work by \citet{Marx10,marx2011stable} on stable matchings with ties and hospital-residents problems with couples established key tractability boundaries. 
This was followed by results such as \citet{Adil18} on stable matchings with ties, \citet{CHSYicalp-par-stable2018} on egalitarian objectives and minimizing blocking pairs in the stable roommates setting, and \citet{Gupta21} on balancing objectives in stable marriage. 
Further contributions, including those by \citet{Chen2019SMsurvey,Meeks20,Gupta20,Gupta22,CSS2025Refugee}, have extended the study of parameterized stable matching to new structural variants, refined objective functions, and emerging application settings.
Looking ahead, the upcoming Dagstuhl Seminar on \emph{Frontiers of Parameterized Algorithmics of Matching under Preferences}\footnote{\url{https://www.dagstuhl.de/en/seminars/seminar-calendar/seminar-details/25342}} (August 17–22, 2025) is expected to catalyze further developments in this vibrant and rapidly evolving area.

 Parameterized complexity in \probResource{} has similarly garnered increasing attention.
 \citet{Bliem16} analyze Pareto-efficient and envy-free allocations of indivisible resources under restricted utilities, parameterized by the number of agents, items, and utility value diversity. Recent works by \citet{deligkas2021parameterized} and \citet{gahlawat2023parameterized} explore connected fair division, while \citet{eiben2023parameterized,bredereck2022envy,limaye2023envy} investigate envy-free allocations and related variants.

 We hope this survey serves both as a comprehensive reference for established results and as an impetus for future research into the intricate parameterized complexity landscape of computational social choice.

\section*{Acknowledgment }
This work is supported by the Vienna Science and Technology Fund (WWTF)~[10.47379/VRG18012].

\bibliographystyle{cas-model2-names}
\bibliography{bib}

\begin{thebibliography}{125}
\expandafter\ifx\csname natexlab\endcsname\relax\def\natexlab#1{#1}\fi
\providecommand{\url}[1]{\texttt{#1}}
\providecommand{\href}[2]{#2}
\providecommand{\path}[1]{#1}
\providecommand{\DOIprefix}{doi:}
\providecommand{\ArXivprefix}{arXiv:}
\providecommand{\URLprefix}{URL: }
\providecommand{\Pubmedprefix}{pmid:}
\providecommand{\doi}[1]{\href{http://dx.doi.org/#1}{\path{#1}}}
\providecommand{\Pubmed}[1]{\href{pmid:#1}{\path{#1}}}
\providecommand{\bibinfo}[2]{#2}
\ifx\xfnm\relax \def\xfnm[#1]{\unskip,\space#1}\fi
\bibitem[{Adil et~al.(2018)Adil, Gupta, Roy, Saurabh and Zehavi}]{Adil18}
\bibinfo{author}{Adil, D.}, \bibinfo{author}{Gupta, S.}, \bibinfo{author}{Roy,
  S.}, \bibinfo{author}{Saurabh, S.}, \bibinfo{author}{Zehavi, M.},
  \bibinfo{year}{2018}.
\newblock \bibinfo{title}{Parameterized algorithms for stable matching with
  ties and incomplete lists}.
\newblock \bibinfo{journal}{Theoretical Computer Science}
  \bibinfo{volume}{723}, \bibinfo{pages}{1--10}.
\bibitem[{Ageev and Sviridenko(1999)}]{AS1999}
\bibinfo{author}{Ageev, A.A.}, \bibinfo{author}{Sviridenko, M.},
  \bibinfo{year}{1999}.
\newblock \bibinfo{title}{Approximation algorithms for maximum coverage and max
  cut with given sizes of parts}, in: \bibinfo{booktitle}{Proceedings of the
  7th International conference on Integer Programming and Combinatorial
  Optimization}, pp. \bibinfo{pages}{17--30}.
\bibitem[{Alon et~al.(1995)Alon, Yuster and Zwick}]{Alon1995Colorcoding}
\bibinfo{author}{Alon, N.}, \bibinfo{author}{Yuster, R.},
  \bibinfo{author}{Zwick, U.}, \bibinfo{year}{1995}.
\newblock \bibinfo{title}{Color-coding}.
\newblock \bibinfo{journal}{Journal of the {ACM}} \bibinfo{volume}{42},
  \bibinfo{pages}{844--856}.
\bibitem[{Arkin et~al.(2009)Arkin, Bae, Efrat, Okamoto, Mitchell and
  Polishchuk}]{arkin2009geometric}
\bibinfo{author}{Arkin, E.M.}, \bibinfo{author}{Bae, S.W.},
  \bibinfo{author}{Efrat, A.}, \bibinfo{author}{Okamoto, K.},
  \bibinfo{author}{Mitchell, J.S.}, \bibinfo{author}{Polishchuk, V.},
  \bibinfo{year}{2009}.
\newblock \bibinfo{title}{Geometric stable roommates}.
\newblock \bibinfo{journal}{Information Processing Letters}
  \bibinfo{volume}{109}, \bibinfo{pages}{219--224}.
\bibitem[{Ayadi et~al.(2019)Ayadi, Amor, Lang and Peters}]{Ayadi19}
\bibinfo{author}{Ayadi, M.}, \bibinfo{author}{Amor, N.B.},
  \bibinfo{author}{Lang, J.}, \bibinfo{author}{Peters, D.},
  \bibinfo{year}{2019}.
\newblock \bibinfo{title}{Single transferable vote: {I}ncomplete knowledge and
  communication issues}, in: \bibinfo{booktitle}{Proceedings of the 18th
  International Conference on Autonomous Agents and Multiagent Systems
  (AAMAS~19)}, pp. \bibinfo{pages}{1288--1296}.
\bibitem[{Aziz et~al.(2019)Aziz, Brandl, Brandt, Harrenstein, Olsen and
  Peters}]{aziz2019fractional}
\bibinfo{author}{Aziz, H.}, \bibinfo{author}{Brandl, F.},
  \bibinfo{author}{Brandt, F.}, \bibinfo{author}{Harrenstein, P.},
  \bibinfo{author}{Olsen, M.}, \bibinfo{author}{Peters, D.},
  \bibinfo{year}{2019}.
\newblock \bibinfo{title}{Fractional hedonic games}.
\newblock \bibinfo{journal}{ACM Transactions on Economics and Computation
  (TEAC)} \bibinfo{volume}{7}, \bibinfo{pages}{1--29}.
\bibitem[{Aziz et~al.(2013)Aziz, Brandt and Seedig}]{aziz2013computing}
\bibinfo{author}{Aziz, H.}, \bibinfo{author}{Brandt, F.},
  \bibinfo{author}{Seedig, H.G.}, \bibinfo{year}{2013}.
\newblock \bibinfo{title}{Computing desirable partitions in additively
  separable hedonic games}.
\newblock \bibinfo{journal}{Artificial Intelligence} \bibinfo{volume}{195},
  \bibinfo{pages}{316--334}.
\bibitem[{Aziz et~al.(2015)Aziz, Gaspers, Gudmundsson, Mackenzie, Mattei and
  Walsh}]{Aziz15}
\bibinfo{author}{Aziz, H.}, \bibinfo{author}{Gaspers, S.},
  \bibinfo{author}{Gudmundsson, J.}, \bibinfo{author}{Mackenzie, S.},
  \bibinfo{author}{Mattei, N.}, \bibinfo{author}{Walsh, T.},
  \bibinfo{year}{2015}.
\newblock \bibinfo{title}{Computational aspects of multi-winner approval
  voting}, in: \bibinfo{booktitle}{Proceedings of the 14th International
  Conference on Autonomous Agents and Multiagent Systems (AAMAS~15)}, pp.
  \bibinfo{pages}{107--115}.
\bibitem[{Aziz and Savani(2016)}]{AS2016HedonicGamesChapter}
\bibinfo{author}{Aziz, H.}, \bibinfo{author}{Savani, R.}, \bibinfo{year}{2016}.
\newblock \bibinfo{title}{Hedonic games}, in: \bibinfo{editor}{Brandt, F.},
  \bibinfo{editor}{Conitzer, V.}, \bibinfo{editor}{Endriss, U.},
  \bibinfo{editor}{Lang, J.}, \bibinfo{editor}{Procaccia, A.D.} (Eds.),
  \bibinfo{booktitle}{Handbook of Computational Social Choice}.
  \bibinfo{publisher}{Cambridge University Press}, pp.
  \bibinfo{pages}{356--376}.
\bibitem[{Ballester and Haeringer(2011)}]{BaHa2011}
\bibinfo{author}{Ballester, M.{\'A}.}, \bibinfo{author}{Haeringer, G.},
  \bibinfo{year}{2011}.
\newblock \bibinfo{title}{A characterization of the single-peaked domain}.
\newblock \bibinfo{journal}{Social Choice and Welfare} \bibinfo{volume}{36},
  \bibinfo{pages}{305--322}.
\bibitem[{Barrot et~al.(2019)Barrot, Ota, Sakurai and
  Yokoo}]{barrot2019unknown}
\bibinfo{author}{Barrot, N.}, \bibinfo{author}{Ota, K.},
  \bibinfo{author}{Sakurai, Y.}, \bibinfo{author}{Yokoo, M.},
  \bibinfo{year}{2019}.
\newblock \bibinfo{title}{Unknown agents in friends oriented hedonic games:
  Stability and complexity}, in: \bibinfo{booktitle}{Proceedings of the 33rd
  AAAI Conference on Artificial Intelligence (AAAI~'19)}, pp.
  \bibinfo{pages}{1756--1763}.
\bibitem[{Barrot and Yokoo(2019)}]{barrot2019stable}
\bibinfo{author}{Barrot, N.}, \bibinfo{author}{Yokoo, M.},
  \bibinfo{year}{2019}.
\newblock \bibinfo{title}{Stable and envy-free partitions in hedonic games.},
  in: \bibinfo{booktitle}{IJCAI}, pp. \bibinfo{pages}{67--73}.
\bibitem[{{Bartholdi~III} et~al.(1989){Bartholdi~III}, Tovey and
  Trick}]{BarTovTri1989}
\bibinfo{author}{{Bartholdi~III}, J.J.}, \bibinfo{author}{Tovey, C.A.},
  \bibinfo{author}{Trick, M.A.}, \bibinfo{year}{1989}.
\newblock \bibinfo{title}{Voting schemes for which it can be difficult to tell
  who won the election}.
\newblock \bibinfo{journal}{Social Choice and Welfare} \bibinfo{volume}{6},
  \bibinfo{pages}{157--165}.
\bibitem[{Bartholdi~III and Trick(1986)}]{BarTri1986}
\bibinfo{author}{Bartholdi~III, J.J.}, \bibinfo{author}{Trick, M.},
  \bibinfo{year}{1986}.
\newblock \bibinfo{title}{Stable matching with preferences derived from a
  psychological model}.
\newblock \bibinfo{journal}{Operations Research Letters} \bibinfo{volume}{5},
  \bibinfo{pages}{165--169}.
\bibitem[{Betzler et~al.(2012)Betzler, Bredereck, Chen and
  Niedermeier}]{BetBreCheNie2012}
\bibinfo{author}{Betzler, N.}, \bibinfo{author}{Bredereck, R.},
  \bibinfo{author}{Chen, J.}, \bibinfo{author}{Niedermeier, R.},
  \bibinfo{year}{2012}.
\newblock \bibinfo{title}{Studies in computational aspects of voting---{A}
  parameterized complexity perspective}, in: \bibinfo{booktitle}{The
  Multivariate Algorithmic Revolution and Beyond}.
  \bibinfo{publisher}{Springer}. volume \bibinfo{volume}{7370} of
  \textit{\bibinfo{series}{Lecture Notes in Computer Science}}, pp.
  \bibinfo{pages}{318--363}.
\bibitem[{Betzler et~al.(2013)Betzler, Slinko and Uhlmann}]{BetzlerMW2013}
\bibinfo{author}{Betzler, N.}, \bibinfo{author}{Slinko, A.},
  \bibinfo{author}{Uhlmann, J.}, \bibinfo{year}{2013}.
\newblock \bibinfo{title}{On the computation of fully proportional
  representation}.
\newblock \bibinfo{journal}{Journal of Artificial Intelligence Research}
  \bibinfo{volume}{47}, \bibinfo{pages}{475--519}.
\bibitem[{Bil{\`o} et~al.(2022)Bil{\`o}, Monaco and
  Moscardelli}]{bilo2022hedonic}
\bibinfo{author}{Bil{\`o}, V.}, \bibinfo{author}{Monaco, G.},
  \bibinfo{author}{Moscardelli, L.}, \bibinfo{year}{2022}.
\newblock \bibinfo{title}{Hedonic games with fixed-size coalitions}, in:
  \bibinfo{booktitle}{Proceedings of the 36th AAAI Conference on Artificial
  Intelligence (AAAI~'22)}, pp. \bibinfo{pages}{9287--9295}.
\bibitem[{Bliem et~al.(2016)Bliem, Bredereck and Niedermeier}]{Bliem16}
\bibinfo{author}{Bliem, B.}, \bibinfo{author}{Bredereck, R.},
  \bibinfo{author}{Niedermeier, R.}, \bibinfo{year}{2016}.
\newblock \bibinfo{title}{Complexity of efficient and envy-free resource
  allocation: Few agents, resources, or utility levels}, in:
  \bibinfo{booktitle}{Proceedings of the 25th International Joint Conference on
  Artificial Intelligence (IJCAI~'16)}, pp. \bibinfo{pages}{102--108}.
\bibitem[{Bogomolnaia and Jackson(2002)}]{bogomolnaia2002stability}
\bibinfo{author}{Bogomolnaia, A.}, \bibinfo{author}{Jackson, M.O.},
  \bibinfo{year}{2002}.
\newblock \bibinfo{title}{The stability of hedonic coalition structures}.
\newblock \bibinfo{journal}{Games and Economic Behavior} \bibinfo{volume}{38},
  \bibinfo{pages}{201--230}.
\bibitem[{Bonnet et~al.(2016)Bonnet, Paschos and Sikora}]{BPS2016}
\bibinfo{author}{Bonnet, {\'{E}}.}, \bibinfo{author}{Paschos, V.T.},
  \bibinfo{author}{Sikora, F.}, \bibinfo{year}{2016}.
\newblock \bibinfo{title}{Parameterized exact and approximation algorithms for
  maximum ${k}$-set cover and related satisfiability problems}.
\newblock \bibinfo{journal}{RAIRO--Theoretical Informatics and Applications}
  \bibinfo{volume}{50}, \bibinfo{pages}{227--240}.
\bibitem[{Booth and Lueker(1976)}]{BLConsecutiveOnes76}
\bibinfo{author}{Booth, K.S.}, \bibinfo{author}{Lueker, G.S.},
  \bibinfo{year}{1976}.
\newblock \bibinfo{title}{Testing for the consecutive ones property, interval
  graphs, and graph planarity using {PQ}-tree algorithms}.
\newblock \bibinfo{journal}{Journal of Computer System Sciences}
  \bibinfo{volume}{13}, \bibinfo{pages}{335--379}.
\bibitem[{Brandt and Bullinger(2022)}]{brandt2022finding}
\bibinfo{author}{Brandt, F.}, \bibinfo{author}{Bullinger, M.},
  \bibinfo{year}{2022}.
\newblock \bibinfo{title}{Finding and recognizing popular coalition
  structures}.
\newblock \bibinfo{journal}{Journal of Artificial Intelligence Research}
  \bibinfo{volume}{74}, \bibinfo{pages}{569--626}.
\bibitem[{Brandt et~al.(2024)Brandt, Bullinger and Tappe}]{brandt2022single}
\bibinfo{author}{Brandt, F.}, \bibinfo{author}{Bullinger, M.},
  \bibinfo{author}{Tappe, L.}, \bibinfo{year}{2024}.
\newblock \bibinfo{title}{Stability based on single-agent deviations in
  additively separable hedonic games}.
\newblock \bibinfo{journal}{Artificial Intelligence} \bibinfo{volume}{334},
  \bibinfo{pages}{104--160}.
\bibitem[{Bredereck et~al.(2014)Bredereck, Chen, Faliszewski, Guo, Niedermeier
  and Woeginger}]{BreCheFalGuoNieWoe2014}
\bibinfo{author}{Bredereck, R.}, \bibinfo{author}{Chen, J.},
  \bibinfo{author}{Faliszewski, P.}, \bibinfo{author}{Guo, J.},
  \bibinfo{author}{Niedermeier, R.}, \bibinfo{author}{Woeginger, G.J.},
  \bibinfo{year}{2014}.
\newblock \bibinfo{title}{Parameterized algorithmics for computational social
  choice: {N}ine research challenges}.
\newblock \bibinfo{journal}{Tsinghua Science and Technology}
  \bibinfo{volume}{19}, \bibinfo{pages}{358--373}.
\bibitem[{Bredereck et~al.(2020a)Bredereck, Chen, Finnendahl and
  Niedermeier}]{BreCheFinNie2020-spscSM-jaamas}
\bibinfo{author}{Bredereck, R.}, \bibinfo{author}{Chen, J.},
  \bibinfo{author}{Finnendahl, U.P.}, \bibinfo{author}{Niedermeier, R.},
  \bibinfo{year}{2020}a.
\newblock \bibinfo{title}{Stable roommate with narcissistic, single-peaked, and
  single-crossing preferences}.
\newblock \bibinfo{journal}{Autonomous Agents and Multi-Agent Systems}
  \bibinfo{volume}{34}, \bibinfo{pages}{1--29}.
\bibitem[{Bredereck et~al.(2013)Bredereck, Chen and Woeginger}]{BCW12}
\bibinfo{author}{Bredereck, R.}, \bibinfo{author}{Chen, J.},
  \bibinfo{author}{Woeginger, G.J.}, \bibinfo{year}{2013}.
\newblock \bibinfo{title}{A characterization of the single-crossing domain}.
\newblock \bibinfo{journal}{Social Choice and Welfare} \bibinfo{volume}{41},
  \bibinfo{pages}{989--998}.
\bibitem[{Bredereck et~al.(2016)Bredereck, Chen and Woeginger}]{BreCheWoe2016}
\bibinfo{author}{Bredereck, R.}, \bibinfo{author}{Chen, J.},
  \bibinfo{author}{Woeginger, G.J.}, \bibinfo{year}{2016}.
\newblock \bibinfo{title}{Are there any nicely structured preference profiles
  nearby?}
\newblock \bibinfo{journal}{Mathematical Social Sciences} \bibinfo{volume}{79},
  \bibinfo{pages}{61--73}.
\bibitem[{Bredereck et~al.(2021a)Bredereck, Faliszewski, Niedermeier and
  Talmon}]{BFNT2021SB}
\bibinfo{author}{Bredereck, R.}, \bibinfo{author}{Faliszewski, P.},
  \bibinfo{author}{Niedermeier, R.}, \bibinfo{author}{Talmon, N.},
  \bibinfo{year}{2021}a.
\newblock \bibinfo{title}{Complexity of shift bribery in committee elections}.
\newblock \bibinfo{journal}{ACM Transactions on on Computation Theory}
  \bibinfo{volume}{13}, \bibinfo{pages}{20:1--20:25}.
\bibitem[{Bredereck et~al.(2020b)Bredereck, Heeger, Knop and
  Niedermeier}]{BredHeeKnoNie2020multidimensional}
\bibinfo{author}{Bredereck, R.}, \bibinfo{author}{Heeger, K.},
  \bibinfo{author}{Knop, D.}, \bibinfo{author}{Niedermeier, R.},
  \bibinfo{year}{2020}b.
\newblock \bibinfo{title}{Multidimensional stable roommates with master list},
  in: \bibinfo{booktitle}{Proceedings of the 16th International Conference on
  Web and Internet Economics (WINE~'20)}, pp. \bibinfo{pages}{59--73}.
\bibitem[{Bredereck et~al.(2021b)Bredereck, Kaczmarczyk and
  Niedermeier}]{BKN2021MWManipulationn}
\bibinfo{author}{Bredereck, R.}, \bibinfo{author}{Kaczmarczyk, A.},
  \bibinfo{author}{Niedermeier, R.}, \bibinfo{year}{2021}b.
\newblock \bibinfo{title}{On coalitional manipulation for multiwinner
  elections: shortlisting}.
\newblock \bibinfo{journal}{Autonomous Agents and Multi-Agent Systems}
  \bibinfo{volume}{35}, \bibinfo{pages}{38}.
\bibitem[{Bredereck et~al.(2022)Bredereck, Kaczmarczyk and
  Niedermeier}]{bredereck2022envy}
\bibinfo{author}{Bredereck, R.}, \bibinfo{author}{Kaczmarczyk, A.},
  \bibinfo{author}{Niedermeier, R.}, \bibinfo{year}{2022}.
\newblock \bibinfo{title}{Envy-free allocations respecting social networks}.
\newblock \bibinfo{journal}{Artificial Intelligence} \bibinfo{volume}{305},
  \bibinfo{pages}{103664}.
\bibitem[{Bullinger(2020)}]{bullinger2020pareto}
\bibinfo{author}{Bullinger, M.}, \bibinfo{year}{2020}.
\newblock \bibinfo{title}{Pareto-optimality in cardinal hedonic games}, in:
  \bibinfo{booktitle}{Proceedings of the 19th International Conference on
  Autonomous Agents and Multiagent Systems (AAMAS 20)}, pp.
  \bibinfo{pages}{213--221}.
\bibitem[{Bullinger(2022)}]{bullinger2022boundaries}
\bibinfo{author}{Bullinger, M.}, \bibinfo{year}{2022}.
\newblock \bibinfo{title}{Boundaries to single-agent stability in additively
  separable hedonic games}, in: \bibinfo{booktitle}{Proceedings of the 47nd
  International Symposium on Mathematical Foundations of Computer Science
  (MFCS~'22)}, pp. \bibinfo{pages}{26:1--26:15}.
\bibitem[{Cechl{\'a}rov{\'a} and Romero-Medina(2001)}]{cechlarova2001stability}
\bibinfo{author}{Cechl{\'a}rov{\'a}, K.n.}, \bibinfo{author}{Romero-Medina,
  A.}, \bibinfo{year}{2001}.
\newblock \bibinfo{title}{Stability in coalition formation games}.
\newblock \bibinfo{journal}{International Journal of Game Theory}
  \bibinfo{volume}{29}, \bibinfo{pages}{487--494}.
\bibitem[{Ceylan et~al.(2023)Ceylan, Chen and Roy}]{CCR2023OptimalSeat}
\bibinfo{author}{Ceylan, E.}, \bibinfo{author}{Chen, J.}, \bibinfo{author}{Roy,
  S.}, \bibinfo{year}{2023}.
\newblock \bibinfo{title}{Optimal seat arrangement: {W}hat are the hard and
  easy cases?}, in: \bibinfo{booktitle}{Proceedings of the 32nd International
  Joint Conference on Artificial Intelligence (IJCAI~'23)}, pp.
  \bibinfo{pages}{2563--2571}.
\bibitem[{Chen(2019)}]{Chen2019SMsurvey}
\bibinfo{author}{Chen, J.}, \bibinfo{year}{2019}.
\newblock \bibinfo{title}{Computational Complexity of Stable Marriage and
  Stable Roommates and Their Variants}.
\newblock \bibinfo{type}{Technical Report}. arXiv:1904.08196.
\bibitem[{Chen et~al.(2023a)Chen, Cs{\'a}ji, Roy and
  Simola}]{CheCsaRoySim2023Verif}
\bibinfo{author}{Chen, J.}, \bibinfo{author}{Cs{\'a}ji, G.},
  \bibinfo{author}{Roy, S.}, \bibinfo{author}{Simola, S.},
  \bibinfo{year}{2023}a.
\newblock \bibinfo{title}{Hedonic games with friends, enemies, and neutrals:
  {R}esolving open questions and fine-grained complexity}, in:
  \bibinfo{booktitle}{Proceedings of the 22nd International Conference on
  Autonomous Agents and Multiagent Systems (AAMAS~23)}, pp.
  \bibinfo{pages}{251--259}.
\bibitem[{Chen et~al.(2020)Chen, Ganian and
  Hamm}]{ChenGanianHamm2020ijcai-diversestable}
\bibinfo{author}{Chen, J.}, \bibinfo{author}{Ganian, R.},
  \bibinfo{author}{Hamm, T.}, \bibinfo{year}{2020}.
\newblock \bibinfo{title}{Stable matchings with diversity constraints:
  {A}ffirmative action is beyond {NP}}, in: \bibinfo{booktitle}{Proceedings of
  the 29th International Joint Conference on Artificial Intelligence
  (IJCAI~'20)}, pp. \bibinfo{pages}{146--152}.
\bibitem[{Chen et~al.(2023b)Chen, Hatschka and Simola}]{chen2023efficient}
\bibinfo{author}{Chen, J.}, \bibinfo{author}{Hatschka, C.},
  \bibinfo{author}{Simola, S.}, \bibinfo{year}{2023}b.
\newblock \bibinfo{title}{Efficient algorithms for {M}onroe and {CC} rules in
  multi-winner elections with (nearly) structured preferences}, in:
  \bibinfo{booktitle}{Proceedings of the 26th European Conference on Artificial
  Intelligence (ECAI '23)}, pp. \bibinfo{pages}{397--404}.
\bibitem[{Chen et~al.(2019)Chen, Hermelin and Sorge}]{CheHerSor2019pnorm}
\bibinfo{author}{Chen, J.}, \bibinfo{author}{Hermelin, D.},
  \bibinfo{author}{Sorge, M.}, \bibinfo{year}{2019}.
\newblock \bibinfo{title}{On computing centroids according to the $p$-norms of
  hamming distance vectors}, in: \bibinfo{booktitle}{Proceedings of the 27th
  Annual European Symposium on Algorithms (ESA~'19)}, pp.
  \bibinfo{pages}{28:1--28:16}.
\bibitem[{Chen et~al.(2018)Chen, Hermelin, Sorge and
  Yedidsion}]{CHSYicalp-par-stable2018}
\bibinfo{author}{Chen, J.}, \bibinfo{author}{Hermelin, D.},
  \bibinfo{author}{Sorge, M.}, \bibinfo{author}{Yedidsion, H.},
  \bibinfo{year}{2018}.
\newblock \bibinfo{title}{How hard is it to satisfy (almost) all roommates?},
  in: \bibinfo{booktitle}{Proceedings of the 45th International Colloquium on
  Automata, Languages, and Programming (ICALP~'18)}, pp.
  \bibinfo{pages}{35:1--35:15}.
\bibitem[{Chen and Manlove(2023)}]{ChenManlove2022}
\bibinfo{author}{Chen, J.}, \bibinfo{author}{Manlove, D.},
  \bibinfo{year}{2023}.
\newblock \bibinfo{title}{Algorithmics of matching markets}, in:
  \bibinfo{editor}{Echenique, F.}, \bibinfo{editor}{Immorlica, N.},
  \bibinfo{editor}{Vazirani, V.V.} (Eds.), \bibinfo{booktitle}{Online and
  Martching-Based Market Design}, pp. \bibinfo{pages}{283--302}.
\bibitem[{Chen and Roy(2022a)}]{Chen2022Euclid2dSR}
\bibinfo{author}{Chen, J.}, \bibinfo{author}{Roy, S.}, \bibinfo{year}{2022}a.
\newblock \bibinfo{title}{Multi-dimensional stable roommates in 2-dimensional
  {E}uclidean space}, in: \bibinfo{booktitle}{Proceedings of the 30th Annual
  European Symposium on Algorithms (ESA~'22)}, pp.
  \bibinfo{pages}{36:1--36:16}.
\bibitem[{Chen and Roy(2022b)}]{CheRoy2022}
\bibinfo{author}{Chen, J.}, \bibinfo{author}{Roy, S.}, \bibinfo{year}{2022}b.
\newblock \bibinfo{title}{Parameterized Intractability for Multi-Winner
  Election under the {C}hamberlin-{C}ourant Rule and the {M}onroe Rule}.
\newblock \bibinfo{type}{Technical Report}. arXiv:2202.12006.
\bibitem[{Chen et~al.(2024)Chen, Schlotter and Simola}]{CSS2025Refugee}
\bibinfo{author}{Chen, J.}, \bibinfo{author}{Schlotter, I.},
  \bibinfo{author}{Simola, S.}, \bibinfo{year}{2024}.
\newblock \bibinfo{title}{Parameterized algorithms for optimal refugee
  resettlement}, in: \bibinfo{booktitle}{Proceedings of the 27th European
  Conference on Artificial Intelligence (ECAI~'24)}, pp.
  \bibinfo{pages}{3413--3420}.
\bibitem[{Christian et~al.(2007a)Christian, Fellows, Rosamond and
  Slinko}]{CFRS07}
\bibinfo{author}{Christian, R.}, \bibinfo{author}{Fellows, M.},
  \bibinfo{author}{Rosamond, F.}, \bibinfo{author}{Slinko, A.},
  \bibinfo{year}{2007}a.
\newblock \bibinfo{title}{On complexity of lobbying in multiple referenda}.
\newblock \bibinfo{journal}{Review of Economic Design} \bibinfo{volume}{11},
  \bibinfo{pages}{217--224}.
\bibitem[{Christian et~al.(2007b)Christian, Fellows, Rosamond and
  Slinko}]{CFRS07lobbying}
\bibinfo{author}{Christian, R.}, \bibinfo{author}{Fellows, M.},
  \bibinfo{author}{Rosamond, F.}, \bibinfo{author}{Slinko, A.},
  \bibinfo{year}{2007}b.
\newblock \bibinfo{title}{On complexity of lobbying in multiple referenda}.
\newblock \bibinfo{journal}{Review of Economic Design} \bibinfo{volume}{11},
  \bibinfo{pages}{217--224}.
\bibitem[{Cornaz et~al.(2012)Cornaz, Galand and Spanjaard}]{Cornaz12}
\bibinfo{author}{Cornaz, D.}, \bibinfo{author}{Galand, L.},
  \bibinfo{author}{Spanjaard, O.}, \bibinfo{year}{2012}.
\newblock \bibinfo{title}{Bounded single-peaked width and proportional
  representation}, in: \bibinfo{booktitle}{Proceedings of the 20th European
  Conference on Artificial Intelligence (ECAI~'12)}, pp.
  \bibinfo{pages}{270--275}.
\bibitem[{Cygan et~al.(2015)Cygan, Fomin, Kowalik, Lokshtanov, Marx, Pilipczuk,
  Pilipczuk and Saurabh}]{CyFoKoLoMaPiPiSa2015}
\bibinfo{author}{Cygan, M.}, \bibinfo{author}{Fomin, F.V.},
  \bibinfo{author}{Kowalik, L.}, \bibinfo{author}{Lokshtanov, D.},
  \bibinfo{author}{Marx, D.}, \bibinfo{author}{Pilipczuk, M.},
  \bibinfo{author}{Pilipczuk, M.}, \bibinfo{author}{Saurabh, S.},
  \bibinfo{year}{2015}.
\newblock \bibinfo{title}{Parameterized Algorithms}.
\newblock \bibinfo{publisher}{Springer}.
\bibitem[{Deligkas et~al.(2021)Deligkas, Eiben, Ganian, Hamm and
  Ordyniak}]{deligkas2021parameterized}
\bibinfo{author}{Deligkas, A.}, \bibinfo{author}{Eiben, E.},
  \bibinfo{author}{Ganian, R.}, \bibinfo{author}{Hamm, T.},
  \bibinfo{author}{Ordyniak, S.}, \bibinfo{year}{2021}.
\newblock \bibinfo{title}{The parameterized complexity of connected fair
  division}, in: \bibinfo{booktitle}{Proceedings of the 30th International
  Joint Conference on Artificial Intelligence (IJCAI~'21)}, pp.
  \bibinfo{pages}{139--145}.
\bibitem[{Dimitrov et~al.(2006)Dimitrov, Borm, Hendrickx and
  Sung}]{dimitrov2006simple}
\bibinfo{author}{Dimitrov, D.}, \bibinfo{author}{Borm, P.},
  \bibinfo{author}{Hendrickx, R.}, \bibinfo{author}{Sung, S.C.},
  \bibinfo{year}{2006}.
\newblock \bibinfo{title}{Simple priorities and core stability in hedonic
  games}.
\newblock \bibinfo{journal}{Social Choice and Welfare} \bibinfo{volume}{26},
  \bibinfo{pages}{421--433}.
\bibitem[{Doignon and Falmagne(1994)}]{DoiFal1994}
\bibinfo{author}{Doignon, J.}, \bibinfo{author}{Falmagne, J.},
  \bibinfo{year}{1994}.
\newblock \bibinfo{title}{A polynomial time algorithm for unidimensional
  unfolding representations}.
\newblock \bibinfo{journal}{Journal of Algorithms} \bibinfo{volume}{16},
  \bibinfo{pages}{218--233}.
\newblock \DOIprefix\doi{https://doi.org/10.1006/jagm.1994.1010}.
\bibitem[{Dorn and Schlotter(2012)}]{DS2012SwapBribery}
\bibinfo{author}{Dorn, B.}, \bibinfo{author}{Schlotter, I.},
  \bibinfo{year}{2012}.
\newblock \bibinfo{title}{Multivariate complexity analysis of swap bribery}.
\newblock \bibinfo{journal}{Algorithmica} \bibinfo{volume}{64},
  \bibinfo{pages}{126--151}.
\bibitem[{Dorn and Schlotter(2017)}]{DS17}
\bibinfo{author}{Dorn, B.}, \bibinfo{author}{Schlotter, I.},
  \bibinfo{year}{2017}.
\newblock \bibinfo{title}{Having a hard time? {E}xplore parameterized
  complexity!}, in: \bibinfo{editor}{Endriss, U.} (Ed.),
  \bibinfo{booktitle}{Trends in Computational Social Choice}.
  \bibinfo{publisher}{AI Access Foundation}, pp. \bibinfo{pages}{209--230}.
\bibitem[{Downey and Fellows(1999)}]{DowneyF99}
\bibinfo{author}{Downey, R.G.}, \bibinfo{author}{Fellows, M.R.},
  \bibinfo{year}{1999}.
\newblock \bibinfo{title}{Parameterized Complexity}.
\newblock Monographs in Computer Science, \bibinfo{publisher}{Springer}.
\bibitem[{Downey and Fellows(2013)}]{DF13}
\bibinfo{author}{Downey, R.G.}, \bibinfo{author}{Fellows, M.R.},
  \bibinfo{year}{2013}.
\newblock \bibinfo{title}{Fundamentals of Parameterized Complexity}.
\newblock \bibinfo{publisher}{Springer}.
\bibitem[{Durand et~al.(2025)Durand, Erlacher, Vistisen and
  Simola}]{durand2024enemies}
\bibinfo{author}{Durand, M.}, \bibinfo{author}{Erlacher, L.},
  \bibinfo{author}{Vistisen, J.M.}, \bibinfo{author}{Simola, S.},
  \bibinfo{year}{2025}.
\newblock \bibinfo{title}{Parameterized complexity of hedonic games with
  enemy-aversion preferences}, in: \bibinfo{booktitle}{Proceedings of the 24th
  International Conference on Autonomous Agents and Multiagent Systems
  (AAMAS~25)}.
\bibitem[{Eiben et~al.(2023)Eiben, Ganian, Hamm and
  Ordyniak}]{eiben2023parameterized}
\bibinfo{author}{Eiben, E.}, \bibinfo{author}{Ganian, R.},
  \bibinfo{author}{Hamm, T.}, \bibinfo{author}{Ordyniak, S.},
  \bibinfo{year}{2023}.
\newblock \bibinfo{title}{Parameterized complexity of envy-free resource
  allocation in social networks}.
\newblock \bibinfo{journal}{Artificial Intelligence} \bibinfo{volume}{315},
  \bibinfo{pages}{103826}.
\bibitem[{Elkind et~al.(2020)Elkind, Fanelli and Flammini}]{elkind2020price}
\bibinfo{author}{Elkind, E.}, \bibinfo{author}{Fanelli, A.},
  \bibinfo{author}{Flammini, M.}, \bibinfo{year}{2020}.
\newblock \bibinfo{title}{Price of pareto optimality in hedonic games}.
\newblock \bibinfo{journal}{Artificial Intelligence} \bibinfo{volume}{288},
  \bibinfo{pages}{103357}.
\bibitem[{Elkind and Lackner(2015)}]{EL2015StructureDicho}
\bibinfo{author}{Elkind, E.}, \bibinfo{author}{Lackner, M.},
  \bibinfo{year}{2015}.
\newblock \bibinfo{title}{Structure in dichotomous preferences}, in:
  \bibinfo{booktitle}{Proceedings of the 24th International Joint Conference on
  Artificial Intelligence (IJCAI~'15)}, pp. \bibinfo{pages}{2019--2025}.
\bibitem[{Elkind and Wooldridge(2009)}]{elkind2009hedonic}
\bibinfo{author}{Elkind, E.}, \bibinfo{author}{Wooldridge, M.J.},
  \bibinfo{year}{2009}.
\newblock \bibinfo{title}{Hedonic coalition nets}, in:
  \bibinfo{booktitle}{Proceedings of International Conference on Autonomous
  Agents and Multiagent Systems}, \bibinfo{organization}{Citeseer}. pp.
  \bibinfo{pages}{417--424}.
\bibitem[{Erd{\'{e}}lyi et~al.(2017)Erd{\'{e}}lyi, Lackner and
  Pfandler}]{ELP17NSP}
\bibinfo{author}{Erd{\'{e}}lyi, G.}, \bibinfo{author}{Lackner, M.},
  \bibinfo{author}{Pfandler, A.}, \bibinfo{year}{2017}.
\newblock \bibinfo{title}{Computational aspects of nearly single-peaked
  electorates}.
\newblock \bibinfo{journal}{Journal of Artificial Intelligence Research}
  \bibinfo{volume}{58}, \bibinfo{pages}{297--337}.
\bibitem[{Escoffier et~al.(2008)Escoffier, Lang and
  {\"O}zt{\"u}rk}]{EscLanOez2008}
\bibinfo{author}{Escoffier, B.}, \bibinfo{author}{Lang, J.},
  \bibinfo{author}{{\"O}zt{\"u}rk, M.}, \bibinfo{year}{2008}.
\newblock \bibinfo{title}{Single-peaked consistency and its complexity}, in:
  \bibinfo{booktitle}{Proceedings of the 18th European Conference on Artificial
  Intelligence (ECAI~'08)}, \bibinfo{publisher}{IOS Press}. pp.
  \bibinfo{pages}{366--370}.
\bibitem[{Faliszewski et~al.(2017a)Faliszewski, Skowron, Slinko and
  Talmon}]{faliszewski2017multiwinner}
\bibinfo{author}{Faliszewski, P.}, \bibinfo{author}{Skowron, P.},
  \bibinfo{author}{Slinko, A.}, \bibinfo{author}{Talmon, N.},
  \bibinfo{year}{2017}a.
\newblock \bibinfo{title}{Multiwinner voting: {A} new challenge for social
  choice theory}, in: \bibinfo{booktitle}{Trends in computational social
  choice}. \bibinfo{publisher}{AI Access Foundation}, pp.
  \bibinfo{pages}{27--47}.
\bibitem[{Faliszewski et~al.(2017b)Faliszewski, Skowron and
  Talmon}]{FST2017MWBribery}
\bibinfo{author}{Faliszewski, P.}, \bibinfo{author}{Skowron, P.},
  \bibinfo{author}{Talmon, N.}, \bibinfo{year}{2017}b.
\newblock \bibinfo{title}{Bribery as a measure of candidate success:
  {C}omplexity results for approval-based multiwinner rules}, in:
  \bibinfo{booktitle}{Proceedings of the 16th International Conference on
  Autonomous Agents and Multiagent Systems (AAMAS~17)}, pp.
  \bibinfo{pages}{6--14}.
\bibitem[{Faliszewski et~al.(2020)Faliszewski, Slinko and
  Talmon}]{Faliszewski20}
\bibinfo{author}{Faliszewski, P.}, \bibinfo{author}{Slinko, A.},
  \bibinfo{author}{Talmon, N.}, \bibinfo{year}{2020}.
\newblock \bibinfo{title}{Multiwinner rules with variable number of winners},
  in: \bibinfo{booktitle}{Proceedings of the 10th Conference on Prestigious
  Applications of Artificial Intelligence (PAIS 2020)}, pp.
  \bibinfo{pages}{67--74}.
\bibitem[{Fellows(2003)}]{Fellows2003}
\bibinfo{author}{Fellows, M.R.}, \bibinfo{year}{2003}.
\newblock \bibinfo{title}{Blow-ups, win/win’s, and crown rules: Some new
  directions in {FPT}}, in: \bibinfo{booktitle}{International Workshop on
  Graph-Theoretic Concepts in Computer Science},
  \bibinfo{organization}{Springer}. pp. \bibinfo{pages}{1--12}.
\bibitem[{Flum and Grohe(2006)}]{FlumG06}
\bibinfo{author}{Flum, J.}, \bibinfo{author}{Grohe, M.}, \bibinfo{year}{2006}.
\newblock \bibinfo{title}{Parameterized Complexity Theory}.
\newblock Texts in Theoretical Computer Science. An {EATCS} Series,
  \bibinfo{publisher}{Springer}.
\bibitem[{Frances and Litman(1997)}]{FraLit1997}
\bibinfo{author}{Frances, M.}, \bibinfo{author}{Litman, A.},
  \bibinfo{year}{1997}.
\newblock \bibinfo{title}{On covering problems of codes}.
\newblock \bibinfo{journal}{Theory of Computing Systems} \bibinfo{volume}{30},
  \bibinfo{pages}{113--119}.
\bibitem[{Gahlawat and Zehavi(2023)}]{gahlawat2023parameterized}
\bibinfo{author}{Gahlawat, H.}, \bibinfo{author}{Zehavi, M.},
  \bibinfo{year}{2023}.
\newblock \bibinfo{title}{{Parameterized Complexity of Incomplete Connected
  Fair Division}}, in: \bibinfo{booktitle}{43rd IARCS Annual Conference on
  Foundations of Software Technology and Theoretical Computer Science (FSTTCS
  2023)}, \bibinfo{publisher}{Schloss Dagstuhl -- Leibniz-Zentrum f{\"u}r
  Informatik}, \bibinfo{address}{Dagstuhl, Germany}. pp.
  \bibinfo{pages}{14:1--14:18}.
\bibitem[{Gairing and Savani(2019)}]{gairing2019computing}
\bibinfo{author}{Gairing, M.}, \bibinfo{author}{Savani, R.},
  \bibinfo{year}{2019}.
\newblock \bibinfo{title}{Computing stable outcomes in symmetric additively
  separable hedonic games}.
\newblock \bibinfo{journal}{Mathematics of Operations Research}
  \bibinfo{volume}{44}, \bibinfo{pages}{1101--1121}.
\bibitem[{Gale and Shapley(1962)}]{GaleShapley1962}
\bibinfo{author}{Gale, D.}, \bibinfo{author}{Shapley, L.S.},
  \bibinfo{year}{1962}.
\newblock \bibinfo{title}{College admissions and the stability of marriage}.
\newblock \bibinfo{journal}{The American Mathematical Monthly}
  \bibinfo{volume}{120}, \bibinfo{pages}{386--391}.
\bibitem[{Garey and Johnson(1979)}]{GJ79}
\bibinfo{author}{Garey, M.R.}, \bibinfo{author}{Johnson, D.S.},
  \bibinfo{year}{1979}.
\newblock \bibinfo{title}{Computers and Intractability---{A} Guide to the
  Theory of {NP}-Completeness}.
\newblock \bibinfo{publisher}{W. H. Freeman and Company}.
\bibitem[{Gonzalez(1985)}]{Gonzalez1985}
\bibinfo{author}{Gonzalez, T.F.}, \bibinfo{year}{1985}.
\newblock \bibinfo{title}{Clustering to minimize the maximum intercluster
  distance}.
\newblock \bibinfo{journal}{Theoretical Computer Science} \bibinfo{volume}{38},
  \bibinfo{pages}{293--306}.
\bibitem[{Gupta et~al.(2020a)Gupta, Jain, Roy, Saurabh and
  Zehavi}]{GuptaJain20}
\bibinfo{author}{Gupta, S.}, \bibinfo{author}{Jain, P.}, \bibinfo{author}{Roy,
  S.}, \bibinfo{author}{Saurabh, S.}, \bibinfo{author}{Zehavi, M.},
  \bibinfo{year}{2020}a.
\newblock \bibinfo{title}{Gehrlein stability in committee selection:
  {P}arameterized hardness and algorithms}.
\newblock \bibinfo{journal}{Autonomous Agents and Multi-Agent Systems}
  \bibinfo{volume}{34}, \bibinfo{pages}{27}.
\bibitem[{Gupta et~al.(2020b)Gupta, Jain, Roy, Saurabh and Zehavi}]{Gupta20}
\bibinfo{author}{Gupta, S.}, \bibinfo{author}{Jain, P.}, \bibinfo{author}{Roy,
  S.}, \bibinfo{author}{Saurabh, S.}, \bibinfo{author}{Zehavi, M.},
  \bibinfo{year}{2020}b.
\newblock \bibinfo{title}{On the (parameterized) complexity of almost stable
  marriage}, in: \bibinfo{booktitle}{Proceedings of the 40th International
  Conference on Foundations of Software Technology and Theoretical Computer
  Science (FSTTCS~'20)}, pp. \bibinfo{pages}{24:1--24:17}.
\bibitem[{Gupta et~al.(2023)Gupta, Jain, Saurabh and Talmon}]{GuptaJain23}
\bibinfo{author}{Gupta, S.}, \bibinfo{author}{Jain, P.},
  \bibinfo{author}{Saurabh, S.}, \bibinfo{author}{Talmon, N.},
  \bibinfo{year}{2023}.
\newblock \bibinfo{title}{Even more effort towards improved bounds and
  fixed-parameter tractability for multiwinner rules}.
\newblock \bibinfo{journal}{Algorithmica} \bibinfo{volume}{85},
  \bibinfo{pages}{3717--3740}.
\bibitem[{Gupta et~al.(2018)Gupta, Roy, Saurabh and Zehavi}]{GRSZ18}
\bibinfo{author}{Gupta, S.}, \bibinfo{author}{Roy, S.},
  \bibinfo{author}{Saurabh, S.}, \bibinfo{author}{Zehavi, M.},
  \bibinfo{year}{2018}.
\newblock \bibinfo{title}{Some hard stable marriage problems: {A} survey on
  multivariate analysis}, in: \bibinfo{editor}{Neogy, S.},
  \bibinfo{editor}{Bapat, R.B.}, \bibinfo{editor}{Dubey, D.} (Eds.),
  \bibinfo{booktitle}{Mathematical Programming and Game Theory}, pp.
  \bibinfo{pages}{141--157}.
\bibitem[{Gupta et~al.(2021)Gupta, Roy, Saurabh and Zehavi}]{Gupta21}
\bibinfo{author}{Gupta, S.}, \bibinfo{author}{Roy, S.},
  \bibinfo{author}{Saurabh, S.}, \bibinfo{author}{Zehavi, M.},
  \bibinfo{year}{2021}.
\newblock \bibinfo{title}{Balanced stable marriage: {H}ow close is close
  enough?}
\newblock \bibinfo{journal}{Theoretical Computer Science}
  \bibinfo{volume}{883}, \bibinfo{pages}{19--43}.
\bibitem[{Gupta et~al.(2022)Gupta, Saurabh and Zehavi}]{Gupta22}
\bibinfo{author}{Gupta, S.}, \bibinfo{author}{Saurabh, S.},
  \bibinfo{author}{Zehavi, M.}, \bibinfo{year}{2022}.
\newblock \bibinfo{title}{On treewidth and stable marriage: {P}arameterized
  algorithms and hardness results (complete characterization)}.
\newblock \bibinfo{journal}{SIAM Journal on Discrete Mathematics}
  \bibinfo{volume}{36}, \bibinfo{pages}{596--681}.
\bibitem[{Hajiaghayi and Ganjali(2002)}]{HGCOSM2002}
\bibinfo{author}{Hajiaghayi, M.T.}, \bibinfo{author}{Ganjali, Y.},
  \bibinfo{year}{2002}.
\newblock \bibinfo{title}{A note on the consecutive ones submatrix problem}.
\newblock \bibinfo{journal}{Inf. Process. Lett.} \bibinfo{volume}{83},
  \bibinfo{pages}{163--166}.
\bibitem[{Hanaka et~al.(2019)Hanaka, Kiya, Maei and
  Ono}]{hanaka2019computational}
\bibinfo{author}{Hanaka, T.}, \bibinfo{author}{Kiya, H.},
  \bibinfo{author}{Maei, Y.}, \bibinfo{author}{Ono, H.}, \bibinfo{year}{2019}.
\newblock \bibinfo{title}{Computational complexity of hedonic games on sparse
  graphs}, in: \bibinfo{booktitle}{Proceedings of the 22nd International
  Conference on Principles and Practice of Multi-Agent Systems (PRIMA 19)}, pp.
  \bibinfo{pages}{576--584}.
\bibitem[{Hanaka et~al.(2024)Hanaka, K\"{o}hler and Lampis}]{hanaka2024core}
\bibinfo{author}{Hanaka, T.}, \bibinfo{author}{K\"{o}hler, N.},
  \bibinfo{author}{Lampis, M.}, \bibinfo{year}{2024}.
\newblock \bibinfo{title}{Core stability in additively separable hedonic games
  of low treewidth}, in: \bibinfo{booktitle}{35th International Symposium on
  Algorithms and Computation (ISAAC 2024)}, \bibinfo{address}{Dagstuhl,
  Germany}. pp. \bibinfo{pages}{39:1--39:17}.
\bibitem[{Hanaka and Lampis(2022)}]{hanaka2022revisited}
\bibinfo{author}{Hanaka, T.}, \bibinfo{author}{Lampis, M.},
  \bibinfo{year}{2022}.
\newblock \bibinfo{title}{Hedonic games and treewidth revisited}, in:
  \bibinfo{booktitle}{Proceedings of the 30th Annual European Symposium on
  Algorithms (ESA~'22)}, pp. \bibinfo{pages}{64:1--64:16}.
\bibitem[{Hemaspaandra et~al.(1997)Hemaspaandra, Hemaspaandra and
  Rothe}]{HHR97}
\bibinfo{author}{Hemaspaandra, E.}, \bibinfo{author}{Hemaspaandra, L.A.},
  \bibinfo{author}{Rothe, J.}, \bibinfo{year}{1997}.
\newblock \bibinfo{title}{Exact analysis of dodgson elections: Lewis carroll's
  1876 voting system is complete for parallel access to {NP}}.
\newblock \bibinfo{journal}{Journal of the {ACM}} \bibinfo{volume}{44},
  \bibinfo{pages}{806--825}.
\bibitem[{Johnson et~al.(1988)Johnson, Papadimitriou and
  Yannakakis}]{johnson1988easy}
\bibinfo{author}{Johnson, D.S.}, \bibinfo{author}{Papadimitriou, C.H.},
  \bibinfo{author}{Yannakakis, M.}, \bibinfo{year}{1988}.
\newblock \bibinfo{title}{How easy is local search?}
\newblock \bibinfo{journal}{Journal of computer and system sciences}
  \bibinfo{volume}{37}, \bibinfo{pages}{79--100}.
\bibitem[{Lackner and Skowron(2023)}]{LacknerSkownron23ABC}
\bibinfo{author}{Lackner, M.}, \bibinfo{author}{Skowron, P.},
  \bibinfo{year}{2023}.
\newblock \bibinfo{title}{Multi-Winner Voting with Approval
  Preferences--{A}rtificial Intelligence, Multiagent Systems, and Cognitive
  Robotics}.
\newblock Springer Briefs in Intelligent Systems,
  \bibinfo{publisher}{Springer}.
\bibitem[{LeGrand(2004)}]{LeGrand04}
\bibinfo{author}{LeGrand, R.}, \bibinfo{year}{2004}.
\newblock \bibinfo{title}{Analysis of the minimax procedure}.
\newblock \bibinfo{type}{Technical Report}.
\bibitem[{Levinger et~al.(2024)Levinger, Hazon, Simola and
  Azaria}]{levinger2024bounded}
\bibinfo{author}{Levinger, C.}, \bibinfo{author}{Hazon, N.},
  \bibinfo{author}{Simola, S.}, \bibinfo{author}{Azaria, A.},
  \bibinfo{year}{2024}.
\newblock \bibinfo{title}{Coalition formation with bounded coalition size}, in:
  \bibinfo{booktitle}{Proceedings of the 23nd International Conference on
  Autonomous Agents and Multiagent Systems (AAMAS~24)}, p.
  \bibinfo{pages}{1119–1127}.
\bibitem[{Li et~al.(2023)Li, Micha, Nikolov and Shah}]{li2023partitioning}
\bibinfo{author}{Li, L.}, \bibinfo{author}{Micha, E.},
  \bibinfo{author}{Nikolov, A.}, \bibinfo{author}{Shah, N.},
  \bibinfo{year}{2023}.
\newblock \bibinfo{title}{Partitioning friends fairly}, in:
  \bibinfo{booktitle}{Proceedings of the 37th AAAI Conference on Artificial
  Intelligence (AAAI~'23)}, pp. \bibinfo{pages}{5747--5754}.
\bibitem[{Li et~al.(2002)Li, Ma and Wang}]{LiMaWa2002}
\bibinfo{author}{Li, M.}, \bibinfo{author}{Ma, B.}, \bibinfo{author}{Wang, L.},
  \bibinfo{year}{2002}.
\newblock \bibinfo{title}{On the closest string and substring problems}.
\newblock \bibinfo{journal}{Journal of the ACM} \bibinfo{volume}{49},
  \bibinfo{pages}{157--171}.
\bibitem[{Limaye(2023)}]{limaye2023envy}
\bibinfo{author}{Limaye, G.}, \bibinfo{year}{2023}.
\newblock \bibinfo{title}{Envy-freeness and relaxed stability for lower-quotas:
  A parameterized perspective}.
\newblock \bibinfo{journal}{Discrete Applied Mathematics}
  \bibinfo{volume}{337}, \bibinfo{pages}{288--302}.
\bibitem[{Lindner and Rothe(2008)}]{LR08}
\bibinfo{author}{Lindner, C.}, \bibinfo{author}{Rothe, J.},
  \bibinfo{year}{2008}.
\newblock \bibinfo{title}{Fixed-parameter tractability and parameterized
  complexity applied to problems from computational social choice}.
\newblock \bibinfo{howpublished}{Supplement in the \emph{Mathematical
  Programming Glossary}}.
\bibitem[{Liu and Guo(2016)}]{LiuGuo2016}
\bibinfo{author}{Liu, H.}, \bibinfo{author}{Guo, J.}, \bibinfo{year}{2016}.
\newblock \bibinfo{title}{Parameterized complexity of winner determination in
  minimax committee elections}, in: \bibinfo{booktitle}{Proceedings of the 15th
  International Conference on Autonomous Agents and Multiagent Systems
  (AAMAS~16)}, \bibinfo{publisher}{{ACM}}. pp. \bibinfo{pages}{341--349}.
\bibitem[{Manlove(2013)}]{Manlove2013}
\bibinfo{author}{Manlove, D.F.}, \bibinfo{year}{2013}.
\newblock \bibinfo{title}{Algorithmics of Matching Under Preferences}.
  volume~\bibinfo{volume}{2} of \textit{\bibinfo{series}{Series on Theoretical
  Computer Science}}.
\newblock \bibinfo{publisher}{WorldScientific}.
\bibitem[{Marx and Schlotter(2010)}]{Marx10}
\bibinfo{author}{Marx, D.}, \bibinfo{author}{Schlotter, I.},
  \bibinfo{year}{2010}.
\newblock \bibinfo{title}{Parameterized complexity and local search approaches
  for the stable marriage problem with ties}.
\newblock \bibinfo{journal}{Algorithmica} \bibinfo{volume}{58},
  \bibinfo{pages}{170--187}.
\bibitem[{Marx and Schlotter(2011)}]{marx2011stable}
\bibinfo{author}{Marx, D.}, \bibinfo{author}{Schlotter, I.},
  \bibinfo{year}{2011}.
\newblock \bibinfo{title}{Stable assignment with couples: Parameterized
  complexity and local search}.
\newblock \bibinfo{journal}{Discrete Optimization} \bibinfo{volume}{8},
  \bibinfo{pages}{25--40}.
\bibitem[{Meeks and Rastegari(2020)}]{Meeks20}
\bibinfo{author}{Meeks, K.}, \bibinfo{author}{Rastegari, B.},
  \bibinfo{year}{2020}.
\newblock \bibinfo{title}{Solving hard stable matching problems involving
  groups of similar agents}.
\newblock \bibinfo{journal}{Theoretical Computer Science}
  \bibinfo{volume}{844}, \bibinfo{pages}{171--194}.
\bibitem[{Meir et~al.(2008)Meir, Procaccia, Rosenschein and
  Zohar}]{RPRZ2008MWManipulation}
\bibinfo{author}{Meir, R.}, \bibinfo{author}{Procaccia, A.D.},
  \bibinfo{author}{Rosenschein, J.S.}, \bibinfo{author}{Zohar, A.},
  \bibinfo{year}{2008}.
\newblock \bibinfo{title}{Complexity of strategic behavior in multi-winner
  elections}.
\newblock \bibinfo{journal}{Journal of Artificial Intelligence Research}
  \bibinfo{volume}{33}, \bibinfo{pages}{149--178}.
\bibitem[{Misra et~al.(2015)Misra, Nabeel and Singh}]{misra2015parameterized}
\bibinfo{author}{Misra, N.}, \bibinfo{author}{Nabeel, A.},
  \bibinfo{author}{Singh, H.}, \bibinfo{year}{2015}.
\newblock \bibinfo{title}{On the parameterized complexity of minimax approval
  voting}, in: \bibinfo{booktitle}{Proceedings of the 2015 International
  Conference on Autonomous Agents and Multiagent Systems}, pp.
  \bibinfo{pages}{97--105}.
\bibitem[{Misra et~al.(2017)Misra, Sonar and
  Vaidyanathan}]{MisraSonarVaid2017MW}
\bibinfo{author}{Misra, N.}, \bibinfo{author}{Sonar, C.},
  \bibinfo{author}{Vaidyanathan, P.R.}, \bibinfo{year}{2017}.
\newblock \bibinfo{title}{On the complexity of {C}hamberlin-{C}ourant on almost
  structured profiles}, in: \bibinfo{booktitle}{Proceedings of the 5th
  International Conference on Algorithmic Decision Theory (ADT '17)}, pp.
  \bibinfo{pages}{124--138}.
\bibitem[{Narayanaswamy and Subashini(2015)}]{NS15}
\bibinfo{author}{Narayanaswamy, N.S.}, \bibinfo{author}{Subashini, R.},
  \bibinfo{year}{2015}.
\newblock \bibinfo{title}{Obtaining matrices with the consecutive ones property
  by row deletions}.
\newblock \bibinfo{journal}{Algorithmica} \bibinfo{volume}{71},
  \bibinfo{pages}{758--773}.
\bibitem[{Niedermeier(2006)}]{Nie06}
\bibinfo{author}{Niedermeier, R.}, \bibinfo{year}{2006}.
\newblock \bibinfo{title}{Invitation to Fixed-Parameter Algorithms}.
\newblock \bibinfo{publisher}{Oxford University Press}.
\bibitem[{Olsen(2009)}]{Olsen09NSsymmetricadd}
\bibinfo{author}{Olsen, M.}, \bibinfo{year}{2009}.
\newblock \bibinfo{title}{Nash stability in additively separable hedonic games
  and community structures}.
\newblock \bibinfo{journal}{Theory of Computing Systems} ,
  \bibinfo{pages}{917--925}.
\bibitem[{Ota et~al.(2017)Ota, Barrot, Ismaili, Sakurai and
  Yokoo}]{ohta2017core}
\bibinfo{author}{Ota, K.}, \bibinfo{author}{Barrot, N.},
  \bibinfo{author}{Ismaili, A.}, \bibinfo{author}{Sakurai, Y.},
  \bibinfo{author}{Yokoo, M.}, \bibinfo{year}{2017}.
\newblock \bibinfo{title}{Core stability in hedonic games among friends and
  enemies: {I}mpact of neutrals}, in: \bibinfo{booktitle}{Proceedings of the
  26th International Joint Conference on Artificial Intelligence (IJCAI~'17)},
  pp. \bibinfo{pages}{359--365}.
\bibitem[{Peters(2016)}]{peters2016graphical}
\bibinfo{author}{Peters, D.}, \bibinfo{year}{2016}.
\newblock \bibinfo{title}{Graphical hedonic games of bounded treewidth}, in:
  \bibinfo{booktitle}{Proceedings of the 30th AAAI Conference on Artificial
  Intelligence (AAAI~'16)}, pp. \bibinfo{pages}{586--593}.
\bibitem[{Peters(2017)}]{peters2017precise}
\bibinfo{author}{Peters, D.}, \bibinfo{year}{2017}.
\newblock \bibinfo{title}{Precise complexity of the core in dichotomous and
  additive hedonic games}, in: \bibinfo{booktitle}{International Conference on
  Algorithmic Decision Theory}, \bibinfo{organization}{Springer}. pp.
  \bibinfo{pages}{214--227}.
\bibitem[{Peters and Lackner(2020)}]{PetersLackner20}
\bibinfo{author}{Peters, D.}, \bibinfo{author}{Lackner, M.},
  \bibinfo{year}{2020}.
\newblock \bibinfo{title}{Preferences single-peaked on a circle}.
\newblock \bibinfo{journal}{Journal of Artificial Intelligence Research}
  \bibinfo{volume}{68}, \bibinfo{pages}{463--502}.
\bibitem[{Procaccia et~al.(2008)Procaccia, Rosenschein and
  Zohar}]{procaccia2008complexity}
\bibinfo{author}{Procaccia, A.D.}, \bibinfo{author}{Rosenschein, J.S.},
  \bibinfo{author}{Zohar, A.}, \bibinfo{year}{2008}.
\newblock \bibinfo{title}{On the complexity of achieving proportional
  representation}.
\newblock \bibinfo{journal}{Social Choice and Welfare} \bibinfo{volume}{30},
  \bibinfo{pages}{353--362}.
\bibitem[{Rey et~al.(2016)Rey, Rothe, Schadrack and Schend}]{rey2016toward}
\bibinfo{author}{Rey, A.}, \bibinfo{author}{Rothe, J.},
  \bibinfo{author}{Schadrack, H.}, \bibinfo{author}{Schend, L.},
  \bibinfo{year}{2016}.
\newblock \bibinfo{title}{Toward the complexity of the existence of wonderfully
  stable partitions and strictly core stable coalition structures in
  enemy-oriented hedonic games}.
\newblock \bibinfo{journal}{Annals of mathematics and artificial intelligence}
  \bibinfo{volume}{77}, \bibinfo{pages}{317--333}.
\bibitem[{Ronn(1990)}]{Ronn1990}
\bibinfo{author}{Ronn, E.}, \bibinfo{year}{1990}.
\newblock \bibinfo{title}{{NP}-complete stable matching problems}.
\newblock \bibinfo{journal}{Journal of Algorithms} \bibinfo{volume}{11},
  \bibinfo{pages}{285--304}.
\bibitem[{Sivarajan(2018)}]{Sivarajan2018MAVpnorm}
\bibinfo{author}{Sivarajan, S.N.}, \bibinfo{year}{2018}.
\newblock \bibinfo{title}{A Generalization of the Minisum and Minimax Voting
  Methods}.
\newblock \bibinfo{type}{Technical Report}. arXiv:1611.01364v2 [cs.GT].
\bibitem[{Skowron et~al.(2015)Skowron, Yu, Faliszewski and Elkind}]{SYFE2015}
\bibinfo{author}{Skowron, P.}, \bibinfo{author}{Yu, L.},
  \bibinfo{author}{Faliszewski, P.}, \bibinfo{author}{Elkind, E.},
  \bibinfo{year}{2015}.
\newblock \bibinfo{title}{The complexity of fully proportional representation
  for single-crossing electorates}.
\newblock \bibinfo{journal}{Theoretical Computer Science}
  \bibinfo{volume}{569}, \bibinfo{pages}{43--57}.
\bibitem[{Skowron and Faliszewski(2015)}]{SF2015}
\bibinfo{author}{Skowron, P.K.}, \bibinfo{author}{Faliszewski, P.},
  \bibinfo{year}{2015}.
\newblock \bibinfo{title}{Fully proportional representation with approval
  ballots: {A}pproximating the maxcover problem with bounded frequencies in
  {FPT} time}, in: \bibinfo{booktitle}{Proceedings of the 29th AAAI Conference
  on Artificial Intelligence (AAAI~'15)}, pp. \bibinfo{pages}{2124--2130}.
\bibitem[{Sornat et~al.(2022)Sornat, Williams and Xu}]{Sornat2022}
\bibinfo{author}{Sornat, K.}, \bibinfo{author}{Williams, V.V.},
  \bibinfo{author}{Xu, Y.}, \bibinfo{year}{2022}.
\newblock \bibinfo{title}{Near-tight algorithms for the chamberlin-courant and
  thiele voting rules}, in: \bibinfo{booktitle}{Proceedings of the 31st
  International Joint Conference on Artificial Intelligence (IJCAI~'22)}, pp.
  \bibinfo{pages}{482--488}.
\bibitem[{Sung and Dimitrov(2007a)}]{sung2007core}
\bibinfo{author}{Sung, S.C.}, \bibinfo{author}{Dimitrov, D.},
  \bibinfo{year}{2007}a.
\newblock \bibinfo{title}{On core membership testing for hedonic coalition
  formation games}.
\newblock \bibinfo{journal}{Operations Research Letters} \bibinfo{volume}{35},
  \bibinfo{pages}{155--158}.
\bibitem[{Sung and Dimitrov(2007b)}]{sung2007myopic}
\bibinfo{author}{Sung, S.C.}, \bibinfo{author}{Dimitrov, D.},
  \bibinfo{year}{2007}b.
\newblock \bibinfo{title}{On myopic stability concepts for hedonic games}.
\newblock \bibinfo{journal}{Theory and Decision} \bibinfo{volume}{62},
  \bibinfo{pages}{31--45}.
\bibitem[{Sung and Dimitrov(2010)}]{sung2010computational}
\bibinfo{author}{Sung, S.C.}, \bibinfo{author}{Dimitrov, D.},
  \bibinfo{year}{2010}.
\newblock \bibinfo{title}{Computational complexity in additive hedonic games}.
\newblock \bibinfo{journal}{European Journal of Operational Research}
  \bibinfo{volume}{203}, \bibinfo{pages}{635--639}.
\bibitem[{Woeginger(2013a)}]{woeginger2013core}
\bibinfo{author}{Woeginger, G.J.}, \bibinfo{year}{2013}a.
\newblock \bibinfo{title}{Core stability in hedonic coalition formation}, in:
  \bibinfo{booktitle}{Proceedings of the 39th International Conference on
  Current Trends in Theory and Practice of Computer Science (SOFSEM~13)}, pp.
  \bibinfo{pages}{33--50}.
\bibitem[{Woeginger(2013b)}]{WOEGINGER2013101}
\bibinfo{author}{Woeginger, G.J.}, \bibinfo{year}{2013}b.
\newblock \bibinfo{title}{A hardness result for core stability in additive
  hedonic games}.
\newblock \bibinfo{journal}{Mathematical Social Sciences} \bibinfo{volume}{65},
  \bibinfo{pages}{101--104}.
\bibitem[{Yang(2019)}]{YangMWManipulationControl2019}
\bibinfo{author}{Yang, Y.}, \bibinfo{year}{2019}.
\newblock \bibinfo{title}{Complexity of manipulating and controlling
  approval-based multiwinner voting}, in: \bibinfo{booktitle}{Proceedings of
  the 28th International Joint Conference on Artificial Intelligence
  (IJCAI~'19)}, pp. \bibinfo{pages}{637--643}.
\bibitem[{Yang(2020)}]{Yang2020MWDestructiveBribery}
\bibinfo{author}{Yang, Y.}, \bibinfo{year}{2020}.
\newblock \bibinfo{title}{On the complexity of destructive bribery in
  approval-based multi-winner voting}, in: \bibinfo{booktitle}{Proceedings of
  the 19th International Conference on Autonomous Agents and Multiagent Systems
  (AAMAS 20)}, pp. \bibinfo{pages}{1584--1592}.
\bibitem[{Yang(2023)}]{tech-Yang2023DControl}
\bibinfo{author}{Yang, Y.}, \bibinfo{year}{2023}.
\newblock \bibinfo{title}{On the Parameterized Complexity of Controlling
  Approval-Based Multiwinner Voting: {D}estructive Model {\&} Sequential
  Rules}.
\newblock \bibinfo{type}{Technical Report}. arXiv:2304.11927v2.
\bibitem[{Yang and Wang(2018)}]{Yang18}
\bibinfo{author}{Yang, Y.}, \bibinfo{author}{Wang, J.}, \bibinfo{year}{2018}.
\newblock \bibinfo{title}{Multiwinner voting with restricted admissible sets:
  Complexity and strategyproofness}, in: \bibinfo{booktitle}{Proceedings of the
  27th International Joint Conference on Artificial Intelligence (IJCAI~'18)},
  pp. \bibinfo{pages}{576--582}.
\bibitem[{Yang and Wang(2023)}]{yang2023parameterized}
\bibinfo{author}{Yang, Y.}, \bibinfo{author}{Wang, J.}, \bibinfo{year}{2023}.
\newblock \bibinfo{title}{Parameterized complexity of multiwinner
  determination: {M}ore effort towards fixed-parameter tractability}.
\newblock \bibinfo{journal}{Autonomous Agents and Multi-Agent Systems}
  \bibinfo{volume}{37}, \bibinfo{pages}{28}.

\end{thebibliography}

\end{document}